\documentclass[aps,prd,twocolumn,showkeys,amsmath,amssymb]{revtex4}
\usepackage[hypertexnames=false,colorlinks,citecolor=blue,anchorcolor=red,menucolor=red,linkcolor=red,filecolor=red,runcolor=red,urlcolor=blue]{hyperref}
\usepackage{graphicx}
\usepackage{subfigure}
\usepackage{epstopdf}
\usepackage{multirow}
\usepackage{enumitem}
\usepackage[vcentermath]{youngtab}
\usepackage{feynmf}
\usepackage{color}
\usepackage{xcolor}
\allowdisplaybreaks[3]

\newcommand{\bb}{\colorbox{black}{\phantom{J}}}

\newcommand{\sn}{0.5}

\begin{document}
\title{Hadrons in group expansion}
\author{Hua-Xing Chen}
\email{hxchen@seu.edu.cn}
\affiliation{School of Physics, Southeast University, Nanjing 210094, China}

\begin{abstract}
We develop a general framework for describing
symmetry-breaking responses in physical systems
with approximate symmetries.
Hadronic systems provide a particularly rich
setting for investigating such symmetries.
For example, the flavor $SU(4)$ symmetry involving
the $u$, $d$, $s$, and $c$ quarks is severely broken,
the flavor $SU(3)$ symmetry involving the
$u$, $d$, and $s$ quarks is moderately broken,
and the isospin $SU(2)$ symmetry involving the
$u$ and $d$ quarks is only slightly broken.
An important question is how a physical system
responds to further symmetry breaking when
its symmetry has already been partially broken.
Motivated by the observed mass nonadditivity
of vector mesons, we formulate this framework
in terms of the group expansion, in which
higher-order responses may modify the coefficients
within the inherited response space or unlock
additional independent response directions
outside this space.
We refer to these two effects as
coefficient evolution and
response unlocking, respectively,
and develop a differential description connecting
the interaction flow, mass flow,
and state flow.
We apply this framework to $J^P=1^-$ vector mesons
and $J^P=3/2^+$ and $1/2^+$ baryons within
flavor $SU(4)$ and $SU(5)$ reference spaces.
These analyses illustrate how the internal
structure and dynamics of a physical system,
together with additional symmetries, can
shape and constrain its symmetry-breaking
responses.
The group expansion thus provides a unified
framework for organizing the evolution,
correlations, and constraints of symmetry-breaking
responses, with potential applications beyond
hadronic systems.
\end{abstract}

\date{\today}
\maketitle

\tableofcontents

\section{Introduction}
\label{sec:intro}

Hadrons are composite particles whose internal structure
and interactions are governed by quantum chromodynamics
(QCD).
In the traditional quark model proposed by
Gell-Mann and Zweig~\cite{Gell-Mann:1964ewy,Zweig:1964ruk},
a conventional meson consists of a quark and an antiquark,
while a conventional baryon consists of three quarks.
Despite its conceptual simplicity, this model provides
a remarkably successful organization of the observed
hadron spectrum~\cite{pdg}.

\begin{figure}[h]
\centering
\includegraphics[width=\columnwidth]{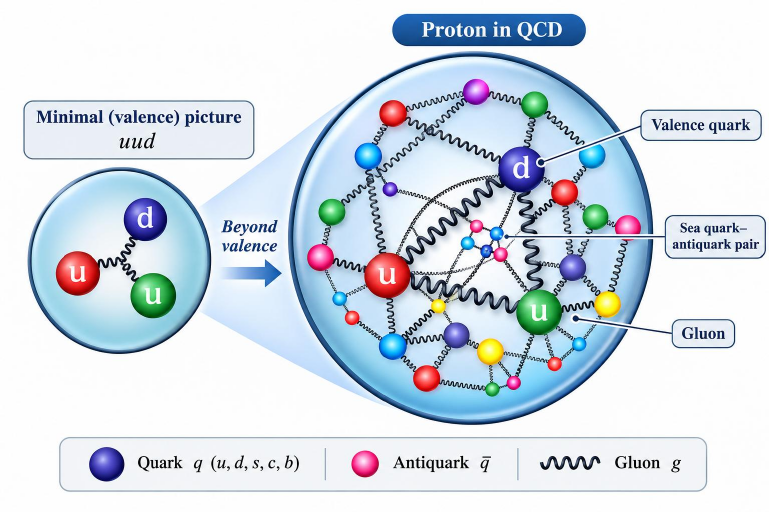}
\caption{
Schematic illustration of the proton in the minimal
valence-quark picture and in a more general QCD picture
involving sea quark--antiquark pairs and gluons.
}
\label{fig:proton}
\end{figure}

Taking the proton as an example, its minimal
valence-quark configuration can be represented
schematically as
\begin{equation}
|p\rangle_{\rm valence}
=
|uud\rangle \, .
\end{equation}
However, the so-called
``proton spin crisis''~\cite{Cheng:1996jr,
Filippone:2001ux,Bass:2004xa,Aidala:2012mv,
Leader:2013jra,Ji:2016djn,Deur:2018roz,
Ji:2020ena,Liu:2021lke},
brought to prominence by early measurements of
the proton spin structure~\cite{EuropeanMuon:1987isl,
EuropeanMuon:1989yki,Hughes:1983kf},
illustrates the limitations of the simplest
valence-quark picture.
The internal structure of the proton also involves
sea quark--antiquark pairs and gluons, as schematically
illustrated in Fig.~\ref{fig:proton}.
Its Fock-space expansion can be expressed symbolically as
\begin{equation}
|p\rangle
=
a_0|uud\rangle
+
a_1|uud q\bar q\rangle
+
a_2|uud g\rangle
+\cdots \, ,
\end{equation}
where the detailed momentum, spin, and color
structures are suppressed.

Why, then, is the traditional quark model so simple
yet still remarkably successful?
One important reason is the role of symmetry.
Under flavor $SU(3)$, the flavor spaces of three
light quarks and of a light quark--antiquark pair
can be decomposed as
\begin{eqnarray}
\mathbf{3}\otimes\mathbf{3}\otimes\mathbf{3}
&=&
\mathbf{10}\oplus\mathbf{8}\oplus\mathbf{8}
\oplus\mathbf{1} \, ,
\nonumber\\
\mathbf{3}\otimes\overline{\mathbf{3}}
&=&
\mathbf{1}\oplus\mathbf{8} \, .
\end{eqnarray}
Accordingly, the light-flavor content of the baryonic
Fock space can be schematically organized as
\begin{equation}
{\cal H}_F
\sim
\bigoplus_{N=0}^{\infty}
\left[
\left(
\mathbf{10}\oplus\mathbf{8}
\oplus\mathbf{8}\oplus\mathbf{1}
\right)
\otimes
\left(
\mathbf{1}\oplus\mathbf{8}
\right)^{\otimes N}
\right]
\, ,
\end{equation}
where $N$ denotes the number of additional light
quark--antiquark pairs, with further restrictions
on the physical configurations imposed by the Pauli
principle and other conserved quantum numbers.
Since both $\mathbf{8}\otimes\mathbf{1}$ and
$\mathbf{8}\otimes\mathbf{8}$ contain octet components,
the proton can transform as a flavor octet in the
exact $SU(3)$ limit despite its complicated internal
structure.
Symmetry thus allows us to organize a complex
hadronic system within a relatively simple
reference space.

In nature, however, flavor symmetry is only approximate.
The isospin $SU(2)$ symmetry is weakly broken,
the light-flavor $SU(3)$ symmetry is more substantially
broken, and its extensions to charm and bottom flavors
are strongly broken.
As flavor symmetry is broken, different reference
representations may become mixed, while the effective
interactions within and between them may also evolve.
Flavor-symmetry-breaking effects have been extensively
investigated in hadron physics~\cite{
Zeppenfeld:1980ex,Gasser:1982ap,Chau:1986jb,
Li:1986iya,Chau:1987tk,Savage:1989ub,Jain:1989kn,
Morpurgo:1989ti,Chau:1990ay,Grozin:1992td,
Praszalowicz:1992gn,Adami:1993xz,Deshpande:1994ii,
Gronau:1995hm,Forkel:1996ty,Kim:1997ip,Zhu:1998ai,
He:1998rq,Yagisawa:2001gz,Durand:2005tb,
Aliev:2010ra,Yang:2010fm,Yang:2010id,Shanahan:2012wa,
Walker-Loud:2012ift,Shanahan:2013xw,Cheng:2014rfa,
Horsley:2014koa,Muller:2015lua,Lu:2016ogy,
Wang:2017azm,Shi:2017dto,He:2018php,Geng:2020tlx,
Matsui:2020wcc,Braghin:2020yri,Braghin:2021hmr,
He:2021qnc,Geng:2022yxb,Geng:2022xfz,Ke:2022gxm,
Xing:2022phq,Deng:2023qaf,Sun:2023noo,
Liu:2023pwr,Liu:2023feb},
including extensions to flavor $SU(4)$~\cite{
Ebrahim:1977iy,Ebrahim:1977hb,Inose:1978qw,
Hallock:1979ar,Ebert:1979ba,Burkitt:1983ca,
Suzuki:1992pa,El-Bennich:2011tme,Fontoura:2017ujf}.
Complementary descriptions and constraints are
provided by the general QCD parametrization,
baryon mass relations, the large-$N_c$
spin--flavor operator expansion~\cite{
Dillon:2009pf,Franklin:1975yu,Dashen:1994qi,
Jenkins:1995td,Jenkins:1998wy},
chiral symmetry~\cite{
Weinberg:1969hw,Ioffe:1981kw,Chung:1981cc,
Espriu:1983hu,Detar:1987kae,Detar:1988kn,
Weinberg:1990xn,Nowak:1992um,Bardeen:1993ae,
Hatsuda:1994pi,Leinweber:1994nm,Glozman:1995fu,
Cohen:1996sb,Jido:2001nt,Cohen:2002st,
Harada:2003jx,Nagata:2007di,Gallas:2009qp,
Dmitrasinovic:2012zz,Yoshida:2015tia,Aarts:2017rrl,
Ma:2017nik,Yamazaki:2018stk,Kawakami:2019hpp,
Suenaga:2021qri},
and heavy-quark symmetry~\cite{
Isgur:1989vq,Georgi:1990um,Isgur:1991wq,
Neubert:1993mb,Korner:1994nh,Jenkins:1996de}.
We have also investigated symmetry properties
and representation mixing in light and heavy
baryons in Refs.~\cite{
Chen:2008qv,Chen:2009sf,Chen:2010ba,Chen:2011rh,
Chen:2013aga,Dmitrasinovic:2016hup,
Dmitrasinovic:2020wye}.

An important question in this context is how
a hadronic system responds to further symmetry
breaking when its symmetry has already been
partially broken.
In particular, the effective response to one
symmetry-breaking direction may depend on the
breaking already present in another.
A simple example is provided by the ground-state
vector mesons.
If the mass increment associated with replacing
a light constituent by a charm constituent were
independent of the flavor of the other constituent,
one would obtain
\begin{equation}
M_{B_c^*}-M_{B^*}
=
M_{D^*}-M_\rho \, .
\end{equation}
However, using the experimental mass inputs adopted
in this work, including the reported $B_c^*$
mass~\cite{ATLAS:2026ubk}, we find
\begin{equation}
M_{B_c^*}-M_{B^*}-M_{D^*}+M_\rho
\simeq -219~{\rm MeV} \, .
\end{equation}
This substantial nonadditivity motivates us to
investigate not only individual symmetry-breaking
responses but also their evolution and correlations
as the symmetry is progressively broken.

To organize these effects, we develop a framework
that we refer to as the \emph{group expansion}.
The basic idea is to distinguish the system under
consideration, represented by a symmetry-adapted
reference space, from the degrees of freedom outside
this space.
Rather than describing all these degrees of freedom
explicitly, we incorporate their effects into the
effective Hamiltonian of the system and organize
their influence on symmetry breaking in terms of
effective responses.
We then investigate how these responses evolve
as the symmetry is progressively broken.

Starting from a specified symmetric reference,
we expand the effective Hamiltonian in the
symmetry-breaking coordinates $\boldsymbol{\xi}$:
\begin{equation}
{\cal H}(\boldsymbol{\xi})
=
{\cal H}^{(0)}
+
{\cal H}^{(1)}(\boldsymbol{\xi})
+
{\cal H}^{(2)}(\boldsymbol{\xi})
+\cdots \, .
\end{equation}
The parent symmetry organizes the allowed response
structures at each order.
In an additional symmetry-breaking background,
higher-order terms may modify the coefficients
within the inherited response space or unlock
independent response directions outside this space.
We refer to these two effects as \emph{coefficient
evolution} and \emph{response unlocking},
respectively.
They can be characterized locally through the
response operators
\begin{equation}
B_a=\partial_a{\cal H},
\qquad
K_{ab}=\partial_b B_a \, ,
\end{equation}
which describe the response to a symmetry-breaking
coordinate and its variation along another.
Within the usual symmetry-based expansions,
the group expansion thus emphasizes the evolution
of response spaces and their dynamical realizations
in different symmetry-breaking backgrounds.
We refer to the evolution of the effective Hamiltonian
and its response operators as the \emph{interaction
flow}, which in turn generates the \emph{mass flow}
and \emph{state flow} within the effective description.
The latter two describe the evolution of mass
eigenvalues and the reorganization of eigenstates,
respectively, including possible mixing between
different flavor representations.

In this work, we apply the group expansion to
$J^P=1^-$ vector mesons and ground-state
$J^P=3/2^+$ and $1/2^+$ baryons within flavor
$SU(4)$ and $SU(5)$ reference spaces.
For the fully symmetric and mixed-symmetry three-quark
flavor representations, we construct the complete
homogeneous flavor-response spaces through second order,
investigate their dynamical realization through common
spin--flavor operators and spectator-dependent
correlations, and examine the constraints imposed
by heavy-quark spin symmetry (HQSS).
Throughout this work, we primarily consider the
isospin limit and the breaking of vector flavor
symmetry associated with quark-mass differences.
The flavor representations serve as reference
spaces for effective Hamiltonians rather than
complete descriptions of physical hadron wave
functions.
Given the strong breaking of flavor $SU(4)$ and
$SU(5)$, the selected low-order Hamiltonians
are phenomenological realizations of the group
expansion rather than controlled perturbative
approximations to the full QCD dynamics.
Accordingly, the effective coefficients extracted
at finite symmetry breaking need not coincide with
local Taylor coefficients at the exact symmetry
point.

This paper is organized as follows.
In Sec.~\ref{sec:meson}, we investigate the flavor
responses and state mixing of vector mesons.
In Sec.~\ref{sec:group-expansion}, we develop
the group-expansion framework.
In Secs.~\ref{sec:20S} and \ref{sec:20M}, we study
the fully symmetric and mixed-symmetry baryon
representations, respectively.
In Sec.~\ref{sec:shared-dynamics}, we investigate
their common spin--flavor dynamics and the constraints
imposed by HQSS.
Finally, we summarize and discuss our results in
Sec.~\ref{sec:summary}.
The flavor tensors and transition matrices are
collected in Appendix~\ref{app:baryon-S-matrices},
while a complementary graphical description
of the state flow using Young tableaux is
presented in Appendix~\ref{sec:young-breaking}.

\section{Vector mesons}
\label{sec:meson}

We begin our investigation of flavor-symmetry breaking with the
ground-state vector mesons. Compared with baryons, the
reference quark--antiquark flavor space has a considerably simpler group
structure, consisting of only a singlet and an adjoint
representation. Moreover, the observed vector-meson spectrum
extends from the light-flavor sector to the charm and bottom
sectors, providing a particularly suitable laboratory for
investigating different directions of flavor-symmetry breaking.

In this section, we first construct the leading
flavor-symmetry-breaking mass Hamiltonian and examine its
realization in flavor $SU(4)$ and $SU(5)$. We then assess
how well this Hamiltonian describes the observed mass
spectra and the associated flavor mixing, and identify the
nonadditive mass contributions that cannot be accommodated
within its leading group-theoretical structure.

\subsection{Leading symmetry-breaking Hamiltonian}
\label{sec:meson-leading}

For a general flavor $SU(N_f)$ group, we introduce the
quark--antiquark flavor components
\begin{equation}
M^N
=
\left[\lambda^N_{(N_f)}\right]_{AB}
\bar q^A q^B,
\label{eq:meson-general-def}
\end{equation}
where the flavor indices run over $A,B=1,\ldots,N_f$,
and $N=0,\ldots,N_f^2-1$.
Here and throughout this section, the Lorentz and spin
structures are suppressed to make the flavor structure explicit.

We introduce
\begin{equation}
\lambda^0_{(N_f)}
=
\sqrt{\frac{2}{N_f}}\,
{\bf 1}_{N_f\times N_f}
\label{eq:meson-lambda0}
\end{equation}
to describe the flavor-singlet direction, while
\begin{equation}
\lambda^N_{(N_f)},
\qquad
N=1\cdots N_f^2-1,
\end{equation}
denote the traceless Hermitian generators of $SU(N_f)$.
We adopt the normalization
\begin{equation}
{\rm Tr}
\left[
\lambda^N_{(N_f)}
\lambda^{N'}_{(N_f)}
\right]
=
2\delta^{NN'},
\label{eq:meson-generator-normalization}
\end{equation}
where $N,N'=0\cdots N_f^2-1$.
Accordingly, the quark--antiquark flavor space is decomposed as
\begin{equation}
\mathbf{N_f}\otimes\overline{\mathbf{N_f}}
=
\mathbf{1}\oplus\mathbf{Adj},
\end{equation}
where $M^0$ denotes the flavor-singlet component, whereas
$M^{N=1\cdots N_f^2-1}$ form the adjoint representation.
In particular, we have
\begin{eqnarray}
SU(4) &:&
\mathbf{4}\otimes\mathbf{\bar 4}
=
\mathbf{1}\oplus\mathbf{15},
\nonumber\\
SU(5) &:&
\mathbf{5}\otimes\mathbf{\bar 5}
=
\mathbf{1}\oplus\mathbf{24}.
\end{eqnarray}

The leading flavor-invariant interactions relevant to the
present analysis take the same general form for $SU(4)$
and $SU(5)$. Here and below, we consider schematic
flavor-invariant contractions rather than explicit
interaction vertices among physical vector mesons.
In constructing the mass Hamiltonian, two field insertions
represent vector-meson fields, while the remaining
insertions represent scalar backgrounds.
In terms of the singlet and adjoint flavor components,
the symmetric cubic flavor structures can be written as
\begin{eqnarray}
{\cal L}_{\rm meson}
&=&
{g_{SSS}\over 3!}\,M^0 M^0 M^0
\nonumber\\
&+&
{g_{SOO}\over 2}\,M^0 M^N M^N
\nonumber\\
&+&
{g_{OOO}\over 3!}\,d_{NPQ}M^N M^P M^Q,
\label{eq:meson-general-L3}
\end{eqnarray}
where $N,P,Q=1\cdots N_f^2-1$ denote adjoint indices,
repeated indices are summed over, and $d_{NPQ}$ is the
totally symmetric invariant tensor of $SU(N_f)$, normalized as
$d_{NPQ}={1\over4}{\rm Tr}(\{\lambda^N,\lambda^P\}\lambda^Q)$.
In general, an antisymmetric $F$-type flavor contraction
proportional to the structure constants $f_{NPQ}$ may also
occur when the participating fields are distinguished by
their Lorentz or derivative structures. Its flavor structure
may be represented schematically as
\begin{equation}
{\cal L}^{\prime}_{\rm meson}
=
i g^\prime_{OOO}\,
f_{NPQ}\,
M_1^N M_2^P M_3^Q,
\label{eq:meson-general-L3F}
\end{equation}
where the subscripts $1,2,3$ distinguish the corresponding
field insertions and do not denote different flavor multiplets.
Such an antisymmetric flavor contraction does not contribute
to the real symmetric quadratic mass operator considered
below. We therefore concentrate on the symmetric flavor
structures in Eq.~(\ref{eq:meson-general-L3}).

A simple realization of flavor-symmetry breaking can be
obtained by introducing Lorentz-scalar background insertions
with nonzero singlet and diagonal adjoint flavor components,
\begin{equation}
\langle M^0\rangle,
\qquad
\langle M^C\rangle,
\label{eq:meson-condensates}
\end{equation}
where $C$ labels the diagonal symmetry-breaking directions,
\begin{eqnarray}
SU(4) &:& C=3,8,15,
\nonumber\\
SU(5) &:& C=3,8,15,24.
\end{eqnarray}
For notational convenience, we use the same flavor
labels for the meson fields and the scalar background
components.
The latter should not be interpreted as vacuum expectation
values of the vector-meson fields.
The cubic and higher-order interaction terms considered
here specify schematic flavor-invariant structures
with the corresponding scalar background insertions.

The singlet background preserves the original flavor symmetry,
whereas a nonzero adjoint background selects a particular
direction in flavor space and generally reduces the symmetry
of the effective interactions.
We collect the diagonal adjoint background
components into a traceless Hermitian matrix,
\begin{equation}
X(\boldsymbol{\xi})
=
\sum_C \xi_C\lambda^C_{(N_f)},
\label{eq:meson-breaking-source}
\end{equation}
where $\boldsymbol{\xi}$ collectively denotes the
symmetry-breaking coordinates, and each component
$\xi_C$ characterizes the strength of the effective
symmetry-breaking source along the direction
$\lambda^C_{(N_f)}$.
In the present construction, $\xi_C$ is proportional to
$\langle M^C\rangle$ with a common normalization,
but need not coincide with the corresponding
component of the current-quark mass matrix.

Replacing one of the three field insertions in
Eq.~(\ref{eq:meson-general-L3}) by the corresponding
background component generates quadratic flavor structures.
Retaining the leading terms, we construct a phenomenological
linear mass Hamiltonian,
\begin{eqnarray}
{\cal H}_{\rm meson}
&=&
{1\over2}m_{\mathbf S}\,M^0 M^0
+
{1\over2}m_{\mathbf O}\,M^N M^N
\nonumber\\
&+&
{1\over2}D^C d_{CPQ}M^P M^Q
+
G^C M^0 M^C,
\label{eq:meson-general-H}
\end{eqnarray}
where repeated adjoint indices are summed over, with $C$
restricted to the diagonal symmetry-breaking directions.
The eigenvalues of the corresponding mass matrix are
identified directly with the physical meson masses.

The parameters $m_{\mathbf S}$ and $m_{\mathbf O}$ characterize
the singlet and adjoint masses in the flavor-symmetric
background, respectively. They are independent parameters
and need not be equal even in the exact flavor-symmetry limit.
The coefficients $D^C$ and $G^C$ characterize two different
responses to the same symmetry-breaking background:
the former acts within the adjoint representation, whereas
the latter induces singlet--adjoint transitions.
Within the minimal background-insertion construction
of Eq.~(\ref{eq:meson-general-L3}), the background-induced
mass contributions take the form
\begin{eqnarray}
m_{\mathbf S}
&=&
g_{SSS}\langle M^0\rangle,
\nonumber\\
m_{\mathbf O}
&=&
g_{SOO}\langle M^0\rangle,
\nonumber\\
G^C
&=&
g_{SOO}\langle M^C\rangle,
\nonumber\\
D^C
&=&
g_{OOO}\langle M^C\rangle.
\label{eq:meson-general-parameters}
\end{eqnarray}
More general flavor-symmetric contributions can be
absorbed into $m_{\mathbf S}$ and $m_{\mathbf O}$,
which are treated as independent parameters throughout
the following analysis. The common use of $g_{SOO}$
in the two background insertions is specific to the
schematic cubic construction.

It follows that, within this leading common-source
construction, the singlet--adjoint transition and the
adjoint $D$-type response induced by the same background
direction satisfy
\begin{equation}
{G^C\over D^C}
=
{g_{SOO}\over g_{OOO}}
\equiv r,
\label{eq:meson-general-r}
\end{equation}
which is independent of the particular symmetry-breaking
direction $C$.
In the isospin-symmetric limit, where the $C=3$ component
is absent, this relation becomes
\begin{eqnarray}
SU(4) &:&
{G^8\over D^8}
=
{G^{15}\over D^{15}}
=
r,
\nonumber\\
SU(5) &:&
{G^8\over D^8}
=
{G^{15}\over D^{15}}
=
{G^{24}\over D^{24}}
=
r.
\end{eqnarray}
This common-ratio relation provides a minimal constraint
on the leading symmetry-breaking responses and will be
used in the vector-meson mass analyses below.

In addition, one may compare the responses along two
different symmetry-breaking directions,
\begin{equation}
{G^C\over G^{C^\prime}}
=
{D^C\over D^{C^\prime}}
=
{\langle M^C\rangle\over\langle M^{C^\prime}\rangle}
\equiv
R_{C/C^\prime}.
\label{eq:meson-general-breaking-ratio}
\end{equation}
Unlike the common ratio $r$, which compares two different
types of responses to the same symmetry-breaking source,
$R_{C/C^\prime}$ compares the responses induced by two
different symmetry-breaking directions.
For example, in flavor $SU(4)$, the ratio
\begin{equation}
R_{15/8}
=
{D^{15}\over D^8}
=
{G^{15}\over G^8}
=
{\langle M^{15}\rangle\over\langle M^8\rangle}
\end{equation}
characterizes the relative strengths of the leading
responses along the $\lambda_{15}$ and $\lambda_8$
directions.
Similarly, in flavor $SU(5)$, one may introduce
$R_{15/8}$ and $R_{24/15}$, with
\begin{equation}
R_{24/8}=R_{24/15}\times R_{15/8}.
\end{equation}
These relations illustrate an important distinction between
symmetry-breaking sources and the responses induced by them.
Group theory determines the allowed flavor structures of
the responses, whereas their actual strengths depend on
both the background components and the corresponding
reduced couplings.
The above ratio relations apply for nonzero denominators
and follow from the leading common-source construction.
These relations are preserved for the linear coefficients
in the restricted common-source models considered below.
More general higher-order effective responses, however,
need not satisfy the same proportionality constraints.

Indeed, the three-meson interactions in
Eq.~(\ref{eq:meson-general-L3}) represent only the leading
flavor structures retained in the present analysis and do
not exhaust the most general effective interactions allowed
by the flavor symmetry.
At higher orders in the field expansion, additional
flavor-invariant interactions can be constructed from four
or more field insertions. For example, a generic
four-adjoint interaction may be written as
\begin{equation}
{\cal L}^{(4)}_{\rm meson}
=
g\,
{\cal T}_{NPQR}\,
M_1^N M_2^P M_3^Q M_4^R,
\label{eq:meson-general-L4}
\end{equation}
where ${\cal T}_{NPQR}$ represents a rank-four invariant
tensor allowed by the corresponding flavor symmetry.
Replacing one of the field insertions by the corresponding
background component generates an effective three-meson
interaction,
\begin{eqnarray}
{\cal L}^{(4)}_{\rm meson}
&\longrightarrow&
g\,
{\cal T}_{NPQR}\,
\langle M_1^N\rangle
M_2^P M_3^Q M_4^R
+\cdots
\nonumber\\
&\longrightarrow&
g\,
{\cal T}_{PQR,{\rm eff}}\,
M_2^P M_3^Q M_4^R
+\cdots,
\label{eq:meson-L4-to-L3}
\end{eqnarray}
where the ellipses denote other possible background
insertions, and the induced rank-three tensor depends
explicitly on the background,
\begin{equation}
{\cal T}_{PQR,{\rm eff}}
=
{\cal T}_{NPQR}\langle M_1^N\rangle.
\end{equation}
Although the original rank-four tensor is invariant under
the parent flavor group, the induced rank-three tensor
need no longer preserve the full parent symmetry.
It remains invariant under the subgroup that preserves
the inserted background.

A further background insertion generates additional
contributions to the quadratic mass Hamiltonian.
The hierarchy can therefore be represented schematically as
\begin{equation}
{\cal L}^{(4)}_{\rm meson}
\longrightarrow
{\cal L}^{(3)}_{\rm meson,eff}
\longrightarrow
{\cal H}^{(2)}_{\rm meson}.
\label{eq:meson-hierarchy}
\end{equation}
Such contributions may involve products of two different
symmetry-breaking background components, schematically,
\begin{equation}
\delta{\cal H}_{\rm meson}
\sim
g\,{\cal T}_{NPQR}\,
\langle M_1^N\rangle
\langle M_2^P\rangle
M_3^QM_4^R.
\label{eq:meson-quadratic-response}
\end{equation}
Consequently, higher-order flavor invariants may either
modify the effective couplings already present in
Eq.~(\ref{eq:meson-general-H}) or generate additional
independent structures in the mass Hamiltonian.

The above construction illustrates a general physical
feature of symmetry breaking. When a symmetry is broken,
not only the masses and compositions of the physical
states but also the effective interactions governing
these states may change.
Consequently, the response of a physical system to one
symmetry-breaking source may depend on the presence of
other symmetry-breaking sources or on additional
symmetry-preserving backgrounds.
In particular, a background that breaks a parent symmetry
may preserve one of its subgroups.
For example, the $\lambda_{15}$ direction in flavor $SU(4)$
breaks the parent symmetry while preserving the light-flavor
$SU(3)$ subgroup.
Although the light-flavor $SU(3)$ symmetry remains exact
in this background, the corresponding effective interactions
need not remain unchanged.
When the $\lambda_8$ source is further introduced, the
response to light-flavor symmetry breaking may therefore
depend on the background established by the
$\lambda_{15}$ source.
Within the leading construction, the contributions of
different symmetry-breaking directions to the mass
Hamiltonian are additive, as described by
Eq.~(\ref{eq:meson-general-H}).
Higher-order interactions, however, allow correlations
between these directions and thereby provide a natural
mechanism for nonadditive symmetry-breaking responses.

In the following subsections, we examine the leading
mass Hamiltonian in flavor $SU(4)$ and $SU(5)$.
The resulting mass spectra, flavor mixing, and
parameter-independent mass relations will allow us to
identify both the success and the limitations of this
leading construction.

\subsection{Vector mesons in flavor $SU(4)$}
\label{sec:meson4}

We first apply the leading mass Hamiltonian introduced above
to flavor $SU(4)_Q$, where $Q=c$ or $b$ denotes a single
heavy flavor in addition to the three light flavors
$u$, $d$, and $s$.
For notational convenience, from here on we use subscripts
instead of superscripts for the flavor indices, such as
$\lambda_C$, $M_C$, $D_C$, and $G_C$.
Throughout the following analysis, we work in the
isospin-symmetric limit, so that
\begin{equation}
\langle M_3\rangle=0,
\qquad
D_3=G_3=0.
\end{equation}
The remaining two diagonal symmetry-breaking directions
are characterized by the generators
\begin{eqnarray}
\lambda_8
&=&
\frac{1}{\sqrt3}
{\rm diag}(1,1,-2,0),
\nonumber\\
\lambda_{15}
&=&
\frac{1}{\sqrt6}
{\rm diag}(1,1,1,-3).
\end{eqnarray}
The $\lambda_{15}$ direction breaks flavor $SU(4)_Q$
while preserving light-flavor $SU(3)$, whereas the
$\lambda_8$ direction further breaks light-flavor
$SU(3)$ while preserving isospin symmetry.
Within the leading common-source construction,
Eq.~(\ref{eq:meson-general-r}) gives
\begin{equation}
\frac{G_8}{D_8}
=
\frac{G_{15}}{D_{15}}
=
r.
\label{eq:meson-SU4-r}
\end{equation}

For the neutral isoscalar vector mesons, we work in the
flavor-$SU(4)_Q$ basis
\begin{equation}
\left(
M_0,M_8,M_{15}
\right).
\end{equation}
The corresponding matrix representation of the linear
mass Hamiltonian in Eq.~(\ref{eq:meson-general-H}) is
\begin{equation}
{\cal H}_{0,8,15}^{SU(4)_Q}
=
\left(
\begin{array}{ccc}
m_{\mathbf S}
&
rD_8
&
rD_{15}
\\[1mm]
rD_8
&
m_{\mathbf O}
-\frac{D_8}{\sqrt3}
+\frac{D_{15}}{\sqrt6}
&
\frac{D_8}{\sqrt6}
\\[1mm]
rD_{15}
&
\frac{D_8}{\sqrt6}
&
m_{\mathbf O}
-\frac{2D_{15}}{\sqrt6}
\end{array}
\right).
\label{eq:meson-SU4-neutral}
\end{equation}
After diagonalization, its three eigenvalues are associated
with the masses of the physical $\omega$, $\phi$, and
$J/\psi$ states for $Q=c$, or those of the $\omega$,
$\phi$, and $\Upsilon$ states for $Q=b$.

The mass structure becomes particularly transparent
for the off-diagonal flavor configurations.
For a vector meson with the flavor configuration
$q_i\bar q_j$, where $i\neq j$, the leading mass
Hamiltonian gives
\begin{equation}
M_{i\bar j}
=
m_{\mathbf O}
+
\frac{\sigma_i+\sigma_j}{2},
\label{eq:meson-additive-mass}
\end{equation}
where
\begin{equation}
\sigma_i
=
\sum_C D_C
\left[\lambda_C\right]_{ii},
\label{eq:meson-single-flavor-response}
\end{equation}
with the sum running over $C=8,15$ in the present case.
This expression exhibits a simple additive structure:
the leading mass shifts associated with the two flavor
indices enter independently.
Consequently, the mass shift induced by replacing one
flavor with another is independent of the flavor carried
by the other constituent within this leading construction.

For the isovector and open-flavor vector mesons in the
charm system, we obtain
\begin{eqnarray}
M_\rho
&=&
m_{\mathbf O}
+
\frac{D_8}{\sqrt3}
+
\frac{D_{15}}{\sqrt6},
\nonumber\\
M_{K^*}
&=&
m_{\mathbf O}
-
\frac{D_8}{2\sqrt3}
+
\frac{D_{15}}{\sqrt6},
\nonumber\\
M_{D^*}
&=&
m_{\mathbf O}
+
\frac{D_8}{2\sqrt3}
-
\frac{D_{15}}{\sqrt6},
\nonumber\\
M_{D_s^*}
&=&
m_{\mathbf O}
-
\frac{D_8}{\sqrt3}
-
\frac{D_{15}}{\sqrt6}.
\label{eq:meson-SU4-open}
\end{eqnarray}
The corresponding expressions for the bottom sector
are obtained by replacing $D^*$ and $D_s^*$ with
$B^*$ and $B_s^*$, respectively.
Together with the three neutral isoscalar masses,
these expressions provide seven mass observables
for testing the five-parameter realization of
the leading flavor-$SU(4)_Q$ Hamiltonian,
\begin{equation}
m_{\mathbf S},
\quad
m_{\mathbf O},
\quad
D_8,
\quad
D_{15},
\quad
r.
\end{equation}

We perform two independent unweighted least-squares
fits in the flavor-$SU(4)_c$ and flavor-$SU(4)_b$
systems, treating these five quantities as independent
effective parameters in each system.
For the charm system, we use the masses of
$\rho$, $K^*$, $D^*$, $D_s^*$, $\omega$, $\phi$,
and $J/\psi$, whereas for the bottom system,
we replace $D^*$, $D_s^*$, and $J/\psi$ with
$B^*$, $B_s^*$, and $\Upsilon$, respectively.
The four light-vector-meson masses are included
in both fits.
Throughout this paper, the root-mean-square (RMS)
mass deviation for a fit to $n$ mass observables
is defined as
\begin{equation}
{\rm RMS}_n
=
\sqrt{
\frac{1}{n}
\sum_{H=1}^{n}
\left(
M_H^{\rm fit}-M_H^{\rm exp}
\right)^2
}.
\label{eq:meson-rms-definition}
\end{equation}
The mass spectrum is invariant under $r\to-r$,
since the corresponding mass matrices are related
by a sign change of the singlet basis vector.
We report the positive-$r$ solutions throughout
the vector-meson analysis.

\begin{table}[t]
\caption{Parameters and masses obtained by fitting
the leading five-parameter Hamiltonian to seven
vector mesons in each of the flavor-$SU(4)_c$ and
flavor-$SU(4)_b$ systems.
The four light-vector-meson masses are included
in both fits.
All dimensional quantities are given in units
of MeV; $r$ is dimensionless.}
\label{tab:meson-SU4-fit}
\begin{ruledtabular}
\begin{tabular}{lrrr}
& Experiment & $SU(4)_c$ & $SU(4)_b$ \\
\hline
$m_{\mathbf S}$ & -- & $1326.60$ & $2724.24$ \\
$m_{\mathbf O}$ & -- & $1448.20$ & $3103.63$ \\
$D_8$           & -- & $-141.94$ & $-138.33$ \\
$D_{15}$        & -- & $-1447.32$ & $-5503.46$ \\
$r$             & -- & $0.62889$ & $0.64379$ \\
RMS             & -- & $6.225$ & $8.583$ \\
\hline
$\rho$          & $775.260$ & $775.389$ & $776.990$ \\
$K^*$           & $893.610$ & $898.314$ & $896.787$ \\
$\omega$        & $782.660$ & $785.774$ & $786.976$ \\
$\phi$          & $1019.460$ & $1012.288$ & $1010.490$ \\
\hline
$D^*$           & $2008.555$ & $1998.095$ & -- \\
$D_s^*$         & $2112.200$ & $2121.020$ & -- \\
$J/\psi$        & $3096.900$ & $3097.765$ & -- \\
\hline
$B^*$           & $5324.940$ & -- & $5310.475$ \\
$B_s^*$         & $5416.337$ & -- & $5430.273$ \\
$\Upsilon$      & $9460.400$ & -- & $9460.676$ \\
\end{tabular}
\end{ruledtabular}
\end{table}

The resulting parameters and fitted masses are
summarized in Table~\ref{tab:meson-SU4-fit},
together with the adopted experimental masses.
The leading five-parameter Hamiltonian yields
${\rm RMS}^{SU(4)_c}_7=6.225$~MeV and
${\rm RMS}^{SU(4)_b}_7=8.583$~MeV.
These deviations are small compared with the mass
differences between the light and heavy sectors,
indicating that the leading Hamiltonian captures
the dominant mass pattern in both systems.
They characterize the fit quality on the adopted
data sets, rather than uncertainties of individual
mass predictions.

Despite the overall success of these fits, the
additive structure in Eq.~(\ref{eq:meson-additive-mass})
imposes certain restrictions on the mass spectra that
are not exactly satisfied by the experimental data.
In particular, Eq.~(\ref{eq:meson-SU4-open}) gives
\begin{eqnarray}
M_{K^*}-M_\rho
&=&
-\frac{\sqrt3}{2}D_8,
\nonumber\\
M_{Q_s^*}-M_{Q^*}
&=&
-\frac{\sqrt3}{2}D_8,
\end{eqnarray}
where $Q=c,b$, and $Q^*$ and $Q_s^*$ denote the
corresponding nonstrange and strange heavy-light
vector mesons, respectively.
We therefore obtain the parameter-independent
mass relation
\begin{equation}
{\cal N}_Q
\equiv
M_{Q_s^*}
-
M_{Q^*}
-
M_{K^*}
+
M_\rho
=
0,
\qquad
Q=c,b.
\label{eq:meson-SU4-null}
\end{equation}
This relation holds independently of the values
of the five parameters in the leading
flavor-$SU(4)_Q$ Hamiltonian.
Using the experimental masses listed in
Table~\ref{tab:meson-SU4-fit}, we obtain
\begin{eqnarray}
{\cal N}_c^{\rm exp}
&=&
-14.705~{\rm MeV},
\nonumber\\
{\cal N}_b^{\rm exp}
&=&
-26.953~{\rm MeV}.
\label{eq:meson-SU4-null-exp}
\end{eqnarray}
These nonzero values reveal that the mass change
associated with the replacement of a light flavor
by a strange one depends on the flavor of the other
constituent.
Indeed, the corresponding mass differences decrease
from $118.350$~MeV in the light sector to
$103.645$~MeV in the charm sector and further to
$91.397$~MeV in the bottom sector.

Since Eq.~(\ref{eq:meson-SU4-null}) is independent of
all five parameters, these departures cannot be
eliminated by readjusting the parameters of the
leading Hamiltonian.
Rather, they indicate the need for additional
contributions beyond its additive flavor structure.
Higher-order flavor structures, such as
those discussed in Sec.~\ref{sec:meson-leading},
provide a possible realization of this nonadditivity.
We will further examine these nonadditive contributions
in Sec.~\ref{sec:meson5}, where both heavy flavors
are incorporated into a common flavor space.

\subsection{Ideal mixing and state flow}
\label{sec:meson-ideal-mixing}

In addition to the mass spectra, the flavor-$SU(4)_Q$ analyses
yield the eigenvectors of the neutral isoscalar
mass matrices.
To examine their flavor structures, we introduce the
ideal flavor basis
\begin{equation}
{\left(u\bar u+d\bar d\right)/\sqrt2},
\qquad
s\bar s,
\qquad
Q\bar Q,
\label{eq:meson-ideal-basis}
\end{equation}
where $Q=c$ or $b$.
These three configurations correspond predominantly to
the physical $\omega$, $\phi$, and $J/\psi$ ($\Upsilon$)
states in the charm (bottom) sector.

For the positive-$r$ solutions reported in the preceding
subsection, the squared overlaps of the three neutral
eigenstates with the corresponding ideal flavor
configurations are approximately
\begin{equation}
99.85\%,
\qquad
99.88\%,
\qquad
99.76\%
\end{equation}
in the flavor-$SU(4)_c$ system, and
\begin{equation}
99.90\%,
\qquad
99.94\%,
\qquad
99.84\%
\end{equation}
in the flavor-$SU(4)_b$ system.
These results indicate that, within the present mass
Hamiltonian, the neutral vector-meson eigenstates are
already strongly localized in the ideal flavor basis.
These overlaps refer to the effective reference-space eigenvectors,
not to probabilities in the full QCD Fock space.

This observation motivates us to investigate how the
physical states may reorganize as flavor symmetry
breaking develops.
A simple physical picture of such a state flow
can be obtained from the familiar isospin system.
In the exact isospin limit, the $u$ and $d$ quarks
are degenerate, and the configurations $u\bar u$
and $d\bar d$ are not distinguished by their diagonal
masses.
In the two-configuration nonstrange subspace, the
symmetry basis is given by
\begin{equation}
{\left(u\bar u+d\bar d\right)/\sqrt2},
\qquad
{\left(u\bar u-d\bar d\right)/\sqrt2},
\end{equation}
corresponding to isospin $I=0$ and $I=1$, respectively.
The isoscalar configuration may also mix with other
isoscalar configurations.
Within the two-configuration subspace, as long as
the isospin-breaking contributions remain small
compared with the energy separation between the two
symmetry eigenstates, their flavor compositions are
only weakly modified.

The situation may change qualitatively when the
symmetry breaking becomes sufficiently strong.
To illustrate this, we consider two flavor configurations
described by the schematic mass Hamiltonian
\begin{equation}
{\cal H}
=
\left(
\begin{array}{cc}
E_1 & V_{\rm mix}\\
V_{\rm mix} & E_2
\end{array}
\right),
\label{eq:meson-state-flow-H}
\end{equation}
where
\begin{equation}
\Delta E_{\rm flavor}
=
|E_1-E_2|
\end{equation}
denotes the diagonal flavor splitting, while
$V_{\rm mix}$ describes the interaction between
the two flavor configurations.
The mixing angle $\theta$ of the orthogonal
eigenvector rotation satisfies
\begin{equation}
\tan 2\theta
=
{2V_{\rm mix}\over E_1-E_2}.
\label{eq:meson-state-flow-angle}
\end{equation}
When $\Delta E_{\rm flavor}\ll 2|V_{\rm mix}|$,
the two flavor configurations are strongly mixed,
and the physical eigenstates approach symmetric
and antisymmetric combinations of them.
In contrast, when
$\Delta E_{\rm flavor}\gg 2|V_{\rm mix}|$,
the residual mixing becomes relatively unimportant,
and the physical eigenstates approach the individual
flavor configurations.
Between these two limiting regimes, the eigenvectors
may undergo a substantial reorganization as the
symmetry-breaking strength changes.

A similar state separation may occur during the
breaking of the larger flavor symmetries considered
here.
Near the flavor-symmetric point, the physical states
can naturally be organized into irreducible
representations of the parent flavor group.
As the symmetry breaking increases, different
flavor branches may gradually separate.
If the diagonal flavor splittings become sufficiently
large compared with the residual off-diagonal
interactions, the physical eigenstates may become
localized in the quark-flavor basis.
In this sense, ideal flavor localization can be viewed
as a possible limiting behavior of the state flow.
The near-ideal mixing obtained in the present
flavor-$SU(4)_Q$ analyses is consistent with this
physical picture.
However, the mass Hamiltonians determined at the
physical point do not by themselves specify the
complete evolution of the eigenstates from the
exact flavor-symmetric limit.

It is important to distinguish ideal flavor
localization from stronger assumptions concerning
the structure of the mass Hamiltonian.
In particular, the near-ideal flavor compositions
of the $\omega$, $\phi$, $J/\psi$, and $\Upsilon$
states do not by themselves imply
\begin{equation}
m_{\mathbf S}=m_{\mathbf O},
\end{equation}
nor do they require an exactly additive mass structure.
Ideal flavor mixing concerns the compositions of
the physical eigenstates, whereas relations among
the Hamiltonian coefficients impose additional
constraints on the effective interactions.

Moreover, the mass spectrum may become insensitive
to certain combinations of the Hamiltonian coefficients.
For example, in the two-state Hamiltonian of
Eq.~(\ref{eq:meson-state-flow-H}), the mass shifts
induced by a small off-diagonal interaction are
of order
\begin{equation}
{V_{\rm mix}^2\over\Delta E_{\rm flavor}},
\end{equation}
when the diagonal energies are held fixed and
$\Delta E_{\rm flavor}\gg 2|V_{\rm mix}|$.
Consequently, even when the physical states are
strongly localized in the flavor basis, the
underlying effective interactions need not be
uniquely determined from the mass spectrum.
We therefore do not impose exact ideal mixing
as an additional algebraic constraint on the
mass Hamiltonian.

Although the individual effective coefficients
remain dependent on the adopted mass Hamiltonian,
an interesting pattern emerges when the leading
symmetry-breaking coefficients of the two
flavor-$SU(4)_Q$ systems are compared.
The fitted coefficients associated with the
light-flavor $SU(3)$ breaking direction are
very similar,
\begin{equation}
D_8^{SU(4)_c}
=
-141.94~{\rm MeV},
\quad
D_8^{SU(4)_b}
=
-138.33~{\rm MeV}.
\label{eq:meson-D8-cb}
\end{equation}
Thus, within the leading construction, the
effective coefficient along the $\lambda_8$
direction changes only weakly when charm is
replaced by bottom as the heavy flavor.
In contrast, the magnitude of the fitted coefficient
along the heavy-flavor $\lambda_{15}$ direction
increases substantially,
\begin{equation}
D_{15}^{SU(4)_c}
=
-1447.32~{\rm MeV},
\quad
D_{15}^{SU(4)_b}
=
-5503.46~{\rm MeV}.
\end{equation}
The corresponding ratios between the two leading
symmetry-breaking directions are
\begin{eqnarray}
R_{15/8}^{SU(4)_c}
&=&
{D_{15}^{SU(4)_c}\over D_8^{SU(4)_c}}
=
10.20,
\nonumber\\
R_{15/8}^{SU(4)_b}
&=&
{D_{15}^{SU(4)_b}\over D_8^{SU(4)_b}}
=
39.79.
\label{eq:meson-R158-cb}
\end{eqnarray}
These results exhibit a clear hierarchy between
the leading light-flavor and heavy-flavor
symmetry-breaking contributions.
The approximate stability of the fitted $D_8$
coefficient, however, does not preclude the
nonadditive mass contributions associated with
different flavor environments, as demonstrated
in the preceding subsection.

The above analyses illustrate how the evolution of
the effective interactions can lead to a reorganization
of the physical eigenstates.
The state flow is thus a physical manifestation of
the underlying interaction flow.
A general description of these flows will be
developed in Sec.~\ref{sec:group-expansion}.

\subsection{Vector mesons in flavor $SU(5)$}
\label{sec:meson5}

The preceding analyses treat the charm and bottom sectors
within two separate flavor-$SU(4)_Q$ frameworks.
Although both realizations provide reasonably accurate
descriptions of the corresponding vector-meson spectra,
neither includes mesons containing both heavy flavors.
To investigate the interplay between the charm and
bottom flavors, we now extend the same leading mass
Hamiltonian to flavor $SU(5)$, in which the five quark
flavors, $u$, $d$, $s$, $c$, and $b$, are incorporated
into a common flavor space.

In addition to the $\lambda_8$ and $\lambda_{15}$
directions introduced above, the enlarged flavor space
contains the additional diagonal generator
\begin{equation}
\lambda_{24}
=
\frac{1}{\sqrt{10}}
{\rm diag}(1,1,1,1,-4).
\end{equation}
The $\lambda_{24}$ direction breaks flavor $SU(5)$
while preserving the $SU(4)_c$ subgroup.
In the isospin-symmetric limit, the three diagonal
symmetry-breaking directions are associated with
the coefficients
\begin{equation}
D_8,\quad
D_{15},\quad
D_{24},
\qquad
G_8,\quad
G_{15},\quad
G_{24}.
\end{equation}
Within the leading common-source construction,
Eq.~(\ref{eq:meson-general-r}) gives
\begin{equation}
\frac{G_8}{D_8}
=
\frac{G_{15}}{D_{15}}
=
\frac{G_{24}}{D_{24}}
=
r.
\label{eq:meson-SU5-r}
\end{equation}

For the neutral isoscalar vector mesons, we work in the
flavor-$SU(5)$ basis
\begin{equation}
\left(
M_0,M_8,M_{15},M_{24}
\right).
\end{equation}
The corresponding matrix representation of the linear
mass Hamiltonian in Eq.~(\ref{eq:meson-general-H}) is
\begin{widetext}
\begin{equation}
{\cal H}_{0,8,15,24}^{SU(5)}
=
\left(
\begin{array}{cccc}
m_{\mathbf S}
&
rD_8
&
rD_{15}
&
rD_{24}
\\[1mm]
rD_8
&
m_{\mathbf O}
-\frac{D_8}{\sqrt3}
+\frac{D_{15}}{\sqrt6}
+\frac{D_{24}}{\sqrt{10}}
&
\frac{D_8}{\sqrt6}
&
\frac{D_8}{\sqrt{10}}
\\[1mm]
rD_{15}
&
\frac{D_8}{\sqrt6}
&
m_{\mathbf O}
-\frac{2D_{15}}{\sqrt6}
+\frac{D_{24}}{\sqrt{10}}
&
\frac{D_{15}}{\sqrt{10}}
\\[1mm]
rD_{24}
&
\frac{D_8}{\sqrt{10}}
&
\frac{D_{15}}{\sqrt{10}}
&
m_{\mathbf O}
-\frac{3D_{24}}{\sqrt{10}}
\end{array}
\right).
\label{eq:meson-SU5-neutral}
\end{equation}
\end{widetext}
After diagonalization, its four eigenvalues are associated
with the masses of the physical $\omega$, $\phi$,
$J/\psi$, and $\Upsilon$ states.

The isovector and open-flavor vector-meson masses follow
from the same additive mass structure introduced in
Eq.~(\ref{eq:meson-additive-mass}), with the sum in
Eq.~(\ref{eq:meson-single-flavor-response}) now running
over $C=8,15,24$.
Explicitly, we obtain
\begin{eqnarray}
M_\rho
&=&
m_{\mathbf O}
+
\frac{D_8}{\sqrt3}
+
\frac{D_{15}}{\sqrt6}
+
\frac{D_{24}}{\sqrt{10}},
\nonumber\\
M_{K^*}
&=&
m_{\mathbf O}
-
\frac{D_8}{2\sqrt3}
+
\frac{D_{15}}{\sqrt6}
+
\frac{D_{24}}{\sqrt{10}},
\nonumber\\
M_{D^*}
&=&
m_{\mathbf O}
+
\frac{D_8}{2\sqrt3}
-
\frac{D_{15}}{\sqrt6}
+
\frac{D_{24}}{\sqrt{10}},
\nonumber\\
M_{D_s^*}
&=&
m_{\mathbf O}
-
\frac{D_8}{\sqrt3}
-
\frac{D_{15}}{\sqrt6}
+
\frac{D_{24}}{\sqrt{10}},
\nonumber\\
M_{B^*}
&=&
m_{\mathbf O}
+
\frac{D_8}{2\sqrt3}
+
\frac{D_{15}}{2\sqrt6}
-
\frac{3D_{24}}{2\sqrt{10}},
\nonumber\\
M_{B_s^*}
&=&
m_{\mathbf O}
-
\frac{D_8}{\sqrt3}
+
\frac{D_{15}}{2\sqrt6}
-
\frac{3D_{24}}{2\sqrt{10}},
\nonumber\\
M_{B_c^*}
&=&
m_{\mathbf O}
-
\frac{3D_{15}}{2\sqrt6}
-
\frac{3D_{24}}{2\sqrt{10}}.
\label{eq:meson-SU5-open}
\end{eqnarray}
Together with the four neutral isoscalar eigenvalues,
these expressions provide eleven mass observables
for testing the six-parameter realization of the
leading flavor-$SU(5)$ Hamiltonian,
\begin{equation}
m_{\mathbf S},
\quad
m_{\mathbf O},
\quad
D_8,
\quad
D_{15},
\quad
D_{24},
\quad
r.
\end{equation}

We first perform an unweighted least-squares fit to the
adopted masses of all eleven vector mesons,
including the reported $B_c^*$ mass~\cite{ATLAS:2026ubk}.
To examine the impact of the mixed-heavy state,
we perform a second fit in which $B_c^*$ is
excluded from the minimization.
As in the separate flavor-$SU(4)_Q$ analyses,
we report the positive-$r$ solutions.

The resulting parameters and fitted masses are
summarized in Table~\ref{tab:meson-SU5-fit},
together with the adopted experimental masses.
The ten-state fit uses the remaining ten mesons,
while its $B_c^*$ entry is a holdout result rather
than an input to the fit or its RMS deviation.
The eleven-state fit yields
${\rm RMS}^{SU(5)}_{11}=40.376$~MeV,
considerably larger than the deviations obtained
in the separate flavor-$SU(4)_c$ and
flavor-$SU(4)_b$ analyses.
Excluding $B_c^*$ reduces the RMS deviation
for the remaining ten states to
${\rm RMS}^{SU(5)}_{10}=8.624$~MeV,
comparable to those of the separate
flavor-$SU(4)_Q$ fits.
However, the ten-state fit gives the holdout mass
\begin{equation}
M_{B_c^*}^{\rm fit}
=
6528.137~{\rm MeV},
\label{eq:meson-SU5-Bc-holdout}
\end{equation}
which lies $189.137$~MeV above the adopted
experimental value of $6339.000$~MeV.
This discrepancy is not included in
${\rm RMS}^{SU(5)}_{10}$.

\begin{table}[t]
\caption{Parameters and masses obtained from the
six-parameter flavor-$SU(5)$ fits.
The eleven-state fit includes all listed vector
mesons, whereas the ten-state fit excludes $B_c^*$
from the minimization and RMS calculation.
The $B_c^*$ mass in the ten-state fit is a
holdout result.
All dimensional quantities are given in units
of MeV; $r$ is dimensionless.}
\label{tab:meson-SU5-fit}
\begin{ruledtabular}
\begin{tabular}{lrrr}
& Experiment & 11 states & 10 states \\
\hline
$m_{\mathbf S}$ & -- & $2748.36$ & $2628.17$ \\
$m_{\mathbf O}$ & -- & $3103.41$ & $3127.15$ \\
$D_8$           & -- & $-138.50$ & $-136.05$ \\
$D_{15}$        & -- & $-1385.54$ & $-1440.25$ \\
$D_{24}$        & -- & $-5224.97$ & $-5310.56$ \\
$r$             & -- & $0.58814$ & $0.55949$ \\
RMS             & -- & $40.376$ & $8.624$ \\
\hline
$\rho$          & $775.260$  & $805.523$  & $781.274$ \\
$K^*$           & $893.610$  & $925.469$  & $899.096$ \\
$\omega$        & $782.660$  & $785.428$  & $786.321$ \\
$\phi$          & $1019.460$ & $1030.303$ & $1007.887$ \\
\hline
$D^*$           & $2008.555$ & $1976.793$ & $1996.509$ \\
$D_s^*$         & $2112.200$ & $2096.739$ & $2114.330$ \\
$J/\psi$        & $3096.900$ & $3074.202$ & $3102.108$ \\
\hline
$B^*$           & $5324.940$ & $5259.032$ & $5312.903$ \\
$B_s^*$         & $5416.337$ & $5378.978$ & $5430.724$ \\
$\Upsilon$      & $9460.400$ & $9466.553$ & $9459.170$ \\
\hline
$B_c^*$         & $6339.000$ & $6430.302$ & $6528.137$ \\
\end{tabular}
\end{ruledtabular}
\end{table}

These results show that the leading Hamiltonian
can describe the light, strange, charm, and bottom
vector mesons simultaneously with an RMS deviation
of order $10$~MeV when $B_c^*$ is withheld.
The neutral eigenstates in both positive-$r$
solutions remain close to their corresponding
ideal flavor configurations, and the fitted
coefficients exhibit a hierarchy similar to
that in the separate flavor-$SU(4)_Q$ analyses.
The inclusion of the mixed-heavy state, however,
reveals a substantial tension in the leading
flavor-$SU(5)$ description.

This tension can be understood more directly through
the parameter-independent mass relations implied by
the additive structure in
Eq.~(\ref{eq:meson-additive-mass}).
The two mass relations in
Eq.~(\ref{eq:meson-SU4-null}) remain valid in the
leading flavor-$SU(5)$ construction.
The simultaneous inclusion of charm and bottom
leads to an additional relation involving
the mixed-heavy state.
In particular, the mass change associated with
replacing a light flavor by a charm flavor is
independent of the flavor carried by the other
constituent within the leading additive construction.
We therefore obtain
\begin{equation}
M_{B_c^*}-M_{B^*}
=
M_{D^*}-M_\rho,
\end{equation}
which leads to the parameter-independent mass relation
\begin{equation}
{\cal N}_{cb}
\equiv
M_{B_c^*}
-
M_{D^*}
-
M_{B^*}
+
M_\rho
=
0.
\label{eq:meson-SU5-null-cb}
\end{equation}
Equivalently, this relation compares the mass change
associated with replacing a light flavor by a bottom
flavor in the light and charm environments.
Using the experimental masses listed in
Table~\ref{tab:meson-SU5-fit}, we obtain
\begin{equation}
{\cal N}_{cb}^{\rm exp}
=
-219.235~{\rm MeV}.
\label{eq:meson-SU5-null-cb-exp}
\end{equation}
Its magnitude is considerably larger than those
found in the single-heavy sectors,
given in Eq.~(\ref{eq:meson-SU4-null-exp}).
As in the flavor-$SU(4)_Q$ analyses, this departure
cannot be eliminated by readjusting the parameters
of the leading Hamiltonian.

To visualize the dominant mass structure more directly,
we introduce the effective flavor increments
\begin{eqnarray}
m_{u,d}^{\rm eff}
&=&
0,
\nonumber\\
m_s^{\rm eff}
&=&
M_{K^*}-M_\rho,
\nonumber\\
m_c^{\rm eff}
&=&
M_{D^*}-M_\rho,
\nonumber\\
m_b^{\rm eff}
&=&
M_{B^*}-M_\rho.
\label{eq:meson-effective-increments}
\end{eqnarray}
These quantities are defined directly from the experimental
hadron masses and should not be identified with the
corresponding current-quark masses.
For a vector meson with the flavor configuration
$q_i\bar q_j$, we introduce the additive flavor coordinate
\begin{equation}
x_{ij}
=
m_i^{\rm eff}+m_j^{\rm eff}.
\label{eq:meson-additive-coordinate}
\end{equation}
For the neutral isoscalar vector mesons, the corresponding
coordinates are assigned according to their dominant
ideal flavor configurations discussed in
Sec.~\ref{sec:meson-ideal-mixing}.

\begin{figure}[t]
\centering
\includegraphics[width=\columnwidth]{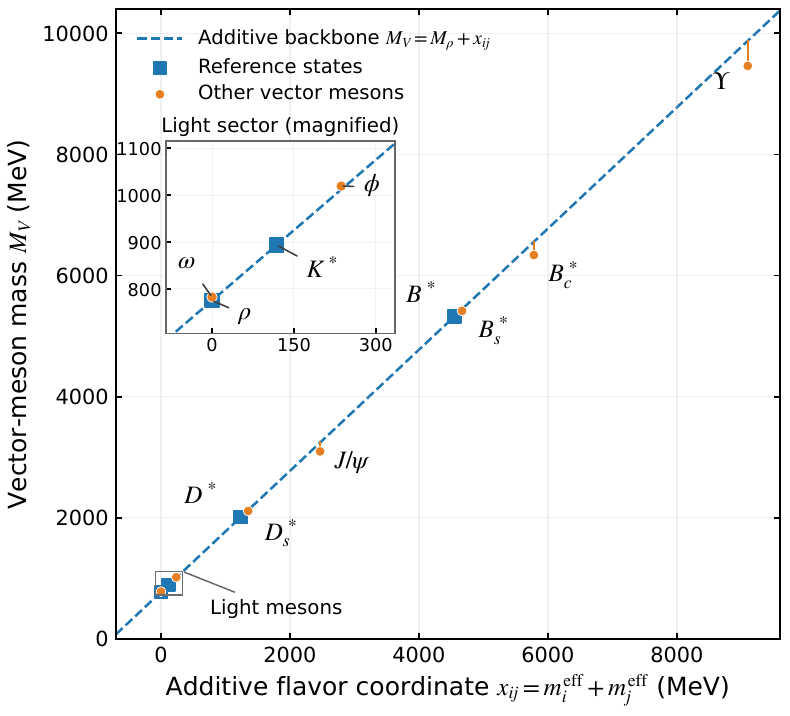}
\caption{
Experimental vector-meson masses plotted against the
effective additive flavor coordinate $x_{ij}$ defined
in Eq.~(\ref{eq:meson-additive-coordinate}).
The $\rho$, $K^*$, $D^*$, and $B^*$ states define the
additive reference line $M_V=M_\rho+x_{ij}$ by construction.
The vertical segments indicate the deviations of the
remaining states from this reference line.
The coordinates of the neutral isoscalar states are
assigned according to their dominant ideal flavor
configurations.
The inset shows an enlarged view of the light-flavor
sector, including the $\rho$, $\omega$, $K^*$, and
$\phi$ mesons.
}
\label{fig:meson-mass-flow}
\end{figure}

As illustrated in Fig.~\ref{fig:meson-mass-flow},
the $\rho$, $K^*$, $D^*$, and $B^*$ states define,
by construction, an additive reference line,
\begin{equation}
M_V^{\rm add}
=
M_\rho+x_{ij}.
\label{eq:meson-additive-backbone}
\end{equation}
Despite the large mass differences between the light
and heavy sectors, the remaining vector-meson masses
also lie relatively close to this reference line.
Their deviations, however, contain important physical
information beyond the leading additive mass structure.
In particular, the deviations of the $D_s^*$, $B_s^*$,
and $B_c^*$ masses from the additive reference line
are precisely the three mass combinations
${\cal N}_c$, ${\cal N}_b$, and ${\cal N}_{cb}$,
respectively.
They show that the mass change associated with replacing
one flavor by another depends on the flavor of the
other constituent.
For the neutral isoscalar states, the deviations may
also contain contributions from the singlet--adjoint
mass difference and flavor mixing.

The vector-meson analyses in this section thus reveal
two complementary aspects of flavor-symmetry breaking:
the reorganization of physical eigenstates and the
emergence of nonadditive mass contributions beyond
the leading flavor structures.
Both phenomena reflect the evolution of the
underlying effective Hamiltonian.
As discussed in Sec.~\ref{sec:meson-leading},
higher-order flavor-invariant interactions provide
a possible mechanism through which the response to
one symmetry-breaking source may depend on the
presence of another.
These observations motivate us to develop a more
general framework for flavor-symmetry breaking,
which we refer to as group expansion.

\section{Generalized Symmetry Breaking}
\label{sec:group-expansion}

The vector-meson analyses in the preceding section illustrate
that flavor-symmetry breaking can change both the masses and
flavor compositions of physical hadrons.
Within the present framework, these changes originate from
the evolution of the effective mass Hamiltonian.
We therefore describe symmetry breaking through the
eigenvalue equation
\begin{equation}
{\cal H}(\boldsymbol{\xi})
|\psi_i(\boldsymbol{\xi})\rangle
=
M_i(\boldsymbol{\xi})
|\psi_i(\boldsymbol{\xi})\rangle ,
\label{eq:gsb-eigen}
\end{equation}
where $\boldsymbol{\xi}$ denotes the symmetry-breaking
coordinates introduced in Eq.~(\ref{eq:meson-breaking-source}).

We refer to the evolution of the effective Hamiltonian
${\cal H}(\boldsymbol{\xi})$ as the \textit{interaction flow},
that of its eigenvalues $M_i(\boldsymbol{\xi})$ as the
\textit{mass flow}, and that of its eigenstates
$|\psi_i(\boldsymbol{\xi})\rangle$ as the \textit{state flow}.
The mass flow and state flow are thus two complementary
manifestations of the interaction flow.
In a fixed symmetry-adapted basis, diagonal components
of the evolving Hamiltonian shift the diagonal mass
levels, whereas off-diagonal components can mix
different flavor configurations.
Although the separation into diagonal and off-diagonal
contributions depends on the choice of basis, the
physical spectrum and eigenspaces do not.

The central question is then how the effective Hamiltonian
itself evolves as the symmetry is broken.
As discussed in Sec.~\ref{sec:meson-leading},
higher-order flavor-invariant interactions may modify
existing symmetry-breaking responses or generate
additional independent structures.
In particular, the response to one symmetry-breaking
source may depend on the presence of another.
In the baryon sector to be discussed later, the state flow
can be difficult to disentangle because several flavor
representations may coexist and the symmetry-breaking
Hamiltonian can connect them.
Group theory determines which transitions are allowed,
whereas their strengths are determined dynamically.
Representation mixing is therefore a possible, but not
inevitable, manifestation of state flow.

The vector-meson sector provides a complementary example.
As shown in Sec.~\ref{sec:meson-ideal-mixing}, the neutral
vector mesons are already very close to ideal quark-flavor
configurations within the present mass-matrix description.
This nearly flavor-localized endpoint does not uniquely
determine the preceding state evolution, but provides a
relatively simple flavor structure at the physical point.
Moreover, as shown in Sec.~\ref{sec:meson5}, the mixed-heavy
sector exhibits the largest departure from the leading
additive mass structure.
We therefore use the vector mesons in flavor $SU(4)_{udcb}$
to investigate the interplay between two strong
symmetry-breaking directions and subsequently develop
a general description of the evolving Hamiltonian.

\subsection{Two strong breaking directions in $SU(4)_{udcb}$}
\label{sec:gsb-udcb}

In the ordered flavor basis $(u,d,c,b)$, we introduce
the two diagonal generators
\begin{equation}
\lambda_8
=
{1\over\sqrt3}
{\rm diag}(1,1,-2,0),
\quad
\lambda_{15}
=
{1\over\sqrt6}
{\rm diag}(1,1,1,-3).
\label{eq:gsb-generators}
\end{equation}
Following Eq.~(\ref{eq:meson-breaking-source}), we
consider the symmetry-breaking source
\begin{equation}
X(\boldsymbol{\xi})
=
b\lambda_{15}
+
c\lambda_8,
\qquad
[\lambda_{15},\lambda_8]=0,
\label{eq:gsb-Sigma}
\end{equation}
where $\boldsymbol{\xi}=(b,c)$, with $b=\xi_{15}$ and
$c=\xi_8$.
Here $b$ parameterizes the breaking that separates
the bottom flavor from the $(u,d,c)$ sector,
while $c$ parameterizes the breaking that further
separates the charm flavor from the $(u,d)$ sector.
These two symmetry-breaking coordinates should
not be identified directly with the bottom-
and charm-quark masses.

Starting from the exact flavor-$SU(4)_{udcb}$
limit, we may first introduce the symmetry-breaking
source along the $\lambda_{15}$ direction and
then introduce that along the $\lambda_8$ direction.
The corresponding sequence of non-Abelian flavor
symmetries is
\begin{equation}
SU(4)_{udcb}
\xrightarrow{\ b\lambda_{15}\ }
SU(3)_{udc}
\xrightarrow{\ c\lambda_8\ }
SU(2)_{ud}.
\label{eq:gsb-udcb-breaking-sequence}
\end{equation}
The unbroken Abelian flavor symmetries are implicit
in this sequence.
Although the $SU(3)_{udc}$ symmetry remains exact
when only the first source is introduced, the
effective interactions governing the mesons may
already change.
Consequently, the response to the subsequent
breaking of $SU(3)_{udc}$ may depend on the
background established by the $\lambda_{15}$ source.

We describe this process by an effective mass
Hamiltonian ${\cal H}(b,c)$ defined in a common
reference flavor space, with all other control
parameters held fixed.
Assuming a regular expansion around $b=c=0$, we write
\begin{eqnarray}
{\cal H}(b,c)
&=&
{\cal H}_{00}+b{\cal H}_{10}+c{\cal H}_{01}
\nonumber\\
&&
+b^2{\cal H}_{20}+bc{\cal H}_{11}
+c^2{\cal H}_{02}+\cdots .
\label{eq:gsb-bc-expansion}
\end{eqnarray}
Here ${\cal H}_{00}$ denotes the Hamiltonian in the
exact flavor-$SU(4)_{udcb}$ limit.
The terms $b{\cal H}_{10}$ and $c{\cal H}_{01}$
describe the leading responses along the two
symmetry-breaking directions.
At the next order, the terms $b^2{\cal H}_{20}$
and $c^2{\cal H}_{02}$ describe nonlinear responses
along the individual directions, whereas the mixed
term $bc{\cal H}_{11}$ accounts for the leading
correlation between them.

At a given physical point, the numerical values
of $b$ and $c$ may be absorbed into the
corresponding fitted effective coefficients.
We keep them explicit here to identify the order
and direction of the symmetry-breaking responses.
Since the physical point lies far from the symmetric
limit, the effective coefficients extracted from a
finite-order fit need not faithfully determine the
local expansion coefficients at $b=c=0$.
The accuracy of the truncation must therefore be
examined independently.

Keeping terms through first order in the two
symmetry-breaking coordinates, we obtain
\begin{equation}
{\cal H}_{\rm Lead}(b,c)
=
{\cal H}_{00}
+
b{\cal H}_{10}
+
c{\cal H}_{01}.
\label{eq:gsb-leading-H}
\end{equation}
Within this leading construction, the contributions
of the two symmetry-breaking sources enter
additively at the Hamiltonian level: the matrix
contribution of either source is independent of
the other.
This does not in general imply additive eigenvalues
in a mixing sector.

The corresponding leading construction was
developed in detail in Sec.~\ref{sec:meson}.
For the vector mesons, the mass Hamiltonian
contains the two symmetric-limit mass parameters
$m_{\mathbf S}$ and $m_{\mathbf O}$, together
with the adjoint mass-response coefficients
$D_8$ and $D_{15}$, which are proportional to
$c$ and $b$, respectively, at leading order.
The singlet--adjoint transition coefficients
satisfy the leading common-source relation
\begin{equation}
{G_8\over D_8}
=
{G_{15}\over D_{15}}
=
r.
\label{eq:gsb-r}
\end{equation}

In the neutral isoscalar basis $(M_0,M_8,M_{15})$,
the mass Hamiltonian is
\begin{equation}
{\cal H}_{0,8,15}^{SU(4)_{udcb}}
=
\left(
\begin{array}{@{}c@{\hspace{4pt}}c@{\hspace{4pt}}c@{}}
m_{\mathbf S}
&
rD_8
&
rD_{15}
\\[1mm]
rD_8
&
m_{\mathbf O}
-\dfrac{D_8}{\sqrt3}
+\dfrac{D_{15}}{\sqrt6}
&
\dfrac{D_8}{\sqrt6}
\\[3mm]
rD_{15}
&
\dfrac{D_8}{\sqrt6}
&
m_{\mathbf O}
-\dfrac{2D_{15}}{\sqrt6}
\end{array}
\right).
\label{eq:gsb-udcb-neutral}
\end{equation}
Its three eigenvalues are associated with the
masses of the physical $\omega$, $J/\psi$,
and $\Upsilon$ states.
The corresponding isovector and open-flavor
vector-meson masses are
\begin{eqnarray}
M_\rho
&=&
m_{\mathbf O}
+
{D_8\over\sqrt3}
+
{D_{15}\over\sqrt6},
\nonumber\\
M_{D^*}
&=&
m_{\mathbf O}
-
{D_8\over2\sqrt3}
+
{D_{15}\over\sqrt6},
\nonumber\\
M_{B^*}
&=&
m_{\mathbf O}
+
{D_8\over2\sqrt3}
-
{D_{15}\over\sqrt6},
\nonumber\\
M_{B_c^*}
&=&
m_{\mathbf O}
-
{D_8\over\sqrt3}
-
{D_{15}\over\sqrt6}.
\label{eq:gsb-udcb-open}
\end{eqnarray}
Using the seven mass inputs
$\rho$, $D^*$, $B^*$, $B_c^*$, $\omega$,
$J/\psi$, and $\Upsilon$, we perform an unweighted
least-squares fit with the five parameters
$m_{\mathbf S}$, $m_{\mathbf O}$, $D_8$,
$D_{15}$, and $r$.
The resulting RMS deviation is
\begin{equation}
{\rm RMS}^{SU(4)_{udcb}}_7
\simeq
44.97~{\rm MeV}.
\label{eq:gsb-leading-rms}
\end{equation}

The large RMS deviation can be understood from the
additive mass structure of the leading Hamiltonian.
As in the flavor-$SU(5)$ analysis in
Sec.~\ref{sec:meson5}, Eq.~(\ref{eq:gsb-udcb-open})
implies the parameter-independent mass relation
\begin{equation}
{\cal N}_{cb}
\equiv
M_{B_c^*}
-
M_{D^*}
-
M_{B^*}
+
M_\rho
=
0
\label{eq:gsb-Ncb-leading}
\end{equation}
within the leading construction.
For the experimental masses adopted in the
present analysis, we have
\begin{equation}
{\cal N}_{cb}^{\rm exp}
=
-219.235~{\rm MeV},
\label{eq:gsb-Ncb-exp}
\end{equation}
revealing a substantial nonadditive mass contribution
that cannot be accommodated by readjusting the
parameters of the leading Hamiltonian.

We next examine whether the daughter symmetry
allows additional response structures that are
absent from the leading parent construction.

\subsection{Higher-order symmetry-breaking responses}
\label{sec:gsb-higher}

Consider the $\lambda_8$ source in the presence of a fixed
$\lambda_{15}$ background. From Eq.~(\ref{eq:gsb-bc-expansion}),
the part of ${\cal H}(b,c)-{\cal H}(b,0)$ linear in $c$ is
\begin{equation}
\Delta {\cal H}_{c,\mathrm{lin}}(b,c)
=
c\left({\cal H}_{01}+b{\cal H}_{11}
+b^2{\cal H}_{21}+\cdots\right).
\label{eq:gsb-daughter-linear-response}
\end{equation}
The first term is inherited from the leading parent Hamiltonian,
whereas the mixed terms may modify the inherited response
or generate additional response structures.
We first determine the structures allowed by the daughter symmetry
and then examine how they can arise from the parent interactions.

The reduction from the parent $SU(4)_{udcb}$ group
to its daughter $SU(3)_{udc}$ subgroup is particularly
transparent for mesons.
The first three flavor indices correspond to $u,d,c$,
while the bottom flavor transforms as a singlet
under $SU(3)_{udc}$.
Consequently, the adjoint representation of
$SU(4)_{udcb}$ decomposes as
\begin{equation}
\mathbf{15}
\longrightarrow
\mathbf{8}
\oplus
\mathbf{3}
\oplus
\mathbf{\bar 3}
\oplus
\mathbf{1}
\qquad
[SU(4) \to SU(3)] .
\label{eq:meson-SU4-SU3-decomposition}
\end{equation}
At the level of the meson fields, this decomposition
may be written schematically as
\begin{equation}
M_{P=1,\ldots,15}
\longrightarrow
\left\{
\begin{array}{l}
L_{P=1,\ldots,8},
\\
H^{B=1,2,3} \oplus H^\dagger_{B=1,2,3},
\\
M_{15},
\end{array}
\right.
\label{eq:gsb-meson-field-decomposition}
\end{equation}
where $L_{P=1,\ldots,8}$ forms the $SU(3)_{udc}$
octet, while $H^{B=1,2,3}$ and
$H^\dagger_{B=1,2,3}$ form the $SU(3)_{udc}$
triplet and antitriplet, respectively, and are
constructed from $M_{9,\ldots,14}$.
We use canonically normalized complex triplet components,
$H^B=(M_{2B+7}-iM_{2B+8})/\sqrt{2}$ for $B=1,2,3$,
so that the invariant triplet mass term is $m_H H^\dagger H$.
The field $M_{15}$ is the $SU(3)_{udc}$ singlet
contained in the parent $SU(4)_{udcb}$ adjoint,
while the parent singlet $M_0$ remains an
independent singlet under $SU(3)_{udc}$.

In this fixed meson flavor space, the daughter
$SU(3)_{udc}$ symmetry permits four independent structures in
the $C$-even quadratic mass contribution linear in the octet
source, with bottom flavor number conserved:
\begin{eqnarray}
{\cal H}^{(c)}_{SU(3)_{udc}}
&=&
{1\over2}D_L d_{8PQ}L_P L_Q
+
{1\over2}D_H H^\dagger\lambda_8 H
\nonumber\\
&&
+
K_0M_0L_8
+
K_{15}M_{15}L_8 .
\label{eq:udcb-daughter-H}
\end{eqnarray}
Here $P,Q=1,\ldots,8$ are summed over, and
$\lambda_8$ in the triplet term denotes
the corresponding $3\times3$ generator of
$SU(3)_{udc}$.
The four coefficients $D_L$, $D_H$, $K_0$, and
$K_{15}$ characterize the response to the
$\lambda_8$ source in the background established
by the $\lambda_{15}$ source.
They are linear in $c$, while their reduced coefficients
may depend on $b$.
The first two terms describe the responses
within the octet and triplet sectors,
respectively, whereas the last two terms
describe the transitions between the octet
and the two singlet components.

By contrast, the leading parent construction
inherited from the three-meson interaction
contains only two independent reduced structures
in the $\lambda_8$ breaking direction,
represented by $D_8$ and $G_8$.
Their embedding into the daughter response space is
\begin{equation}
\left(
\begin{array}{c}
D_L
\\
D_H
\\
K_0
\\
K_{15}
\end{array}
\right)
=
\left(
\begin{array}{cc}
1&0\\
1&0\\
0&1\\
1/\sqrt6&0
\end{array}
\right)
\left(
\begin{array}{c}
D_8
\\
G_8
\end{array}
\right).
\label{eq:udcb-response-embedding}
\end{equation}
Thus,
\begin{equation}
D_L=D_H=D_8,
\quad
K_0=G_8,
\quad
K_{15}={D_8/\sqrt6}.
\label{eq:udcb-parent-relations}
\end{equation}
The leading parent construction generated by the
three-meson interaction therefore spans only a
two-dimensional subspace of the four-dimensional
response space allowed by the daughter
$SU(3)_{udc}$ symmetry.

The reduced symmetry therefore allows two independent response
directions that cannot be generated by merely changing $D_8$
and $G_8$ in the leading construction.
Two convenient combinations that vanish
throughout the inherited subspace are
\begin{equation}
U_1
=
D_H-D_L,
\quad
U_2
=
K_{15}-{D_L/\sqrt6}.
\label{eq:udcb-U12}
\end{equation}
Hence the leading parent construction imposes
$U_1=U_2=0$.

Within the linear daughter-response construction
in Eq.~(\ref{eq:udcb-daughter-H}), the mass
differences in the octet and triplet sectors are
given by
\begin{eqnarray}
M_{D^*}-M_\rho
&=&
-{\sqrt3\over2}D_L,
\nonumber\\
M_{B_c^*}-M_{B^*}
&=&
-{\sqrt3\over2}D_H.
\label{eq:udcb-daughter-mass-differences}
\end{eqnarray}
Consequently, $U_1$ is related to the mixed-heavy
mass combination through
\begin{equation}
{\cal N}_{cb}
\equiv
M_{B_c^*}
-
M_{D^*}
-
M_{B^*}
+
M_\rho
=
-{\sqrt3\over2}U_1 .
\label{eq:udcb-Ncb-U1}
\end{equation}
By contrast, $U_2$ characterizes an independent
response direction involving the neutral
$M_{15}$--$L_8$ mixing.
Unlike $U_1$, it is not isolated by a simple
parameter-independent combination of the
open-flavor masses.
An independent determination of $U_2$ would
therefore generally require additional information
on the neutral mass matrix, including the
corresponding state mixing.

The two missing daughter-response directions
may emerge at the next order in the meson
interaction.
As discussed in Sec.~\ref{sec:meson-leading},
at the four-meson level, two insertions of the
symmetry-breaking background generate responses
quadratic in the breaking coordinates, namely
$b^2$, $bc$, and $c^2$.
The mixed $bc$ sector describes the interplay
between the two breaking directions, whereas
the pure $b^2$ and $c^2$ sectors describe
nonlinear responses along the individual
directions.

At second order in the adjoint breaking source,
the mass sector of $\mathbf{1}\oplus\mathbf{15}$
admits six independent flavor-covariant structures,
four of which belong to the inherited singlet
and adjoint channels.
To investigate the additional response directions,
we consider two representative quartic flavor
contractions,
\begin{eqnarray}
{\cal L}^{(4)}_1
&\sim&
{\rm Tr}(\Phi^\dagger S\Phi S),
\nonumber\\
{\cal L}^{(4)}_2
&\sim&
{1\over4}
{\rm Tr}(\Phi^\dagger S)
{\rm Tr}(S\Phi),
\label{eq:gsb-quartic-contractions}
\end{eqnarray}
where $\Phi=M_P\lambda_P$ denotes the adjoint
meson flavor matrix and $S$ is an adjoint
Lorentz-scalar source, with Lorentz contractions
suppressed.
Setting $S=X=b\lambda_{15}+c\lambda_8$,
these two contractions provide a minimal
realization of the additional response directions,
rather than the most general quartic interaction.

Expanding the two contractions gives six
sectorwise contributions,
\begin{eqnarray}
{\cal O}_{b^2}^{(1)}
&=&
\kappa_{b^2}^{(1)} M_P^\dagger M_Q
{\rm Tr}
\left(
\lambda_P\lambda_{15}
\lambda_Q\lambda_{15}
\right),
\label{eq:gsb-Ob2}
\\[1mm]
{\cal O}_{b^2}^{(2)}
&=&
{\kappa_{b^2}^{(2)}\over4}
M_P^\dagger M_Q
{\rm Tr}(\lambda_P\lambda_{15})
{\rm Tr}(\lambda_{15}\lambda_Q),
\label{eq:gsb-Ob2-2}
\\[1mm]
{\cal O}_{bc}^{(1)}
&=&
\kappa_{bc}^{(1)}\,M_P^\dagger M_Q
\Big[
{\rm Tr}
\left(
\lambda_P\lambda_{15}
\lambda_Q\lambda_8
\right)
\nonumber\\
&&\hspace{17mm}
+
{\rm Tr}
\left(
\lambda_P\lambda_8
\lambda_Q\lambda_{15}
\right)
\Big],
\label{eq:gsb-Obc1}
\\[1mm]
{\cal O}_{bc}^{(2)}
&=&
{\kappa_{bc}^{(2)}\over4}M_P^\dagger M_Q
\Big[
{\rm Tr}(\lambda_P\lambda_{15})
{\rm Tr}(\lambda_8\lambda_Q)
\nonumber\\
&&\hspace{17mm}
+
{\rm Tr}(\lambda_P\lambda_8)
{\rm Tr}(\lambda_{15}\lambda_Q)
\Big],
\label{eq:gsb-Obc2}
\\[1mm]
{\cal O}_{c^2}^{(1)}
&=&
\kappa_{c^2}^{(1)} M_P^\dagger M_Q
{\rm Tr}
\left(
\lambda_P\lambda_8
\lambda_Q\lambda_8
\right),
\label{eq:gsb-Oc21}
\\[1mm]
{\cal O}_{c^2}^{(2)}
&=&
{\kappa_{c^2}^{(2)}\over4}M_P^\dagger M_Q
{\rm Tr}(\lambda_P\lambda_8)
{\rm Tr}(\lambda_8\lambda_Q).
\label{eq:gsb-Oc22}
\end{eqnarray}
Here $P,Q=1,\ldots,15$ denote the adjoint flavor
indices and are summed over.
The coefficient multiplying $M_P^\dagger M_Q$
in each expression gives its contribution to the
mass-matrix element $\delta H_{PQ}$.
The corresponding quadratic mass form in the real-field
convention of Eq.~(\ref{eq:meson-general-H}) is thus one half
of the expression displayed above.
The physical values of $b^2$, $bc$, and $c^2$
have been absorbed into the corresponding
coefficients $\kappa$, which have mass dimension one.
The two $b^2$ contributions are not independent
modulo the inherited lower-order matrix structures.
After separating the inherited contributions,
the $b^2$, $bc$, and $c^2$ sectors contain one,
two, and two additional response directions,
respectively, as summarized in
Sec.~\ref{sec:group-expansion-picture}.

\begin{table}[hbt]
\renewcommand{\arraystretch}{1.25}
\setlength{\tabcolsep}{3pt}
\caption{
Selected higher-order tests in the flavor-$SU(4)_{udcb}$
and $SU(5)_{udscb}$ systems.
The leading Hamiltonians contain five and six parameters,
respectively, and each higher-order extension introduces
one additional fitted parameter.
ST and DT denote the complete single-trace and
double-trace quartic contributions in
Eq.~(\ref{eq:gsb-quartic-general}).
For individual operators, $\kappa_{\rm rep}$ is the
corresponding fitted coefficient.
For the complete ST and DT models, it denotes the
induced $\kappa_{bc}^{(1)}$ and
$\kappa_{bc}^{(2)}$, respectively.
For the $SU(5)$ mixed-response model, it denotes
$\kappa_{bc}^{(1)}$ within the common one-parameter
ansatz of Eq.~(\ref{eq:SU5-mixed-ansatz}).
All reported $\kappa_{\rm rep}$ values are in MeV.
They represent either the additional fitted coefficient
or a coefficient induced by the fitted quartic coupling,
without increasing the stated parameter count.
One representative optimal branch is shown
where multiple parameter sets yield the same spectrum.
}
\label{tab:udcb-expansion}
\begin{ruledtabular}
\begin{tabular}{lrrr}
Contribution
&
$\kappa_{\rm rep}$
&
RMS
&
${\cal N}_{cb}^{\rm fit}$
\\
&
(MeV)
&
(MeV)
&
(MeV)
\\
\hline
\multicolumn{4}{l}{$SU(4)_{udcb}$}
\\
Leading
&
---
&
44.97
&
0
\\
$+{\cal O}_{b^2}^{(1)}$
&
$-102.00$
&
41.43
&
0
\\
$+{\cal O}_{b^2}^{(2)}$
&
$-725.93$
&
41.43
&
0
\\
$+{\cal O}_{bc}^{(1)}$
&
$-37.21$
&
3.60
&
$-210.48$
\\
$+{\cal O}_{bc}^{(2)}$
&
$-179.66$
&
41.43
&
0
\\
$+{\cal O}_{c^2}^{(1)}$
&
$-11.09$
&
44.31
&
$-22.18$
\\
$+{\cal O}_{c^2}^{(2)}$
&
$+160.45$
&
41.43
&
0
\\
$+{\rm ST}$
&
$-36.78$
&
5.77
&
$-225.87$
\\
$+{\rm DT}$
&
$-86.56$
&
41.43
&
0
\\
\hline
\multicolumn{4}{l}{$SU(5)_{udscb}$}
\\
Leading
&
---
&
40.38
&
0
\\
$+{\cal O}_{bc,bs,cs}^{(1)}$
&
$-37.45$
&
5.50
&
$-202.31$
\\
$+{\rm ST}$
&
$-36.19$
&
4.85
&
$-206.35$
\end{tabular}
\end{ruledtabular}
\end{table}

The six contributions are sectorwise representatives,
rather than six independent quartic invariants of
the parent group.
A common quartic invariant can generate correlated
contributions to the $b^2$, $bc$, and $c^2$ sectors.
Here we use independent effective coefficients to test
the individual structures at the physical point.
For each extension, the leading five parameters,
retaining their common-ratio constraint, and one
additional response coefficient are fitted
simultaneously to the seven masses in
Sec.~\ref{sec:gsb-udcb}.
The numerical results are collected in
Table~\ref{tab:udcb-expansion}:
\begin{itemize}

\item The two pure $b^2$ responses in
Eqs.~(\ref{eq:gsb-Ob2}) and (\ref{eq:gsb-Ob2-2})
preserve the daughter $SU(3)_{udc}$ symmetry
and therefore give $\delta{\cal N}_{cb}=0$.
They are equivalent modulo the inherited
lower-order matrix structures.
Both separate fits reach the same lower bound,
giving an identical modest improvement,
\begin{eqnarray}
{\rm RMS}_{7}^{{\cal O}_{b^2}^{(1)}}
&=&
41.43~{\rm MeV},
\label{eq:udcb-b2-rms}
\\
{\rm RMS}_{7}^{{\cal O}_{b^2}^{(2)}}
&=&
41.43~{\rm MeV}.
\label{eq:udcb-b2-2-rms}
\end{eqnarray}

\item The first pure $c^2$ response in
Eq.~(\ref{eq:gsb-Oc21}) modifies the mixed-heavy
mass combination according to
\begin{equation}
\delta{\cal N}_{cb}
=
2\,\kappa_{c^2}^{(1)}.
\label{eq:udcb-c21-Ncb}
\end{equation}
The corresponding one-parameter extension
improves the fit only slightly,
\begin{equation}
{\rm RMS}_{7}^{{\cal O}_{c^2}^{(1)}}
=
44.31~{\rm MeV}.
\label{eq:udcb-c21-rms}
\end{equation}

\item The second pure $c^2$ response in
Eq.~(\ref{eq:gsb-Oc22}) contributes only to
the neutral isoscalar sector and therefore
gives $\delta{\cal N}_{cb}=0$.
The corresponding fit gives
\begin{equation}
{\rm RMS}_{7}^{{\cal O}_{c^2}^{(2)}}
=
41.43~{\rm MeV}.
\label{eq:udcb-c22-rms}
\end{equation}

\item The first mixed $bc$ response in
Eq.~(\ref{eq:gsb-Obc1}) generates the isovector
and open-flavor mass contributions
\begin{eqnarray}
&&
\left(
\delta M_\rho,\,
\delta M_{D^*},\,
\delta M_{B^*},\,
\delta M_{B_c^*}
\right)
\nonumber\\
&&\hspace{5mm}
=
\kappa_{bc}^{(1)}
\left(
{2\sqrt2\over3},\,
-{\sqrt2\over3},\,
-\sqrt2,\,
2\sqrt2
\right),
\label{eq:udcb-bc1-pattern}
\end{eqnarray}
and hence
\begin{equation}
\delta{\cal N}_{cb}
=
4\sqrt2\,
\kappa_{bc}^{(1)}.
\label{eq:udcb-bc1-Ncb}
\end{equation}
This response directly opens the $U_1$ direction
identified above.
Its inclusion substantially improves the fit,
\begin{equation}
{\rm RMS}_{7}^{{\cal O}_{bc}^{(1)}}
=
3.60~{\rm MeV}.
\label{eq:udcb-bc1-fit}
\end{equation}

\item The second mixed $bc$ response in
Eq.~(\ref{eq:gsb-Obc2}) reduces to
\begin{equation}
{\cal O}_{bc}^{(2)}
=
\kappa_{bc}^{(2)}
\left(
M_{15}^\dagger M_8
+
M_8^\dagger M_{15}
\right).
\label{eq:udcb-bc2-neutral}
\end{equation}
It therefore generates an independent neutral
$M_8$--$M_{15}$ response and opens the $U_2$
direction, while leaving the isovector and
open-flavor masses unchanged directly.
Hence, $\delta{\cal N}_{cb}=0$.
The corresponding fit gives
\begin{equation}
{\rm RMS}_{7}^{{\cal O}_{bc}^{(2)}}
=
41.43~{\rm MeV}.
\label{eq:udcb-bc2-rms}
\end{equation}

\end{itemize}

The two mixed responses provide an explicit
realization of the additional directions
allowed by the daughter symmetry.
Evaluating their contributions to the daughter coefficients gives
\begin{equation}
\delta\left(
\begin{array}{c}
D_L\\ D_H\\ K_0\\ K_{15}
\end{array}
\right)
=
\left(
\begin{array}{cc}
2\sqrt6/3&0\\
-2\sqrt6&0\\
0&0\\
2/3&1
\end{array}
\right)
\left(
\begin{array}{c}
\kappa_{bc}^{(1)}\\
\kappa_{bc}^{(2)}
\end{array}
\right).
\label{eq:udcb-mixed-embedding}
\end{equation}
The first operator also contributes within the inherited subspace.
The independent new directions are isolated by the combinations
in Eq.~(\ref{eq:udcb-U12}), which give
\begin{eqnarray}
\delta U_1
&=&
-{8\sqrt6\over3}\kappa_{bc}^{(1)},
\nonumber\\
\delta U_2
&=&
\kappa_{bc}^{(2)}.
\label{eq:udcb-mixed-U12}
\end{eqnarray}
The two contributions to $(U_1,U_2)$ are linearly independent.
Together with the two inherited directions, they span the complete
four-dimensional daughter linear-response space.
Thus, a response structure that is absent at
first order in the parent symmetry-breaking
expansion may emerge at second order, while
contributing linearly to the subsequent breaking
of the daughter symmetry.

As summarized in Table~\ref{tab:udcb-expansion}, the first mixed
$bc$ response gives the largest improvement among the
six individual operator tests. It directly accesses
the nonadditive mass direction associated with $U_1$. By contrast,
the second mixed response opens $U_2$ without contributing directly
to the four masses in ${\cal N}_{cb}$.
The four extensions retaining ${\cal N}_{cb}=0$ saturate
the same lower bound,
${\rm RMS}_7\geq |{\cal N}_{cb}^{\rm exp}|/(2\sqrt7)$,
to the quoted precision, although their mass-matrix
structures are different.
Note also that the pure $c^2$ response
${\cal O}_{c^2}^{(1)}$ generates a nonzero ${\cal N}_{cb}$.
The observed nonadditivity therefore need not arise
exclusively from mixed responses.

The six quadratic contributions considered above
are not necessarily independent when they originate
from common quartic interactions.
To examine their relations, we consider the most general
quartic flavor polynomial of a single traceless
Hermitian adjoint matrix $A$,
\begin{equation}
{\cal L}^{(4)}_{\rm meson}
=
\underbrace{g_1{\rm Tr}(A^4)}_{\rm ST}
+
\underbrace{g_2\left[{\rm Tr}(A^2)\right]^2}_{\rm DT},
\label{eq:gsb-quartic-general}
\end{equation}
where ST and DT denote the single-trace and
double-trace contributions, respectively,
and the flavor singlet is treated separately.
Writing $A=X+\Phi$, with
$X=b\lambda_{15}+c\lambda_8$ and
$\Phi=M_P\lambda_P$, the quadratic contributions
are
\begin{eqnarray}
\delta{\cal H}^{(4)}_{\rm meson}
&\sim&
g_1\left[
4{\rm Tr}(\Phi^2X^2)
+
2{\rm Tr}(\Phi X\Phi X)
\right]
\nonumber\\
&+&
g_2\left[
2{\rm Tr}(X^2){\rm Tr}(\Phi^2)
+
4\left[{\rm Tr}(X\Phi)\right]^2
\right].
\label{eq:gsb-quartic-expansion}
\end{eqnarray}
Here the background insertion is a flavor-space
construction for the effective mass kernel, rather
than a Lorentz-vector condensate. The resulting
fixed contraction weights define a restricted
common-source ansatz, not the most general
vector-meson quartic interaction.
This expansion contains both modifications of
the inherited responses and additional response
directions.

In the normalization of the six operators above,
their coefficients satisfy
\begin{eqnarray}
\left(
\kappa_{b^2}^{(1)},
\kappa_{bc}^{(1)},
\kappa_{c^2}^{(1)}
\right)
&=&
4g_1\left(b^2,bc,c^2\right),
\nonumber\\
\left(
\kappa_{b^2}^{(2)},
\kappa_{bc}^{(2)},
\kappa_{c^2}^{(2)}
\right)
&=&
32g_2\left(b^2,bc,c^2\right).
\label{eq:gsb-quartic-coefficient-relations}
\end{eqnarray}
Consequently, for nonzero denominators,
\begin{equation}
\frac{\kappa_{b^2}^{(1)}}{\kappa_{bc}^{(1)}}
=
\frac{\kappa_{bc}^{(1)}}{\kappa_{c^2}^{(1)}}
=
\frac{\kappa_{b^2}^{(2)}}{\kappa_{bc}^{(2)}}
=
\frac{\kappa_{bc}^{(2)}}{\kappa_{c^2}^{(2)}}
=
\frac{b}{c}.
\label{eq:gsb-quartic-common-ratio}
\end{equation}
These relations refer to the displayed operator
basis before absorbing inherited contributions.
They follow from the chosen common quartic
structure rather than from flavor symmetry alone.
Following the leading common-source relation
in Eq.~(\ref{eq:meson-general-breaking-ratio}),
we adopt
\begin{equation}
\frac{b}{c}
=
R_{15/8}
=
\frac{D_{15}}{D_8}
\label{eq:gsb-quartic-source-ratio}
\end{equation}
for the following fits, with the common source
normalization absorbed into $g_1$ and $g_2$.
Here $D_{15}$ and $D_8$ denote the leading-source
parameters before any higher-order redefinition.
This identification is an additional assumption
of the restricted model.

We next perform simultaneous fits using the complete
quadratic contributions in
Eq.~(\ref{eq:gsb-quartic-expansion}),
including their inherited components.
We obtain
\begin{eqnarray}
{\rm RMS}_7^{\rm ST}
&=&
5.77~{\rm MeV},
\nonumber\\
{\rm RMS}_7^{\rm DT}
&=&
41.43~{\rm MeV}.
\label{eq:gsb-quartic-common-fits}
\end{eqnarray}
The complete ST interaction gives a substantially
better fit than the leading Hamiltonian,
although its RMS remains above that of the isolated
${\cal O}_{bc}^{(1)}$ response.
The combined ST+DT model reproduces all seven
input masses with seven fitted parameters,
but this exact interpolation does not constitute
an independent validation of the quartic interactions.

As a final numerical test, we extend the first
mixed response to the flavor-$SU(5)_{udscb}$
system, where the three diagonal breaking
directions are characterized by $\lambda_8$, $\lambda_{15}$,
and $\lambda_{24}$ in the ordered basis $(u,d,s,c,b)$.
We use the same single-trace structure with separated
source insertions as in Eq.~(\ref{eq:gsb-Obc1}),
with the generator pairs $(15,24)$,
$(8,24)$, and $(8,15)$ for the $bc$, $bs$, and $cs$ sectors,
respectively. The three coefficients are related through the
common one-parameter ansatz
\begin{eqnarray}
\kappa_{bc}^{(1)}
&=&
\kappa D_{15}D_{24},
\nonumber\\
\kappa_{bs}^{(1)}
&=&
\kappa D_8D_{24},
\nonumber\\
\kappa_{cs}^{(1)}
&=&
\kappa D_8D_{15}.
\label{eq:SU5-mixed-ansatz}
\end{eqnarray}
Here $\kappa$ is a common phenomenological
coefficient with inverse mass dimension.
The resulting seven-parameter fit to the eleven
vector-meson masses reduces the RMS deviation
from $40.38$~MeV to
\begin{equation}
{\rm RMS}_{11}^{{\cal O}_{bc,bs,cs}^{(1)}}
=
5.50~{\rm MeV}.
\end{equation}
In particular, the three null combinations
are shifted from their leading values
${\cal N}_c={\cal N}_b={\cal N}_{cb}=0$
to approximately
\begin{eqnarray}
{\cal N}_c^{\rm fit}
&=&
-5.45~{\rm MeV},
\nonumber\\
{\cal N}_b^{\rm fit}
&=&
-21.25~{\rm MeV},
\nonumber\\
{\cal N}_{cb}^{\rm fit}
&=&
-202.31~{\rm MeV},
\end{eqnarray}
to be compared with the experimental values
$-14.705$~MeV, $-26.953$~MeV, and
$-219.235$~MeV, respectively.
The improvement is especially pronounced
in the mixed-heavy $bc$ sector, indicating
that this minimal mixed-response extension
captures part of the nonadditive mass structure
inaccessible to the leading construction.

For comparison, we also consider the complete ST
quadratic response generated by ${\rm Tr}(A^4)$,
with the $SU(5)$ breaking source
$X=D_8\lambda_8+D_{15}\lambda_{15}
+D_{24}\lambda_{24}$.
Using the same eleven input masses and seven
fitted parameters, this restricted common-source
model gives
${\rm RMS}_{11}^{\rm ST}=4.85~{\rm MeV}$,
slightly below the mixed-only result of
$5.50~{\rm MeV}$.
The corresponding nonadditive combinations are
${\cal N}_c^{\rm fit}=-4.67~{\rm MeV}$,
${\cal N}_b^{\rm fit}=-19.88~{\rm MeV}$, and
${\cal N}_{cb}^{\rm fit}=-206.35~{\rm MeV}$.
Although the overall fit improves, the first
two combinations are described slightly less
accurately than in the mixed-only model.
The complete ST response includes correlated
mixed, pure-square, and inherited contributions,
whose relative weights are fixed by the
common-source ansatz.
These results illustrate the role of higher-order
responses without establishing a unique
microscopic origin for them.

\subsection{Group expansion}
\label{sec:group-expansion-picture}

The preceding examples illustrate two ways in which the response
of an effective Hamiltonian may evolve as the symmetry is broken.
Higher-order interactions may modify the strengths of structures
already present at lower orders, or may generate additional
independent response directions. We refer to these effects as
\textit{coefficient evolution} and \textit{response unlocking},
respectively.
For the $\lambda_8$ response in the $SU(4)_{udcb}$ example,
coefficient evolution corresponds to changes in the two inherited
coefficients $D_8$ and $G_8$. These changes alone cannot release
the constraints $U_1=U_2=0$ in Eq.~(\ref{eq:udcb-U12}).
The two mixed responses in Eq.~(\ref{eq:udcb-mixed-U12}), on the
other hand, provide independent contributions outside this
inherited subspace. They thus realize response unlocking, raising
the accessible daughter linear-response rank from two to four.

More generally, suppose a leading response has the form
\begin{equation}
{\cal R}_{\rm Lead}=a_1{\cal O}_1+a_2{\cal O}_2,
\end{equation}
where ${\cal O}_1$ and ${\cal O}_2$ are independent flavor
matrices. In an additional symmetry-breaking background, it may
acquire contributions of the form
\begin{equation}
{\cal R}
=
a^\prime_1{\cal O}_1+a^\prime_2{\cal O}_2
+a^\prime_3{\cal O}_3+a^\prime_4{\cal O}_4,
\label{eq:gsb-response-unlocking}
\end{equation}
where the last two matrices are independent modulo the leading
response space. Changes within that space represent coefficient
evolution, while a nonzero component outside it represents
response unlocking. The two effects can occur simultaneously.
Their distinction is relative to the inherited response space,
not to a particular choice of basis for it. Symmetry determines
which structures are allowed, whereas their strengths are
determined dynamically.

To organize these effects, we regroup Eq.~(\ref{eq:gsb-bc-expansion})
according to the total order in the breaking coordinates,
\begin{equation}
{\cal H}(b,c)
={\cal H}^{(0)}+{\cal H}^{(1)}+{\cal H}^{(2)}+\cdots ,
\label{eq:gsb-order-expansion}
\end{equation}
where
\begin{eqnarray}
{\cal H}^{(0)}&=&{\cal H}_{00},\nonumber\\
{\cal H}^{(1)}&=&b{\cal H}_{10}+c{\cal H}_{01},\nonumber\\
{\cal H}^{(2)}&=&b^2{\cal H}_{20}+bc{\cal H}_{11}
+c^2{\cal H}_{02}.
\label{eq:gsb-order-components}
\end{eqnarray}

\begin{table}[hbt]
\caption{
Order-by-order response structure in the $SU(4)_{udcb}$ example.
The last column gives the numbers of matrix directions at zeroth
and first orders and the additional directions in each quadratic
sector after removing inherited structures in that sector.
The two $b^2$ operator structures are equivalent modulo
the inherited lower-order matrix structures.
These sectorwise counts are not numbers of independent global
couplings.
}
\label{tab:gsb-order-counting}
\begin{ruledtabular}
\begin{tabular}{ccc}
Order & Representative coefficients & Directions \\
\hline
$O(1)$ & $m_{\mathbf S},\,m_{\mathbf O}$ & 2 \\
$O(b)$ & $D_{15},\,G_{15}$ & 2 \\
$O(c)$ & $D_8,\,G_8$ & 2 \\
$O(b^2)$ & $\kappa_{b^2}^{(1)},\,\kappa_{b^2}^{(2)}$ & 1 \\
$O(bc)$ & $\kappa_{bc}^{(1)},\,\kappa_{bc}^{(2)}$ & 2 \\
$O(c^2)$ & $\kappa_{c^2}^{(1)},\,\kappa_{c^2}^{(2)}$ & 2
\end{tabular}
\end{ruledtabular}
\end{table}

Table~\ref{tab:gsb-order-counting} summarizes the meson response
structures. The zeroth- and first-order matrices span six
independent directions if their coefficients are allowed to vary
independently. In the leading common-source mass model, however,
the relation
\begin{equation}
G_{15}=rD_{15},\qquad G_8=rD_8
\label{eq:gsb-common-r}
\end{equation}
reduces the number of independent effective parameters to five:
\begin{equation}
m_{\mathbf S},\quad m_{\mathbf O},\quad D_{15},\quad D_8,\quad r.
\label{eq:gsb-leading-five}
\end{equation}
This parameter restriction is distinct from a linear dependence
among the flavor matrices.

At quadratic order, the dimensions of the inherited and extended
matrix spaces in the individual sectors are $4\to5$ for $b^2$,
$2\to4$ for $bc$, and $4\to6$ for $c^2$. In this counting the
inherited space consists of the singlet and adjoint mass tensors
available in that sector, including their background-dependent
coefficients. The increments $1,2,2$ cannot simply be added to
obtain the global number of independent quadratic couplings.
For example, at a fixed generic point with $b,c\ne0$, the six
lower-order matrices and the six quadratic contributions in
Sec.~\ref{sec:gsb-higher} span ten, rather than twelve,
independent matrix directions.
The reduction reflects linear dependencies within and
between the quadratic sectors.
Moreover, the common ST and DT interactions impose additional
relations among the quadratic coefficients.
This fixed-point rank does not specify the number of parameters
identifiable from the physical mass spectrum.

We use the term ``group expansion'' for the successive expansion
of the effective Hamiltonian away from a symmetry limit, with
particular emphasis on coefficient evolution and response
unlocking. Schematically,
\begin{eqnarray}
{\cal H}^{(0)}
&\longrightarrow&
{\cal H}^{(0)}+{\cal H}^{(1)}
\nonumber\\
&\longrightarrow&
{\cal H}^{(0)}+{\cal H}^{(1)}+{\cal H}^{(2)}
\nonumber\\
&\longrightarrow&\cdots .
\label{eq:gsb-group-expansion}
\end{eqnarray}
For a general set of symmetry-breaking coordinates, we write the
formal expansion as
\begin{equation}
{\cal H}(\boldsymbol{\xi})
=\sum_{n=0}^{\infty}{\cal H}^{(n)}(\boldsymbol{\xi}),
\label{eq:gsb-general-expansion}
\end{equation}
where ${\cal H}^{(n)}$ is homogeneous of total order $n$ in
$\boldsymbol{\xi}$. The parent symmetry organizes the allowed
structures at each order, while the reduced symmetry can admit
a larger response space than that inherited at leading order.
The expansion does not presume that every allowed structure is
generated, nor that a low-order truncation converges at the
physical point.

The group-theoretical foundation of this expansion can be
formulated in terms of the covariance of the Hamiltonian
family. Let $G$ denote the parent symmetry group, and let
$U(g)$ be its representation on the reference space $R_H$.
For an adjoint breaking source $X$, transforming as
$X\to gXg^\dagger$, covariance requires
\begin{equation}
{\cal H}(gXg^\dagger)
=
U(g){\cal H}(X)U^\dagger(g).
\label{eq:gsb-parent-covariance}
\end{equation}
Here the reference representation is chosen covariantly,
and the other fixed control parameters preserve $G$.
Before restricting the source to diagonal directions, the
allowed homogeneous structures at order $n$ are classified by
\begin{equation}
{\cal C}_n
=
{\rm Hom}_G\!\left({\rm Sym}^n R_X,
{\rm End}(R_H)\right),
\label{eq:gsb-covariant-space}
\end{equation}
where $R_X$ is the representation of the breaking source
(the adjoint in the present example), and the group acts on
${\rm End}(R_H)$ by conjugation.
For Hermitian sources and mass matrices, the physical
structures are taken in the corresponding real Hermitian
space, subject also to further symmetries, such as charge
conjugation when applicable.
The dimension of this covariant space after imposing the
relevant restrictions counts independent homogeneous
structures before evaluation at a particular source.
It need not coincide with the rank of the matrices at a
fixed source, or with the number of independently
identifiable parameters in the mass spectrum.

For example, in the present $SU(4)$ meson system,
$R_X=\mathbf{15}$ and
$R_H=\mathbf{1}\oplus\mathbf{15}$.
The symmetric products decompose as
\begin{eqnarray}
{\rm Sym}^2(\mathbf{15})
&=&
\mathbf{1}\oplus\mathbf{15}
\oplus\mathbf{20}^\prime\oplus\mathbf{84},
\nonumber\\
{\rm Sym}^2(\mathbf{1}\oplus\mathbf{15})
&=&
2\cdot\mathbf{1}\oplus2\cdot\mathbf{15}
\oplus\mathbf{20}^\prime\oplus\mathbf{84}.
\label{eq:gsb-meson-covariant-decomposition}
\end{eqnarray}
Restricting ${\cal C}_n$ to the real symmetric mass
structures considered here, denoted by
${\cal C}_n^{\rm phys}$, we obtain
\begin{equation}
\dim{\cal C}_0^{\rm phys}=2,\qquad
\dim{\cal C}_1^{\rm phys}=2,\qquad
\dim{\cal C}_2^{\rm phys}=6.
\label{eq:gsb-meson-covariant-counting}
\end{equation}
At quadratic order, four structures are associated
with the singlet and adjoint channels, while two
additional structures are associated with the
$\mathbf{20}^\prime$ and $\mathbf{84}$ channels.
The latter provide new flavor tensor structures
that are absent at lower orders.

\begin{figure}[hbt]
\centering
\includegraphics[width=\columnwidth]{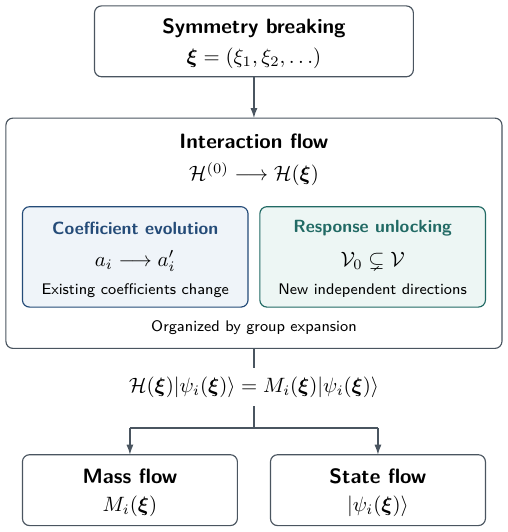}
\caption{
Schematic illustration of the group expansion framework.
As the symmetry is broken, the effective Hamiltonian may 
evolve through coefficient evolution and response unlocking.
The former modifies the coefficients of existing response
structures, while the latter opens additional independent
response directions, extending the inherited response space
$\mathcal V_0$ to a larger space $\mathcal V$.
Through the eigenvalue equation, the interaction flow
can give rise to both the mass flow and the state flow.
}
\label{fig:group-expansion}
\end{figure}

Within the present framework, coefficient evolution and
response unlocking describe two aspects of the interaction
flow, namely the evolution of the effective Hamiltonian
as the symmetry is broken.
Through the eigenvalue equation in Eq.~(\ref{eq:gsb-eigen}),
this evolution can change both the eigenvalues and
eigenstates of the Hamiltonian, giving rise to the mass
flow and state flow, respectively.
The interaction flow is thus the common origin of
the mass flow and the state flow.
The overall physical picture of group expansion is illustrated
in Fig.~\ref{fig:group-expansion}.

In particular, the near-ideal flavor compositions of the
neutral vector mesons do not imply an additive mass
Hamiltonian, as demonstrated by the nonadditive masses
in Sec.~\ref{sec:gsb-higher}.
Having established the order-by-order picture,
we now turn to a continuous description of the
interaction flow and its relation to the mass
and state flows.

\subsection{Differential description of breaking responses}
\label{sec:gsb-differential}

Having established the order-by-order picture of group expansion,
we now develop a differential description of the interaction flow.
In particular, we wish to examine how the response to one
symmetry-breaking source changes in the presence of another.
Let us return to the flavor-$SU(4)_{udcb}$ example
discussed in Sec.~\ref{sec:gsb-udcb}.
Suppose that the system is already at a point $(b,c)$
in the symmetry-breaking plane, and that an additional
infinitesimal source $\delta c\lambda_8$ is introduced.
The change in the mass Hamiltonian depends not only on
$\delta c$ but also, in general, on the existing background.
We therefore define the response operator associated with
the $\lambda_8$ direction as
\begin{equation}
B_c(b,c)
\equiv
\lim_{\delta c\to0}
\frac{{\cal H}(b,c+\delta c)-{\cal H}(b,c)}{\delta c}
=
\frac{\partial{\cal H}(b,c)}{\partial c},
\label{eq:gsb-response-c}
\end{equation}
where the Hamiltonians at different parameter points
are represented in the same reference space,
and all other control parameters are held fixed.
Similarly, we define
\begin{equation}
B_b(b,c)
=
\frac{\partial{\cal H}(b,c)}{\partial b}.
\label{eq:gsb-response-b}
\end{equation}
These operators describe the change in the effective
Hamiltonian per unit change in the corresponding
symmetry-breaking coordinate.

Using the expansion in Eq.~(\ref{eq:gsb-bc-expansion}),
we obtain
\begin{eqnarray}
B_c(b,c)
&=&
{\cal H}_{01}
+b{\cal H}_{11}
+2c{\cal H}_{02}
+\cdots,
\nonumber\\
B_b(b,c)
&=&
{\cal H}_{10}
+2b{\cal H}_{20}
+c{\cal H}_{11}
+\cdots .
\label{eq:gsb-response-bc-expansion}
\end{eqnarray}
In particular, the response to the $\lambda_8$
source in the presence of the $\lambda_{15}$
background is given by
\begin{equation}
B_c(b,0)
=
{\cal H}_{01}
+b{\cal H}_{11}
+b^2{\cal H}_{21}
+\cdots .
\label{eq:gsb-daughter-response-differential}
\end{equation}
The linear daughter response introduced in
Eq.~(\ref{eq:gsb-daughter-linear-response})
can thus be expressed as
\begin{equation}
\Delta{\cal H}_{c,\mathrm{lin}}(b,c)
=
cB_c(b,0).
\label{eq:gsb-daughter-differential}
\end{equation}
Thus, the parent contributions proportional to $b^nc$
all enter the first-order daughter response, even though
they belong to different orders of the parent expansion.

Although the $\lambda_{15}$ source breaks the parent
$SU(4)_{udcb}$ symmetry, it preserves the daughter
$SU(3)_{udc}$ symmetry at $c=0$.
Nevertheless, Eq.~(\ref{eq:gsb-daughter-response-differential})
shows that the response to the subsequent breaking of
$SU(3)_{udc}$ may depend on $b$.
More generally, a physical factor may influence the
response to a symmetry-breaking source without itself
breaking the symmetry under consideration.
To characterize this influence, we consider an
infinitesimal change in the background $b$ and
define the corresponding cross-response operator,
\begin{eqnarray}
K_{cb}(b,c)
&\equiv&
\lim_{\delta b\to0}
\frac{
B_c(b+\delta b,c)-B_c(b,c)
}{
\delta b
}
\nonumber\\
&=&
\frac{\partial B_c(b,c)}{\partial b}
=
\frac{\partial^2{\cal H}(b,c)}
{\partial b\,\partial c}.
\label{eq:gsb-cross-response}
\end{eqnarray}
The operator $K_{cb}$ describes how the response
to the $\lambda_8$ source changes as the background
established by the $\lambda_{15}$ source is varied.
The strength of a background and its effect on a given
response are therefore distinct quantities.

More generally, for a set of symmetry-breaking
coordinates $\boldsymbol{\xi}=(\xi_1,\xi_2,\ldots)$,
we introduce the first- and second-order response
operators
\begin{eqnarray}
B_a(\boldsymbol{\xi})
&=&
\frac{\partial{\cal H}(\boldsymbol{\xi})}
{\partial\xi_a},
\nonumber\\
K_{ab}(\boldsymbol{\xi})
&=&
\frac{\partial B_a(\boldsymbol{\xi})}
{\partial\xi_b}
=
\frac{\partial^2{\cal H}(\boldsymbol{\xi})}
{\partial\xi_b\,\partial\xi_a}.
\label{eq:gsb-general-response}
\end{eqnarray}
The pure responses $K_{aa}$ characterize nonlinearities
along individual breaking directions, whereas the mixed
responses $K_{ab}$ with $a\neq b$ characterize how the
response along one direction changes along another.
The same definitions apply to additional control
coordinates that preserve the symmetry under study.
Depending on the background and the symmetry under
consideration, the response operators may contain both
symmetry-preserving and symmetry-breaking contributions.
For a twice continuously differentiable Hamiltonian
in a fixed reference space, we have
\begin{equation}
K_{ab}(\boldsymbol{\xi})
=
K_{ba}(\boldsymbol{\xi}).
\label{eq:gsb-response-symmetry}
\end{equation}
Higher-order responses can be defined by further
differentiation, provided the required derivatives exist.

In the vicinity of the exact symmetry limit,
the Hamiltonian can then be expanded as
\begin{equation}
{\cal H}(\boldsymbol{\xi})
=
{\cal H}(\boldsymbol{0})
+\sum_a\xi_a B_a(\boldsymbol{0})
+\frac{1}{2}
\sum_{ab}\xi_a\xi_b K_{ab}(\boldsymbol{0})
+ \cdots .
\label{eq:gsb-response-taylor-expansion}
\end{equation}
This is the local second-order form of the group expansion
in Eq.~(\ref{eq:gsb-general-expansion}).
For the two-direction example considered above,
the expansion coefficients satisfy
\begin{eqnarray}
B_b(0,0)&=&{\cal H}_{10},
\qquad
B_c(0,0)={\cal H}_{01},
\nonumber\\
K_{bb}(0,0)&=&2{\cal H}_{20},
\nonumber\\
K_{bc}(0,0)&=&K_{cb}(0,0)={\cal H}_{11},
\nonumber\\
K_{cc}(0,0)&=&2{\cal H}_{02}.
\label{eq:gsb-response-coefficient-relations}
\end{eqnarray}
These are relations between local expansion coefficients
and derivatives at the symmetric point, while effective coefficients
fitted at a distant physical point do not, by themselves,
determine these derivatives.

The differential description also allows us to
characterize the coefficient evolution and response
unlocking introduced in
Sec.~\ref{sec:group-expansion-picture}.
For the $\lambda_8$ response in the present meson example,
let ${\cal V}_0$ denote the inherited two-dimensional
space spanned by the flavor matrices associated with
$D_8$ and $G_8$.
Along $c=0$, the daughter symmetry allows a fixed
four-dimensional linear-response space containing
${\cal V}_0$, while the parent symmetry at $b=0$
restricts $B_c(0,0)$ to ${\cal V}_0$.
We introduce the orthogonal projection ${\cal P}_0$
onto ${\cal V}_0$ with respect to a fixed matrix inner
product, for example $(A,B)={\rm Tr}(AB)$ for Hermitian
matrices in an orthonormal reference basis.
The cross-response operator along the background
direction can then be decomposed as
\begin{equation}
K_{cb}(b,0)
=
K_{cb}^{\parallel}(b,0)
+
K_{cb}^{\perp}(b,0),
\label{eq:gsb-cross-response-decomposition}
\end{equation}
where
\begin{eqnarray}
K_{cb}^{\parallel}(b,0)
&=&
{\cal P}_0 K_{cb}(b,0),
\nonumber\\
K_{cb}^{\perp}(b,0)
&=&
(1-{\cal P}_0)K_{cb}(b,0).
\label{eq:gsb-cross-response-projection}
\end{eqnarray}
The parallel component describes changes in the
response within the inherited space, corresponding
to coefficient evolution.
The perpendicular component describes changes in
the additional response outside that space,
providing a differential description of
response unlocking.

Assuming the required smoothness throughout the
interval of integration, we integrate the cross-response
operator along the background direction to obtain
\begin{equation}
B_c(b,0)
=
B_c(0,0)
+
\int_0^b
K_{cb}(b^\prime,0)\,db^\prime .
\label{eq:gsb-background-response-integral}
\end{equation}
Since $B_c(0,0)$ belongs to the inherited response
space ${\cal V}_0$, the additional response generated
in the presence of the background is given by
\begin{equation}
B_c^\perp(b,0)
\equiv
(1-{\cal P}_0)B_c(b,0)
=
\int_0^b
K_{cb}^{\perp}(b^\prime,0)\,db^\prime .
\label{eq:gsb-response-unlocking-integral}
\end{equation}
Similarly, the change of the inherited component is
\begin{equation}
{\cal P}_0B_c(b,0)-B_c(0,0)
=
\int_0^b K_{cb}^{\parallel}(b^\prime,0)\,db^\prime.
\label{eq:gsb-coefficient-evolution-integral}
\end{equation}
A nonzero $K_{cb}^{\perp}$ signals a local change
outside the inherited space. Whether response unlocking
occurs at a finite background is determined by the
accumulated contribution in
Eq.~(\ref{eq:gsb-response-unlocking-integral}), since
contributions from different intervals may cancel.
Coefficient evolution and response unlocking may occur
simultaneously.
At quadratic order, the two mixed responses constructed
in Sec.~\ref{sec:gsb-higher} span the additional response
directions after projection onto the complement of
${\cal V}_0$.

In the vector-meson example, the two combinations
$U_1$ and $U_2$ defined in
Eq.~(\ref{eq:udcb-U12}) vanish throughout the
inherited response space.
To apply this test to $B_c(b,0)$ itself, we remove the
explicit factor of $c$ included in the daughter coefficients.
For a linear daughter Hamiltonian with nonzero $c$,
$U_1/c$ and $U_2/c$ are the corresponding reduced
combinations. Their simultaneous vanishing is equivalent
to $B_c(b,0)\in{\cal V}_0$.
Although the orthogonal decomposition above
depends on the chosen matrix inner product,
whether a response lies outside the inherited
space is independent of the choice of basis
or complementary subspace.

We may further characterize the nonadditivity
between two symmetry-breaking directions by
introducing the finite Hamiltonian difference
\begin{equation}
{\cal I}_{bc}(b,c)
\equiv
{\cal H}(b,c)-{\cal H}(b,0)
-{\cal H}(0,c)+{\cal H}(0,0).
\label{eq:gsb-nonadditive-Hamiltonian}
\end{equation}
This difference removes all contributions depending
only on $b$ or only on $c$. It vanishes identically
for a Hamiltonian that is additive in the two coordinates.
If the Hamiltonian is twice continuously differentiable
throughout the corresponding rectangle in the breaking
plane, the differential response operators give
\begin{equation}
{\cal I}_{bc}(b,c)
=
\int_0^b db^\prime
\int_0^c dc^\prime\,
K_{cb}(b^\prime,c^\prime).
\label{eq:gsb-nonadditive-integral}
\end{equation}
Close to the exact symmetry limit, its leading possible
contribution is
\begin{equation}
{\cal I}_{bc}(b,c)
=
bcK_{cb}(0,0)
+\cdots
=
bc{\cal H}_{11}+\cdots .
\label{eq:gsb-nonadditive-leading}
\end{equation}
The finite nonadditive contribution between the
two breaking directions can therefore be understood
as the accumulated cross response over the
corresponding region of the parameter space.

It is important to distinguish
${\cal I}_{bc}$ from the experimental mass
combination ${\cal N}_{cb}$ introduced in
Eq.~(\ref{eq:gsb-Ncb-leading}).
The former compares Hamiltonians at different
points in the symmetry-breaking parameter space,
whereas the latter compares the masses of
different physical states at a single physical point.
For example, the pure $c^2$ term in
Eq.~(\ref{eq:gsb-Oc21}) can contribute to
${\cal N}_{cb}$ without contributing to ${\cal I}_{bc}$.
A nonzero mass combination ${\cal N}_{cb}$ therefore
does not by itself establish a nonzero cross response
$K_{cb}$.

For the vector-meson example, this connection can be
made explicit. Within the adopted flavor reference space,
the isovector and open-flavor sectors entering
${\cal N}_{cb}$ are invariant under the diagonal sources
and have no additional states with the same conserved
quantum numbers within the retained mass kernel.
We can therefore choose fixed normalized flavor states
for the four masses. For a mass-matrix operator $A$,
we define the linear functional
\begin{eqnarray}
{\cal L}_{cb}(A)
&\equiv&
\langle B_c^*|A|B_c^*\rangle
-\langle D^*|A|D^*\rangle
\nonumber\\
&&
-\langle B^*|A|B^*\rangle
+\langle\rho|A|\rho\rangle,
\label{eq:gsb-mass-functional}
\end{eqnarray}
where the states denote these fixed reference states.
The mass combination along the breaking plane is then
\begin{equation}
{\cal N}_{cb}(b,c)
=
{\cal L}_{cb}\!\left({\cal H}(b,c)\right).
\label{eq:gsb-mass-functional-H}
\end{equation}
In particular, since the functional is independent of
$b$ and $c$, we have
\begin{equation}
\partial_b\partial_c{\cal N}_{cb}(b,c)
=
{\cal L}_{cb}\!\left(K_{cb}(b,c)\right),
\label{eq:gsb-mass-cross-response}
\end{equation}
where $\partial_b=\partial/\partial b$ and
$\partial_c=\partial/\partial c$.

Applying ${\cal L}_{cb}$ to
Eq.~(\ref{eq:gsb-nonadditive-Hamiltonian}), we obtain
\begin{equation}
{\cal L}_{cb}({\cal I}_{bc})
=
{\cal N}_{cb}(b,c)-{\cal N}_{cb}(b,0)
-{\cal N}_{cb}(0,c)+{\cal N}_{cb}(0,0).
\label{eq:gsb-mass-finite-difference}
\end{equation}
Since the daughter $SU(3)_{udc}$ symmetry is exact
at $c=0$, we have
${\cal N}_{cb}(b,0)={\cal N}_{cb}(0,0)=0$.
Using Eq.~(\ref{eq:gsb-nonadditive-integral}),
we thus obtain
\begin{equation}
{\cal N}_{cb}(b,c)
=
{\cal N}_{cb}(0,c)
+\int_0^b db^\prime\int_0^c dc^\prime\,
{\cal L}_{cb}\!\left(K_{cb}(b^\prime,c^\prime)\right).
\label{eq:gsb-mass-nonadditivity-integral}
\end{equation}
The first term describes the mass nonadditivity generated
by the $c$ source alone, while the second describes
its modification by the $b$ background.
The pure $c^2$ and mixed $bc$ contributions studied in
Sec.~\ref{sec:gsb-higher} illustrate these two origins,
respectively, at quadratic order.
This separation is defined for the specified Hamiltonian
family and breaking coordinates. Mass data at the physical
point alone do not determine the two terms separately.

The same functional connects the local mass response
at $c=0$ to the additional daughter-response direction
$U_1$. From Eq.~(\ref{eq:udcb-Ncb-U1}), we find
\begin{equation}
\left.\frac{\partial{\cal N}_{cb}(b,c)}{\partial c}
\right|_{c=0}
=
{\cal L}_{cb}\!\left(B_c(b,0)\right)
=
-\frac{\sqrt3}{2}\lim_{c\to0}\frac{U_1(b,c)}{c},
\label{eq:gsb-U1-mass-response}
\end{equation}
where $U_1(b,c)$ is evaluated in the linear daughter
Hamiltonian. This mass combination probes one of the two
additional daughter-response directions, but does not
probe $U_2$ directly. These identities rely on the fixed
flavor sectors of the present mass kernel.
In a sector with state mixing, the eigenvalues cannot
in general be extracted by a fixed linear functional
of the Hamiltonian.

Given the response operators $B_a$ along a specified
path $\boldsymbol{\xi}(t)$ and an initial Hamiltonian,
the interaction flow follows from
\begin{equation}
\frac{d{\cal H}}{dt}
=
\sum_a B_a(\boldsymbol{\xi}(t))\frac{d\xi_a}{dt},
\label{eq:gsb-interaction-flow-path}
\end{equation}
where $t$ is a path parameter, not physical time.
Having developed a differential description of the
interaction flow, we now turn to its physical
manifestations through the mass flow and the state flow.

\subsection{Mass flow and state flow}
\label{sec:gsb-interaction-state-flow}

To examine how the interaction flow gives rise to
the mass flow and the state flow, we return to
the eigenvalue equation in Eq.~(\ref{eq:gsb-eigen}),
\begin{equation}
{\cal H}(\boldsymbol{\xi})
|\psi_i(\boldsymbol{\xi})\rangle
=
M_i(\boldsymbol{\xi})
|\psi_i(\boldsymbol{\xi})\rangle .
\end{equation}
We assume an energy-independent Hermitian mass Hamiltonian
in a fixed, orthonormal reference space, with a specified
identification of that space at different parameter points.
We choose a complete orthonormal eigenbasis satisfying
\begin{equation}
\langle\psi_j(\boldsymbol{\xi})|
\psi_i(\boldsymbol{\xi})\rangle
=
\delta_{ji}.
\end{equation}
We first consider isolated and nondegenerate
eigenstates that vary smoothly with the symmetry-breaking
coordinates.

Differentiating the eigenvalue equation with respect
to the symmetry-breaking coordinate $\xi_a$, we obtain
\begin{equation}
B_a|\psi_i\rangle
+
{\cal H}|\partial_a\psi_i\rangle
=
(\partial_a M_i)|\psi_i\rangle
+
M_i|\partial_a\psi_i\rangle ,
\label{eq:gsb-differential-eigen}
\end{equation}
where $\partial_a\equiv\partial/\partial\xi_a$,
and $B_a$ is defined in Eq.~(\ref{eq:gsb-general-response}).
Multiplying Eq.~(\ref{eq:gsb-differential-eigen})
by $\langle\psi_i|$ from the left, we obtain
the Hellmann--Feynman relation
\begin{equation}
\frac{\partial M_i}{\partial\xi_a}
=
\langle\psi_i|B_a|\psi_i\rangle .
\label{eq:gsb-mass-flow}
\end{equation}
This relation describes the local mass flow.
It shows that the local mass response per unit change
in the breaking coordinate is determined by
the diagonal matrix element of the corresponding
response operator in the instantaneous physical
eigenstate.

Similarly, multiplying
Eq.~(\ref{eq:gsb-differential-eigen})
by $\langle\psi_j|$ with $j\neq i$, we obtain
\begin{equation}
\langle\psi_j|\partial_a\psi_i\rangle
=
\frac{
\langle\psi_j|B_a|\psi_i\rangle
}{
M_i-M_j
}.
\label{eq:gsb-state-flow-component}
\end{equation}
Choosing a local phase convention in which
$\langle\psi_i|\partial_a\psi_i\rangle=0$
along the direction under consideration, we have
\begin{equation}
\frac{\partial|\psi_i\rangle}{\partial\xi_a}
=
\sum_{j\neq i}
|\psi_j\rangle
\frac{
\langle\psi_j|B_a|\psi_i\rangle
}{
M_i-M_j
}.
\label{eq:gsb-state-flow}
\end{equation}
This relation describes the local state flow.
In contrast to the mass flow in
Eq.~(\ref{eq:gsb-mass-flow}), the state flow
is governed by the off-diagonal matrix elements
of the response operator, together with the
mass separations between the corresponding
physical eigenstates.
At exact degeneracies, including those of the parent multiplets,
Eq.~(\ref{eq:gsb-state-flow}) must be replaced by degenerate
perturbation theory or a description of the corresponding
eigenspaces.
Eqs.~(\ref{eq:gsb-mass-flow}) and
(\ref{eq:gsb-state-flow}) illustrate that the
mass flow and state flow are two complementary
manifestations of the same interaction flow.
In the instantaneous eigenbasis, the diagonal matrix
elements of a symmetry-breaking response determine the
local mass shifts, while its off-diagonal matrix elements,
together with the mass separations, govern the local
rotation of the eigenstates.

The resulting state evolution also depends
on the spectrum of the Hamiltonian.
For a given off-diagonal response matrix element,
the state flow can be enhanced when the mass
separation is small and suppressed when it is
large.
Thus, the strength of the symmetry-breaking
response alone does not determine the extent
to which the physical states are reorganized.
This feature is closely related to the discussion
of ideal flavor mixing in
Sec.~\ref{sec:meson-ideal-mixing}.
When the mass separations grow relative to the
corresponding off-diagonal response matrix elements,
local changes in the eigenstates can be suppressed.
This provides a possible mechanism for the stabilization
of nearly ideal flavor configurations. However, the
compositions at the physical point alone do not determine
the preceding state flow or its local rate of change.

The mass flow in
Eq.~(\ref{eq:gsb-mass-flow}) depends on both
the response operator and the physical state
in which it is evaluated.
Consequently, its variation with another
symmetry-breaking coordinate contains
contributions from both the interaction flow
and the state flow.
Differentiating with respect to $\xi_b$, we obtain
\begin{equation}
\partial_b\partial_a M_i
=
\langle\psi_i|K_{ab}|\psi_i\rangle
+
\langle\partial_b\psi_i|B_a|\psi_i\rangle
+
\langle\psi_i|B_a|\partial_b\psi_i\rangle .
\label{eq:gsb-mass-flow-second}
\end{equation}
Using Eq.~(\ref{eq:gsb-state-flow}), this expression
can be written as
\begin{equation}
\partial_b\partial_a M_i
=
\langle\psi_i|K_{ab}|\psi_i\rangle
+2\,{\rm Re}\sum_{j\neq i}
\frac{\langle\psi_i|B_a|\psi_j\rangle
\langle\psi_j|B_b|\psi_i\rangle}{M_i-M_j}.
\label{eq:gsb-second-order-physical-response}
\end{equation}
The first term is the direct interaction response,
arising from the change in the operator $B_a$ itself.
The second term is the response induced by state evolution,
arising from the change in the state on which $B_a$ acts.
The two terms together determine the second-order mass response.
For the fixed flavor sectors entering ${\cal N}_{cb}$ in
Sec.~\ref{sec:gsb-differential}, the state-evolution
contribution is absent within the adopted reference space.
In sectors with state mixing, both contributions must
in general be taken into account.

In particular, along a single symmetry-breaking
direction, we obtain
\begin{equation}
\frac{\partial^2M_i}{\partial\xi_a^2}
=
\langle\psi_i|K_{aa}|\psi_i\rangle
+
2\sum_{j\neq i}
\frac{
|\langle\psi_j|B_a|\psi_i\rangle|^2
}{
M_i-M_j
}.
\label{eq:gsb-diagonal-mass-response}
\end{equation}
Even when the Hamiltonian depends linearly on
the symmetry-breaking coordinate, so that
$K_{aa}=0$, the physical masses may still
exhibit nonlinear responses as a consequence
of state evolution.
Likewise, the absence of an explicit mixed
response $K_{ab}$ in the Hamiltonian does not,
in general, imply the absence of a mixed
response in its eigenvalues.
It is therefore important to distinguish
higher-order responses of the Hamiltonian
from those of its physical eigenvalues.
The former characterize the evolution of the
effective interactions, whereas the latter
also contain contributions from the
reorganization of the physical states.

The separation in Eq.~(\ref{eq:gsb-second-order-physical-response})
is defined for the specified effective Hamiltonian and fixed
identification of its reference space.
Under a parameter-dependent change of basis,
additional derivative terms enter the individual
contributions, while the physical mass response is unchanged.
Moreover, effects of configurations already eliminated in
constructing the effective Hamiltonian must not be counted again
as explicit state-evolution contributions in the retained space.

To illustrate the distinction between interaction
and physical responses, we consider a simple
two-state system described by
\begin{equation}
{\cal H}(x,y)
=
\left(
\begin{array}{cc}
E_1(y) & xv \\
xv & E_2(y)
\end{array}
\right),
\label{eq:gsb-background-two-state}
\end{equation}
where $x$ denotes a symmetry-breaking coordinate,
$y$ denotes an additional background coordinate,
and $v$ is real and independent of both coordinates.
The two basis states may, for example, belong to different
symmetry sectors at $x=0$, with the term $xv$ connecting them.
The two diagonal energies are allowed to depend
on the background $y$.
We assume that the unperturbed mass separation
\begin{equation}
\Delta(y)=E_2(y)-E_1(y)
\end{equation}
is positive in the parameter region considered.

The response operator associated with the
symmetry-breaking coordinate $x$ is
\begin{equation}
B_x
=
\frac{\partial{\cal H}}{\partial x}
=
\left(
\begin{array}{cc}
0 & v\\
v & 0
\end{array}
\right),
\end{equation}
which is independent of the background $y$.
Consequently, the corresponding cross-response
operator vanishes,
\begin{equation}
K_{xy}
=
\frac{\partial B_x}{\partial y}
=
0.
\label{eq:gsb-two-state-zero-cross-response}
\end{equation}
Nevertheless, the eigenvalues of the Hamiltonian
are given by
\begin{equation}
M_\pm(x,y)
=
\frac{E_1(y)+E_2(y)}{2}
\pm
\frac{1}{2}
\sqrt{\Delta^2(y)+4x^2v^2},
\label{eq:gsb-background-two-state-masses}
\end{equation}
and the corresponding mixing angle satisfies
\begin{equation}
|\tan2\theta|
=
\frac{2|xv|}{\Delta(y)}.
\label{eq:gsb-background-two-state-angle}
\end{equation}
Both the mass shifts $M_\pm(x,y)-M_\pm(0,y)$ and the
state mixing can therefore depend on the background $y$
through $\Delta(y)$, although the operator $B_x$ itself
does not.

In particular, the first-order mass responses
along the $x$ direction vanish at $x=0$,
whereas the corresponding second-order
responses are given by
\begin{equation}
\left.
\frac{\partial^2M_\pm}{\partial x^2}
\right|_{x=0}
=
\pm\frac{2v^2}{\Delta(y)}.
\label{eq:gsb-background-two-state-curvature}
\end{equation}
The entire $x$ curvature in this example originates
from the state-evolution term in
Eq.~(\ref{eq:gsb-diagonal-mass-response}).
Its dependence on the background arises from
the changing mass separation rather than from
a modification of the response operator itself.
At $x=0$, the mixed second-order mass response
$\partial_y\partial_x M_\pm$ also vanishes.
Instead, the background dependence in
Eq.~(\ref{eq:gsb-background-two-state-curvature})
appears in the pure $x$ curvature.

More generally, an additional physical factor can
influence symmetry breaking in two distinct ways.
It may modify the response operators themselves,
as described by the cross-response operators introduced
in Sec.~\ref{sec:gsb-differential}.
It may also modify the masses and compositions of the
states on which these operators act, thereby changing
the physical responses even when the corresponding
interaction response remains unchanged.
The distinction between these two mechanisms
is particularly useful in studying systems in
which several symmetry-breaking mechanisms
or approximate symmetries coexist.
It shows that a suppression of state mixing
does not necessarily imply a corresponding
suppression of the underlying symmetry-breaking
interaction.
Conversely, a small symmetry-breaking interaction
may produce a substantial state reorganization
when the relevant states are nearly degenerate.

In the following sections, we apply this
framework to the baryon sector, where the
additional flavor and spin degrees of freedom
provide further possibilities for the evolution
and restriction of symmetry-breaking responses.
A complementary graphical description of
possible state-flow directions in flavor
representation space is provided by Young
tableaux in Appendix~\ref{sec:young-breaking}.

\section{Fully symmetric baryons}
\label{sec:20S}

Having established a general framework for describing
symmetry-breaking responses, we now extend our investigation
from vector mesons to baryons.
Compared with the quark--antiquark reference configurations
of mesons, the three-quark reference configurations of baryons
provide additional possibilities for internal flavor correlations.
Moreover, the different permutation symmetries of the three
quarks give rise to different flavor structures and,
consequently, different symmetry-breaking responses.

In this section, we investigate the ground-state baryons
with $J^P=3/2^+$, whose lowest three-quark reference
configurations possess fully symmetric flavor structures.
Within this relatively simple reference space, we examine
how higher-order symmetry-breaking responses arise beyond
the leading additive mass structure.
We further investigate the possible connections between
the symmetry-breaking responses of these baryons and
those of the vector mesons studied in Sec.~\ref{sec:meson}.

\subsection{Leading symmetry-breaking Hamiltonian}
\label{sec:baryon-S-leading}

For a general flavor $SU(N_f)$ group, we introduce the
fully symmetric three-quark flavor components
\begin{equation}
\Delta^P
=
\mathbb{S}^P_{ABC}\,
q^A q^B q^C,
\label{def:Delta}
\end{equation}
where the flavor indices run over $A,B,C=1,\ldots,N_f$,
and $\mathbb{S}^P_{ABC}$ is a totally symmetric flavor
tensor, whose normalization and explicit construction
are presented in Appendix~\ref{app:baryon-S-matrices}.
The index $P$ runs over
\begin{equation}
P=1,\ldots,{N_f(N_f+1)(N_f+2)/6},
\end{equation}
where the upper limit is the dimension of the fully
symmetric three-quark flavor representation.
Here and throughout this section, the color, Lorentz,
and spin structures are suppressed to make the flavor
structure explicit.
The corresponding flavor representations are
\begin{eqnarray}
SU(4) &:& \mathbf{20_S},
\nonumber\\
SU(5) &:& \mathbf{35_S}.
\end{eqnarray}
These representations provide the lowest three-quark
flavor reference configurations for the ground-state
baryons with $J^P=3/2^+$, rather than specifying their
complete physical wave functions.

Following the construction in
Sec.~\ref{sec:meson-leading}, we combine the singlet
and adjoint flavor components $M^N$ with the baryon
and antibaryon components $\Delta^P$ and $\bar\Delta^P$
to construct the leading flavor-invariant interactions:
\begin{eqnarray}
{\cal L}_{\rm baryon}^{\mathbf S}
&=&
g_{\bar\Delta\Delta}\,
\delta_{PQ}\,
M^0\bar\Delta^P\Delta^Q
\nonumber\\
&+&
g_{M\bar\Delta\Delta}\,
\left[{\bf F}_{\Delta}\right]^N_{PQ}\,
M^N\bar\Delta^P\Delta^Q,
\label{eq:baryon-S-L3}
\end{eqnarray}
where $N=1,\ldots,N_f^2-1$ denotes the adjoint index,
and repeated indices are summed over.
The matrices ${\bf F}_{\Delta}$ describe the unique
linear flavor structure for an adjoint source within
each of the fully symmetric reference representations.
We refer to this structure as the $F$-type response.
The tensor construction and normalization conventions
for these matrices are given in
Appendix~\ref{app:baryon-S-matrices}.

As in the vector-meson sector, these expressions
specify schematic flavor-invariant contractions.
Replacing the singlet and adjoint insertions in
Eq.~(\ref{eq:baryon-S-L3}) by the corresponding
Lorentz-scalar background components introduced in
Eq.~(\ref{eq:meson-condensates}), we obtain the leading
mass term in the fully symmetric reference space:
\begin{eqnarray}
{\cal H}_{\rm baryon}^{\mathbf S}
&=&
m_{\mathbf S}\,
\delta_{PQ}\,
\bar\Delta^P\Delta^Q
\nonumber\\
&+&
F_C^{\mathbf S}\,
\left[{\bf F}_{\Delta}\right]^C_{PQ}\,
\bar\Delta^P\Delta^Q,
\label{eq:baryon-S-leading-H}
\end{eqnarray}
where $C$ is summed over the diagonal
symmetry-breaking directions.
Within this leading background-insertion construction,
we have
\begin{eqnarray}
m_{\mathbf S}
&=&
g_{\bar\Delta\Delta}\langle M^0\rangle,
\nonumber\\
F_C^{\mathbf S}
&=&
g_{M\bar\Delta\Delta}\langle M^C\rangle.
\label{eq:baryon-S-leading-parameters}
\end{eqnarray}
More general flavor-symmetric contributions can be
absorbed into $m_{\mathbf S}$. The reduced couplings
are not fixed by flavor symmetry.
It follows that the responses along two different
symmetry-breaking directions satisfy
\begin{equation}
{F_C^{\mathbf S}\over F_{C^\prime}^{\mathbf S}}
=
{\langle M^C\rangle\over\langle M^{C^\prime}\rangle}
\equiv
R_{C/C^\prime}
\label{eq:baryon-S-breaking-ratio}
\end{equation}
within the leading common-source construction,
provided that the denominators are nonzero.
This relation is analogous to
Eq.~(\ref{eq:meson-general-breaking-ratio})
for the vector mesons.
The proportionality relation is retained for the
linear coefficients in the common-source Hamiltonians
used below. It does not, however, constrain all
higher-order flavor responses allowed by symmetry.

We first consider flavor $SU(4)_Q$, where $Q=c$ or $b$,
with the ordered flavor basis $(u,d,s,Q)$.
Under the light-flavor $SU(3)$ subgroup, the fully
symmetric representation is decomposed as
\begin{equation}
\mathbf{20_S}
\longrightarrow
\mathbf{10}\oplus\mathbf{6}
\oplus\mathbf{3}\oplus\mathbf{1}
\qquad
[SU(4) \to SU(3)] ,
\end{equation}
where the four representations contain zero, one,
two, and three heavy quarks, respectively.
In the isospin-symmetric limit, the corresponding
mass matrix takes the form
\begin{equation}
{\cal H}_{\mathbf{20_S}}
=
m_{\mathbf S}{\bf 1}_{20}
+
F_{15}^{\mathbf S}
\left[{\bf F}_{\Delta}\right]^{15}
+
F_8^{\mathbf S}
\left[{\bf F}_{\Delta}\right]^8.
\label{eq:baryon-20S-leading-H}
\end{equation}
This matrix is diagonal in the symmetric
flavor-occupation basis. 
Evaluating the matrices ${\bf F}_{\Delta}$
from the normalized flavor-tensor basis and
transition identities presented in
Appendix~\ref{app:baryon-S-matrices},
we obtain the following mass eigenvalues:
\begin{eqnarray}
   \nonumber M_{\Delta}
&=& m_{\mathbf{S}}
    + {F_{15}^{\mathbf{S}}\over\sqrt{6}}
    + {F_8^{\mathbf{S}}\over\sqrt{3}} \, ,
\\ \nonumber M_{\Sigma^*}
&=& m_{\mathbf{S}}
    + {F_{15}^{\mathbf{S}}\over\sqrt{6}} \, ,
\\ \nonumber M_{\Xi^*}
&=& m_{\mathbf{S}}
    + {F_{15}^{\mathbf{S}}\over\sqrt{6}}
    - {F_8^{\mathbf{S}}\over\sqrt{3}} \, ,
\\ \nonumber M_{\Omega}
&=& m_{\mathbf{S}}
    + {F_{15}^{\mathbf{S}}\over\sqrt{6}}
    - {2 F_8^{\mathbf{S}}\over\sqrt{3}} \, ,
\\ \nonumber M_{\Sigma_Q^*}
&=& m_{\mathbf{S}}
    - {F_{15}^{\mathbf{S}}\over\sqrt{54}}
    + {2 F_8^{\mathbf{S}}\over\sqrt{27}} \, ,
\\           M_{\Xi_Q^*}
&=& m_{\mathbf{S}}
    - {F_{15}^{\mathbf{S}}\over\sqrt{54}}
    - {F_8^{\mathbf{S}}\over\sqrt{27}} \, ,
\label{mass:20S}
\\ \nonumber M_{\Omega_Q^*}
&=& m_{\mathbf{S}}
    - {F_{15}^{\mathbf{S}}\over\sqrt{54}}
    - {4 F_8^{\mathbf{S}}\over\sqrt{27}} \, ,
\\ \nonumber M_{\Xi_{QQ}^*}
&=& m_{\mathbf{S}}
    - {5 F_{15}^{\mathbf{S}}\over\sqrt{54}}
    + {F_8^{\mathbf{S}}\over\sqrt{27}} \, ,
\\ \nonumber M_{\Omega_{QQ}^*}
&=& m_{\mathbf{S}}
    - {5 F_{15}^{\mathbf{S}}\over\sqrt{54}}
    - {2 F_8^{\mathbf{S}}\over\sqrt{27}} \, ,
\\ \nonumber M_{\Omega_{QQQ}}
&=& m_{\mathbf{S}}
    - \sqrt{3\over2} F_{15}^{\mathbf{S}} \, ,
\end{eqnarray}
where $Q=c$ or $b$.

We perform two independent unweighted least-squares
fits to the observed baryon masses in the charm and
bottom systems, treating
$m_{\mathbf S}$, $F_{15}^{\mathbf S}$, and
$F_8^{\mathbf S}$ as independent effective parameters
in each system.
For the charm system, we use the four light-baryon
masses and the three singly-charmed baryon masses,
whereas for the bottom system, we use the same four
light-baryon masses and the two available
singly-bottom baryon masses.
We adopt the experimental mass inputs listed in
Table~\ref{tab:baryon-20S-leading-masses}, using arithmetic
averages over the measured charge states where appropriate.
In particular, we adopt a representative Breit--Wigner
mass of $1232$~MeV for the $\Delta$ baryon and use
the average of the two measured charged
$\Sigma_b^*$ masses.

\begin{table*}[t]
\renewcommand{\arraystretch}{1.22}
\setlength{\tabcolsep}{2.6pt}
\small
\caption{
Leading and quadratic (Sym) mass fits for the fully symmetric
$J^P=3/2^+$ baryons in flavor $SU(4)_c$, $SU(4)_b$,
and $SU(5)_{udscb}$. The fits use seven, six, and nine
experimental masses, respectively, including the same four
light baryons. Leading fits have three parameters in $SU(4)$
and four in $SU(5)$; each Sym fit adds one quadratic coupling
and refits the leading parameters. The quoted $\kappa_{\mathbf S}$
denotes $\kappa_{\mathbf S}^{(5)}$ in the $SU(5)$ columns.
All $\kappa_{ab}$ are induced by the common-source Sym model,
not independently fitted; the heavy--heavy curvatures and
masses without experimental inputs are conditional model
extrapolations. All dimensional entries are in MeV except
$\kappa_{\mathbf S}$, quoted in units of $10^{-5}$ MeV$^{-1}$.
}
\label{tab:baryon-20S-leading-masses}
\begin{ruledtabular}
\begin{tabular}{lrrrrrrr}
& Experiment
& \multicolumn{2}{c}{$SU(4)_c$}
& \multicolumn{2}{c}{$SU(4)_b$}
& \multicolumn{2}{c}{$SU(5)_{udscb}$} \\
& & Leading & $+{\rm Sym}$
& Leading & $+{\rm Sym}$
& Leading & $+{\rm Sym}$ \\
\hline
$m_{\mathbf S}$ & -- & $2293.70$ & $2317.93$ & $4783.82$ & $4864.73$ & $4830.77$ & $4875.61$ \\
$F_8^{\mathbf S}$ & -- & $-243.22$ & $-226.13$ & $-250.60$ & $-220.92$ & $-241.02$ & $-206.38$ \\
$F_{15}^{\mathbf S}$ & -- & $-2224.86$ & $-2142.19$ & $-8329.59$ & $-7912.44$ & $-2224.47$ & $-1899.87$ \\
$F_{24}^{\mathbf S}$ & -- & -- & -- & -- & -- & $-8021.42$ & $-7547.67$ \\
$\kappa_{\mathbf S}$ ($10^{-5}$ MeV$^{-1}$) & -- & -- & $-7.845$ & -- & $-2.415$ & -- & $-3.289$ \\
$\kappa_{ss}$ & -- & -- & $-2.67$ & -- & $-0.79$ & -- & $-0.93$ \\
$\kappa_{cs}$ & -- & -- & $-24.78$ & -- & -- & -- & $-8.42$ \\
$\kappa_{bs}$ & -- & -- & -- & -- & $-26.80$ & -- & $-33.51$ \\
$\kappa_{cc}$ & -- & -- & $-229.55$ & -- & -- & -- & $-75.85$ \\
$\kappa_{bb}$ & -- & -- & -- & -- & $-913.80$ & -- & $-1202.74$ \\
$\kappa_{bc}$ & -- & -- & -- & -- & -- & -- & $-302.04$ \\
RMS & -- & $10.80$ & $1.79$ & $7.60$ & $2.51$ & $10.47$ & $6.90$ \\
\hline
$\Delta$ & $1232.00$ & $1244.98$ & $1233.58$ & $1238.59$ & $1234.64$ & $1246.88$ & $1239.91$ \\
$\Sigma^*$ & $1384.61$ & $1385.40$ & $1383.39$ & $1383.27$ & $1382.48$ & $1386.04$ & $1384.32$ \\
$\Xi^*$ & $1533.40$ & $1525.83$ & $1530.52$ & $1527.96$ & $1529.54$ & $1525.19$ & $1527.79$ \\
$\Omega$ & $1672.45$ & $1666.25$ & $1674.98$ & $1672.64$ & $1675.80$ & $1664.35$ & $1670.32$ \\
\hline
$\Sigma_c^*$ & $2518.10$ & $2502.85$ & $2519.13$ & -- & -- & $2504.12$ & $2507.65$ \\
$\Xi_c^*$ & $2645.71$ & $2643.27$ & $2644.16$ & -- & -- & $2643.27$ & $2643.63$ \\
$\Omega_c^*$ & $2766.00$ & $2783.69$ & $2766.52$ & -- & -- & $2782.42$ & $2778.69$ \\
\hline
$\Xi_{cc}^*$ & -- & $3760.71$ & $3575.12$ & -- & -- & $3761.35$ & $3699.53$ \\
$\Omega_{cc}^*$ & -- & $3901.14$ & $3675.38$ & -- & -- & $3900.50$ & $3827.10$ \\
$\Omega_{ccc}$ & -- & $5018.58$ & $4401.58$ & -- & -- & $5018.58$ & $4815.56$ \\
\hline
$\Sigma_b^*$ & $5832.53$ & -- & -- & $5820.87$ & $5832.69$ & $5823.64$ & $5837.75$ \\
$\Xi_b^*$ & $5953.90$ & -- & -- & $5965.56$ & $5953.74$ & $5962.79$ & $5948.64$ \\
$\Omega_b^*$ & -- & -- & -- & $6110.24$ & $6073.99$ & $6101.95$ & $6058.60$ \\
\hline
$\Xi_{bb}^*$ & -- & -- & -- & $10403.16$ & $9516.94$ & $10400.39$ & $9232.84$ \\
$\Omega_{bb}^*$ & -- & -- & -- & $10547.84$ & $9611.19$ & $10539.55$ & $9310.22$ \\
$\Omega_{bbb}$ & -- & -- & -- & $14985.44$ & $12287.38$ & $14977.15$ & $11425.19$ \\
\hline
$\Xi_{bc}^*$ & -- & -- & -- & -- & -- & $7080.87$ & $6803.44$ \\
$\Omega_{bc}^*$ & -- & -- & -- & -- & -- & $7220.02$ & $6905.91$ \\
$\Omega_{ccb}^*$ & -- & -- & -- & -- & -- & $8338.10$ & $7693.28$ \\
$\Omega_{cbb}^*$ & -- & -- & -- & -- & -- & $11657.62$ & $9896.49$ \\
\end{tabular}
\end{ruledtabular}
\end{table*}

The resulting parameters and fitted masses are
also given in Table~\ref{tab:baryon-20S-leading-masses}.
The leading flavor-symmetry-breaking Hamiltonian
provides a reasonable description of the overall
baryon mass spectra, with RMS mass deviations of
$10.80$ and $7.60$~MeV for the charm and bottom systems,
respectively. These deviations characterize the fit
quality on the adopted data set, rather than the
uncertainty of predictions for unmeasured baryons.
Nevertheless, systematic deviations from the
experimental masses remain.

It is interesting to compare the symmetry-breaking
responses of the fully symmetric baryons with
those of the vector mesons.
From the above fitting results, we obtain
\begin{eqnarray}
\left.
{F_{15}^{\mathbf S}\over F_8^{\mathbf S}}
\right|_{SU(4)_c}
&=&
9.15 \, ,
\nonumber\\*
\left.
{F_{15}^{\mathbf S}\over F_8^{\mathbf S}}
\right|_{SU(4)_b}
&=&
33.24 \, .
\label{eq:F15F8-20S}
\end{eqnarray}
These values are to be compared with
$R_{15/8}^{SU(4)_c}=10.20$ and
$R_{15/8}^{SU(4)_b}=39.79$
obtained for the vector mesons in
Eq.~(\ref{eq:meson-R158-cb}).
Although the baryon ratios are smaller than the
corresponding meson ratios by approximately
$10.3\%$ and $16.5\%$, respectively,
both sectors exhibit a similar hierarchy between
the light- and heavy-flavor symmetry-breaking
responses.

Within the leading common-source construction,
these ratios would be identical if the same
symmetry-breaking backgrounds were employed
for both sectors.
The differences between the independently fitted
ratios thus indicate the limitations of the simple
leading common-source description at the physical
flavor-breaking points.
They may reflect different evolutions of the effective
responses in the two sectors, but do not by themselves
determine the underlying dynamics.
We shall further investigate the possible connections
between these responses in the following subsections.

A further limitation of the leading mass Hamiltonian
can be understood from its additive structure.
In analogy with Eq.~(\ref{eq:meson-additive-mass}),
the mass of a fully symmetric baryon with the
reference flavor configuration $q_iq_jq_k$
can be expressed as
\begin{equation}
M_{ijk}
=
m_{\mathbf S}
+
{\sigma_i^{\mathbf S}
+\sigma_j^{\mathbf S}
+\sigma_k^{\mathbf S}\over3},
\label{eq:baryon-additive-mass}
\end{equation}
where
\begin{equation}
\sigma_i^{\mathbf S}
=
\sum_C F_C^{\mathbf S}
\left[\lambda_C\right]_{ii},
\label{eq:baryon-single-flavor-response}
\end{equation}
with the sum running over $C=8,15$ in the present case.
Hence, within the leading construction, the mass
shifts associated with the three flavor slots
enter additively.
Let $n_Q$ and $n_s$ now denote the numbers of heavy and
strange quarks, respectively.
In the isospin-symmetric limit, the ten mass formulas
in Eq.~(\ref{mass:20S}) can therefore be expressed
in a unified form:
\begin{equation}
M_{\mathbf{20_S}}(n_Q,n_s)
=
A_0+B_Qn_Q+B_sn_s \, ,
\label{eq:baryon-20S-mass-plane}
\end{equation}
where
\begin{eqnarray}
   \nonumber A_0
&=& m_{\mathbf S}
+ {F_{15}^{\mathbf S}\over\sqrt6}
+ {F_8^{\mathbf S}\over\sqrt3} \, ,
\\ \nonumber B_Q
&=& -{4F_{15}^{\mathbf S}\over3\sqrt6}
    -{F_8^{\mathbf S}\over3\sqrt3} \, ,
\\           B_s
&=& -{F_8^{\mathbf S}\over\sqrt3} \, .
\end{eqnarray}
The additive mass structure thus gives rise to
an affine mass plane in the flavor-occupation
coordinates $(n_Q,n_s)$. 

Consequently, the leading mass formulas imply the
following equal-spacing relations:
\begin{eqnarray}
   \nonumber M_{\Sigma^*}-M_\Delta
&=& M_{\Xi^*}-M_{\Sigma^*}
\\ \nonumber
&=& M_\Omega-M_{\Xi^*}
\\ \nonumber
&=& M_{\Xi_Q^*}-M_{\Sigma_Q^*}
\\ \nonumber
&=& M_{\Omega_Q^*}-M_{\Xi_Q^*}
\\ \nonumber
&=& M_{\Omega_{QQ}^*}-M_{\Xi_{QQ}^*}
\\
&=& B_s \, .
\label{eq:baryon-20S-equal-spacing}
\end{eqnarray}
The first two equalities reproduce the familiar
light-baryon decuplet equal-spacing rule.
More generally, within the leading
symmetry-breaking Hamiltonian, the mass increment
associated with replacing a light quark by a strange
quark is independent of the flavor composition of
the remaining two quarks.
The analogous mass increment for replacing a light
quark by a heavy quark is $B_Q$.
This is the baryonic counterpart of the leading additive
structure of the off-diagonal flavor configurations
in the vector-meson sector.

However, the experimental baryon masses exhibit
clear departures from these equal-spacing relations.
In the following numerical mass differences, $M_X$
denotes the adopted experimental mass of baryon $X$.
For the light baryons, we obtain
\begin{eqnarray}
   \nonumber M_{\Sigma^*}-M_\Delta
&=& 152.61~{\rm MeV} \, ,
\\ \nonumber M_{\Xi^*}-M_{\Sigma^*}
&=& 148.79~{\rm MeV} \, ,
\\           M_\Omega-M_{\Xi^*}
&=& 139.05~{\rm MeV} \, ,
\label{eq:baryon-light-exp-spacing}
\end{eqnarray}
whereas for the singly-heavy baryons, we obtain
\begin{eqnarray}
   \nonumber M_{\Xi_c^*}-M_{\Sigma_c^*}
&=& 127.61~{\rm MeV} \, ,
\\ \nonumber M_{\Omega_c^*}-M_{\Xi_c^*}
&=& 120.29~{\rm MeV} \, ,
\\           M_{\Xi_b^*}-M_{\Sigma_b^*}
&=& 121.37~{\rm MeV} \, .
\label{eq:baryon-heavy-exp-spacing}
\end{eqnarray}
Although all these mass differences correspond to the
same replacement of a light quark by a strange quark
in the reference configurations, they depend on the
flavor composition of the remaining two slots.
In particular, the increments are smaller for the
singly-heavy baryons considered here than for the
light baryons.

This nonadditive behavior can be quantified by
constructing appropriate combinations of the
experimental baryon masses.
For the light baryons, we obtain
\begin{eqnarray}
{\cal N}_{ss}^{q,{\rm exp}}
&\equiv&
M_{\Xi^*}-2M_{\Sigma^*}+M_\Delta
= -3.82~{\rm MeV} \, ,
\nonumber\\[1mm]
{\cal N}_{ss}^{s,{\rm exp}}
&\equiv&
M_\Omega-2M_{\Xi^*}+M_{\Sigma^*}
= -9.74~{\rm MeV} \, .
\label{eq:baryon-nonadditivity-ss}
\end{eqnarray}
For the singly-charmed baryons, the corresponding
combination is
\begin{equation}
{\cal N}_{ss}^{c,{\rm exp}}
\equiv
M_{\Omega_c^*}-2M_{\Xi_c^*}+M_{\Sigma_c^*}
= -7.32~{\rm MeV} \, .
\label{eq:baryon-nonadditivity-ss-c}
\end{equation}
Moreover, comparing the strange-quark replacement
in the light and singly-heavy baryons, we obtain
\begin{eqnarray}
{\cal N}_{cs}^{q,{\rm exp}}
&\equiv&
M_{\Xi_c^*}-M_{\Sigma_c^*}
-M_{\Sigma^*}+M_\Delta
= -25.00~{\rm MeV} \, ,
\nonumber\\[1mm]
{\cal N}_{bs}^{q,{\rm exp}}
&\equiv&
M_{\Xi_b^*}-M_{\Sigma_b^*}
-M_{\Sigma^*}+M_\Delta
= -31.24~{\rm MeV} \, .
\nonumber\\
\label{eq:baryon-nonadditivity-Qs}
\end{eqnarray}
Here the two subscripts specify the two replacements
of light flavors by the indicated flavors, whereas the
flavor superscript specifies the unchanged third flavor
(the spectator), with $q=u,d$ in the isospin-symmetric
limit.

The corresponding mass combinations in
Eqs.~(\ref{eq:baryon-nonadditivity-ss})--(\ref{eq:baryon-nonadditivity-Qs})
vanish identically when evaluated on the leading
additive spectrum in Eq.~(\ref{eq:baryon-20S-mass-plane}),
irrespective of the values of the three parameters
$m_{\mathbf S}$, $F_{15}^{\mathbf S}$, and
$F_8^{\mathbf S}$.
Their nonzero experimental values therefore cannot
be accommodated simply by readjusting these
leading parameters.
These values are finite differences of the adopted
experimental spectrum at the physical flavor-breaking
point, rather than local derivatives with respect to
continuous symmetry-breaking sources. They do not by
themselves identify a microscopic interaction mechanism.
Their quantitative interpretation should also take into
account the treatment of isospin splittings, electromagnetic
contributions, and resonance mass definitions.

The observed departures from additivity motivate an
effective description in which the flavor response
associated with one slot depends on the flavor
composition of the other slots.
In particular, the mixed strange--heavy double differences
are substantially larger in magnitude than the
strange--strange second differences considered above.
In the following subsection, we investigate how such
nonadditive contributions can be accommodated by
higher-order symmetry-breaking responses and effective
flavor correlations within the framework of group expansion.

\subsection{Higher-order symmetry-breaking responses}
\label{sec:baryon-S-higher}

The nonvanishing experimental mass differences in
Eqs.~(\ref{eq:baryon-nonadditivity-ss})--
(\ref{eq:baryon-nonadditivity-Qs}) show that the
leading additive mass formula cannot fully describe
the observed flavor-dependent mass increments.
As in the vector-meson sector, we consider the
next-order interactions with two adjoint
flavor-source insertions.
For flavor $SU(4)$, the relevant decompositions are
\begin{eqnarray}
\operatorname{Sym}^2(\mathbf{15})
&=&\mathbf{1}\oplus\mathbf{15}
\oplus\mathbf{20'}\oplus\mathbf{84},
\nonumber\\
\overline{\mathbf{20_S}}\otimes\mathbf{20_S}
&=&\mathbf{1}\oplus\mathbf{15}
\oplus\mathbf{84}\oplus\mathbf{300}.
\label{eq:baryon-S-quadratic-irreps}
\end{eqnarray}
Thus, beyond the inherited singlet and adjoint mass
structures, the $\mathbf{84}$ provides the only new
irreducible quadratic response in this representation.

Extending the leading interaction in
Eq.~(\ref{eq:baryon-S-L3}), the next-order
flavor-invariant interactions can be expressed as
\begin{eqnarray}
{\cal L}_{\rm baryon}^{\mathbf S,(4)}
&=&
M^N M^{N'}\bar\Delta^P\Delta^Q
\times
\Big[
g_{4,1}\,\delta_{NN'}\delta_{PQ}
\nonumber\\
&& +
g_{4,2}\,d_{NN'R}
\left[{\bf F}_{\Delta}\right]^R_{PQ}
\nonumber\\
&& +
g_{4,3}
\left\{
\left[{\bf F}_{\Delta}\right]^N,
\left[{\bf F}_{\Delta}\right]^{N'}
\right\}_{PQ}
\Big],
\label{eq:baryon-S-quartic-interactions}
\end{eqnarray}
where $M^N$ denotes the adjoint Lorentz-scalar
flavor source used in the symmetry-breaking
construction. The anticommutator contains the new
$\mathbf{84}$ component together with inherited
singlet and adjoint components. At a fixed diagonal
breaking background, the latter can be absorbed
into the effective constant and linear mass terms.

Consequently, for $n_Q$ heavy and $n_s$ strange quarks,
with $n_Q+n_s\leq3$, the quadratic mass formula takes
the form
\begin{eqnarray}
M_{\mathbf{20_S}}(n_Q,n_s)
&=&A_0+B_Qn_Q+B_sn_s
\label{eq:baryon-S-quadratic-mass-surface}
\\ \nonumber
&+&\kappa_{QQ}{n_Q\choose2}
+\kappa_{Qs}n_Qn_s
+\kappa_{ss}{n_s\choose2}.
\end{eqnarray}
The three $\kappa$ coefficients describe the
effective heavy--heavy, heavy--strange, and
strange--strange occupation-space curvatures.
They are independent coefficients in this general
phenomenological mass surface, but do not represent
three independent irreducible $SU(4)$ quadratic
couplings.
We perform separate one-parameter higher-order tests
using the same experimental masses as in
Table~\ref{tab:baryon-20S-leading-masses}.
Each test refits the three leading coefficients
and adds only one quadratic coefficient.
The isolated $Qs$ and $ss$ terms in
Eq.~(\ref{eq:baryon-S-quadratic-mass-surface})
serve as phenomenological diagnostics.

For comparison, we also fit the complete symmetric-
generator product (Sym) in
Eq.~(\ref{eq:baryon-S-quartic-interactions}),
including its inherited components.
In this four-parameter test, we adopt
\begin{eqnarray}
{\cal H}_{\mathbf{20_S}}^{\rm Sym}
&=&m_{\mathbf S}{\bf1}_{20}
+{\cal X}_{\mathbf S}
+\kappa_{\mathbf S}{\cal X}_{\mathbf S}^2,
\nonumber\\
{\cal X}_{\mathbf S}
&=&F_8^{\mathbf S}
\left[{\bf F}_{\Delta}\right]^8
+F_{15}^{\mathbf S}
\left[{\bf F}_{\Delta}\right]^{15}.
\label{eq:baryon-S-Sym-fit-H}
\end{eqnarray}
For a diagonal flavor configuration, the eigenvalue
of ${\cal X}_{\mathbf S}$ is
$x_q+\beta_Qn_Q+\beta_sn_s$, where
\begin{equation}
\beta_s=-\frac{F_8^{\mathbf S}}{\sqrt3},\qquad
\beta_Q=-\frac{F_8^{\mathbf S}
+2\sqrt2F_{15}^{\mathbf S}}{3\sqrt3}.
\end{equation}
The quadratic curvatures in
Eq.~(\ref{eq:baryon-S-quadratic-mass-surface})
are therefore
\begin{eqnarray}
\kappa_{ss}
&=&\frac{2\kappa_{\mathbf S}}{3}
\left(F_8^{\mathbf S}\right)^2,
\nonumber\\
\kappa_{Qs}
&=&\frac{2\kappa_{\mathbf S}}{9}
F_8^{\mathbf S}
\left(F_8^{\mathbf S}
+2\sqrt2F_{15}^{\mathbf S}\right),
\nonumber\\
\kappa_{QQ}
&=&\frac{2\kappa_{\mathbf S}}{27}
\left(F_8^{\mathbf S}
+2\sqrt2F_{15}^{\mathbf S}\right)^2.
\label{eq:baryon-S-Sym-curvatures}
\end{eqnarray}
The three quadratic curvatures are therefore induced
by the single additional coupling $\kappa_{\mathbf S}$,
rather than fitted independently. In particular,
Eq.~(\ref{eq:baryon-S-Sym-curvatures}) implies
\begin{equation}
\kappa_{Qs}^{\,2}=\kappa_{QQ}\kappa_{ss},
\label{eq:baryon-S-Sym-rank-one}
\end{equation}
so that the quadratic curvature matrix has rank one.
This relation follows from the specific Sym ansatz
in Eq.~(\ref{eq:baryon-S-Sym-fit-H}), in which the
quadratic response is constructed from the square
of the same fitted leading operator
${\cal X}_{\mathbf S}$.
It is not required by flavor symmetry alone.
As illustrated by the general four-field interaction
in Eq.~(\ref{eq:baryon-S-quartic-interactions}),
the singlet, adjoint, and $\mathbf{84}$ channels
admit independent reduced couplings.
At a fixed symmetry-breaking background, the
inherited singlet and adjoint contributions can be
absorbed into the effective constant and linear
mass parameters. Such a redefinition, however,
does not require the remaining quadratic response
to be aligned with the square of the refitted
leading operator.
The Sym model should therefore be regarded as
a restricted common-source description of
correlated quadratic mass responses, rather than
the unique quadratic completion allowed by
flavor symmetry. The constant and linear
contributions generated by
${\cal X}_{\mathbf S}^2$ are included in the
effective coefficients of
Eq.~(\ref{eq:baryon-S-quadratic-mass-surface}).

\begin{table}[t]
\renewcommand{\arraystretch}{1.25}
\setlength{\tabcolsep}{3pt}
\caption{
Selected quadratic-response tests for the
fully symmetric $J^P=3/2^+$ baryons.
The leading fits contain three parameters, and each
higher-order test introduces one additional parameter.
For the isolated occupation-space responses,
$\kappa_{\rm rep}$ is the corresponding fitted
$\kappa_{Qs}$ or $\kappa_{ss}$; for the complete
symmetric-generator product (Sym), it is the
induced $\kappa_{Qs}$.
Sym includes the inherited singlet and adjoint
contributions of the anticommutator in
Eq.~(\ref{eq:baryon-S-quartic-interactions}).
All reported quantities are in MeV.
}
\label{tab:baryon-S-higher-nested}
\begin{ruledtabular}
\begin{tabular}{lrrr}
Contribution
& $\kappa_{\rm rep}$
& RMS
& ${\cal N}_{Qs}^{\rm fit}$
\\
& (MeV) & (MeV) & (MeV)
\\ \hline
\multicolumn{4}{l}{$SU(4)_c$}
\\
Leading
& --- & 10.80 & 0
\\
$+\kappa_{ss}$
& $+5.56$ & 10.48 & 0
\\
$+\kappa_{cs}$
& $-23.06$ & 2.84 & $-23.06$
\\
$+{\rm Sym}$
& $-24.78$ & 1.79 & $-24.78$
\\ \hline
\multicolumn{4}{l}{$SU(4)_b$}
\\
Leading
& --- & 7.60 & 0
\\
$+\kappa_{ss}$
& $+3.35$ & 7.42 & 0
\\
$+\kappa_{bs}$
& $-25.64$ & 2.82 & $-25.64$
\\
$+{\rm Sym}$
& $-26.80$ & 2.51 & $-26.80$
\end{tabular}
\end{ruledtabular}
\end{table}

As shown in Table~\ref{tab:baryon-S-higher-nested},
the isolated heavy--strange quadratic term already
reduces the RMS from $10.80$ to $2.84$~MeV in the
charm sector and from $7.60$ to $2.82$~MeV in the
bottom sector. The complete Sym interaction gives
further improvements to $1.79$ and $2.51$~MeV,
respectively, while introducing the same number
of additional fitted parameters. The corresponding
improvements in the individual baryon masses are
shown in Table~\ref{tab:baryon-20S-leading-masses}.
In the quadratic mass surface,
${\cal N}_{Qs}^{q,\rm fit}=\kappa_{Qs}$,
so the representative mixed curvatures directly
determine the corresponding fitted nonadditivities.
The Sym values, $\kappa_{cs}=-24.78$~MeV and
$\kappa_{bs}=-26.80$~MeV, are comparable in magnitude
to the experimental values
${\cal N}_{cs}^{q,\rm exp}=-25.00$~MeV and
${\cal N}_{bs}^{q,\rm exp}=-31.24$~MeV.

By contrast, an isolated $ss$ curvature produces
only a modest improvement and changes sign when
the complete Sym response is included.
In particular, the positive fitted values of
$\kappa_{ss}$ in the isolated tests do not reproduce
the negative experimental $ss$ double differences.
Instead, they reflect the global compromise of
the restricted least-squares fits, in which the
leading parameters are simultaneously refitted.
Therefore, the isolated $\kappa_{ss}$ values
should not be interpreted as separately determined
physical couplings.
These results demonstrate that quadratic responses,
particularly the mixed heavy--strange response,
substantially improve the description of the
observed nonadditive mass splittings.
Nevertheless, the quadratic occupation surface
is not exact over the measured spectrum.
The experimental third finite differences are
\begin{eqnarray}
{\cal T}_{sss}^{\rm exp}
&=&{\cal N}_{ss}^{s,\rm exp}
-{\cal N}_{ss}^{q,\rm exp}
=-5.92~{\rm MeV},
\nonumber\\
{\cal T}_{ssc}^{\rm exp}
&=&{\cal N}_{ss}^{c,\rm exp}
-{\cal N}_{ss}^{q,\rm exp}
=-3.50~{\rm MeV}.
\label{eq:baryon-S-third-differences}
\end{eqnarray}
Their nonzero values demonstrate deviations
from an exactly quadratic flavor-occupation
mass surface. Such deviations may reflect
higher-order responses and other effects not
resolved in the present flavor description.
These finite differences should not be
identified with local derivatives at the
exact flavor-symmetric point.

Finally, all experimental inputs contain at most
one heavy quark, so ${n_Q\choose2}=0$ throughout
the fitted data. The heavy--heavy curvature
$\kappa_{QQ}$ is therefore not independently
identifiable from present spin-$3/2$ masses.
Any doubly- or triply-heavy mass prediction
obtained by extending the fitted quadratic
operator to $n_Q\geq2$ remains conditional on
its assumed source structure. The improved
in-sample RMS does not establish controlled
convergence in that unmeasured region.

\subsection{Joint charm--bottom analysis in flavor $SU(5)$}
\label{sec:baryon-S-SU5}

We now investigate whether the leading and quadratic
responses can describe the charm and bottom sectors
simultaneously. We consider the flavor-$SU(5)_{udscb}$
system in the ordered basis $(u,d,s,c,b)$, for which
the fully symmetric three-quark reference representation
is $\mathbf{35_S}$. Under the $SU(4)_{udsc}$ subgroup,
this representation decomposes as
\begin{equation}
\mathbf{35_S}
\longrightarrow
\mathbf{20_S}\oplus\mathbf{10}
\oplus\mathbf{4}\oplus\mathbf{1}
\quad [SU(5)\to SU(4)] \, ,
\label{eq:baryon-35S-SU4-branching}
\end{equation}
with the four components containing zero, one, two,
and three bottom quarks, respectively.

The leading symmetry-breaking Hamiltonian extends
Eq.~(\ref{eq:baryon-20S-leading-H}) to three diagonal
flavor directions:
\begin{equation}
{\cal H}_{\mathbf{35_S}}
=m_{\mathbf S}{\bf1}_{35}
+\sum_{C=8,15,24}
F_C^{\mathbf S}
\left[{\bf F}_{\Delta}\right]^C.
\label{eq:baryon-35S-leading-H}
\end{equation}
The $SU(5)$ diagonal generators are defined in the
five-flavor basis, so their coefficients need not
coincide with those of the separately fitted $SU(4)_b$
Hamiltonian. We perform an unweighted joint fit to
nine observed masses: the four light-baryon masses,
three singly-charmed masses, and two singly-bottom
masses used in Sec.~\ref{sec:baryon-S-leading}.
The four leading parameters give an RMS mass deviation
of $10.47$~MeV, as shown in
Table~\ref{tab:baryon-20S-leading-masses}.

At quadratic order, the $SU(5)$ representation
content is
\begin{eqnarray}
\operatorname{Sym}^2(\mathbf{24})
&=&\mathbf1\oplus\mathbf{24}\oplus\mathbf{75}
   \oplus\mathbf{200},
\nonumber\\
\overline{\mathbf{35_S}}\otimes\mathbf{35_S}
&=&\mathbf1\oplus\mathbf{24}\oplus\mathbf{200}
   \oplus\mathbf{1000}.
\label{eq:baryon-35S-quadratic-irreps}
\end{eqnarray}
Thus the $\mathbf{200}$ supplies the only new
irreducible quadratic response beyond the inherited
singlet and adjoint components, in analogy with the
$\mathbf{84}$ of flavor $SU(4)$.
For later reference, the general quadratic mass surface
can be expressed in terms of the strange-, charm-,
and bottom-quark occupation numbers $n_s,n_c,n_b$:
\begin{eqnarray}
M_{\mathbf{35_S}}(n_s,n_c,n_b)
&=& A_0+B_sn_s+B_cn_c+B_bn_b
\label{eq:baryon-35S-quadratic-mass-surface}
\\
&+&\kappa_{ss}{n_s\choose2}
  +\kappa_{cc}{n_c\choose2}
  +\kappa_{bb}{n_b\choose2}
\nonumber \\ \nonumber
&+&\kappa_{cs}n_cn_s
  +\kappa_{bs}n_bn_s
  +\kappa_{bc}n_bn_c,
\end{eqnarray}
where $n_s+n_c+n_b\leq3$. The six occupation-space
curvatures characterize distinct flavor-pair mass
patterns, but are not six independent couplings
in a restricted common-source interaction.

To test the same complete symmetric-generator product
(Sym) as in Sec.~\ref{sec:baryon-S-higher},
we extend Eq.~(\ref{eq:baryon-S-Sym-fit-H}) to
\begin{eqnarray}
{\cal H}_{\mathbf{35_S}}^{\rm Sym}
&=&m_{\mathbf S}{\bf1}_{35}
+{\cal X}_{\mathbf S}^{(5)}
+\kappa_{\mathbf S}^{(5)}
\left({\cal X}_{\mathbf S}^{(5)}\right)^2,
\nonumber\\
{\cal X}_{\mathbf S}^{(5)}
&=&\sum_{C=8,15,24}F_C^{\mathbf S}
\left[{\bf F}_{\Delta}\right]^C.
\label{eq:baryon-35S-Sym-fit-H}
\end{eqnarray}
The complete product includes the inherited singlet
and adjoint components. The leading coefficients
are refitted simultaneously with the single
additional coupling $\kappa_{\mathbf S}^{(5)}$.
Writing the eigenvalue of ${\cal X}_{\mathbf S}^{(5)}$
as $x_q+\beta_sn_s+\beta_cn_c+\beta_bn_b$, we obtain
\begin{eqnarray}
\beta_s&=&-{F_8^{\mathbf S}\over\sqrt3},
\nonumber\\
\beta_c&=&-{F_8^{\mathbf S}\over3\sqrt3}
-{4F_{15}^{\mathbf S}\over3\sqrt6},
\nonumber\\
\beta_b&=&-{F_8^{\mathbf S}\over3\sqrt3}
-{F_{15}^{\mathbf S}\over3\sqrt6}
-{5F_{24}^{\mathbf S}\over3\sqrt{10}}.
\label{eq:baryon-35S-Sym-betas}
\end{eqnarray}
Consequently, all six curvatures in
Eq.~(\ref{eq:baryon-35S-quadratic-mass-surface})
are fixed by
\begin{equation}
\kappa_{ab}=2\kappa_{\mathbf S}^{(5)}\beta_a\beta_b,
\qquad a,b=s,c,b,
\label{eq:baryon-35S-Sym-curvatures}
\end{equation}
including $a=b$ for the coefficients of
${n_a\choose2}$. The constant and linear terms
induced by the square are included in the effective
coefficients $A_0$ and $B_a$.
The resulting quadratic curvature matrix
$(\kappa_{ab})$, with $a,b=s,c,b$, has rank one.
This factorization follows from the restricted
Sym ansatz and is not required by the
$\mathbf{200}$ representation alone.

The resulting five-parameter joint fit gives
$\kappa_{\mathbf S}^{(5)}\simeq
-3.29\times10^{-5}$~MeV$^{-1}$ and reduces
the RMS from $10.47$ to $6.90$~MeV.
The fitted parameters and individual masses are
presented in Table~\ref{tab:baryon-20S-leading-masses}.
The improvement is less pronounced than in the separate
$SU(4)_c$ and $SU(4)_b$ Sym fits, whose RMS deviations
are $1.79$ and $2.51$~MeV, respectively.
In particular, the joint Sym fit induces
\begin{equation}
\kappa_{cs}=-8.42~{\rm MeV},\qquad
\kappa_{bs}=-33.51~{\rm MeV},
\label{eq:baryon-35S-mixed-curvatures}
\end{equation}
whereas the experimental mixed finite differences
are ${\cal N}_{cs}^{q,\rm exp}=-25.00$~MeV and
${\cal N}_{bs}^{q,\rm exp}=-31.24$~MeV.
The common quadratic response therefore reproduces
the bottom--strange nonadditivity more closely than
the charm--strange one in this joint fit.
The separate $SU(4)$ fits allow different quadratic
couplings in the two heavy-flavor sectors, while the
$SU(5)$ fit must accommodate both sectors with a
single quadratic coupling. Moreover, the three RMS
values refer to different fitting samples. The joint
fit thus constitutes a more restrictive test, rather
than a direct comparison of identical statistical models.

Recent observations of doubly charmed baryons provide
a complementary check of the mass extrapolations.
The measured central masses of the two isospin partners
are $M_{\Xi_{cc}^{++}}=3621.55$~MeV~\cite{LHCb:2019epo}
and $M_{\Xi_{cc}^{+}}=3619.97$~MeV~\cite{LHCb:2026pxn}.
For an isospin-symmetric reference we use their
arithmetic mean, $M_{\Xi_{cc}}^{\rm exp}=3620.76$~MeV.
The observed $\Omega_{cc}^{+}$ has
$M_{\Omega_{cc}}^{\rm exp}=3725.90$~MeV~\cite{LHCb:2026lnj},
so that
\begin{equation}
M_{\Omega_{cc}}^{\rm exp}-M_{\Xi_{cc}}^{\rm exp}
=105.14~{\rm MeV}.
\label{eq:baryon-doubly-charm-exp-spacing}
\end{equation}
These observed states are conventionally assigned
to the $J^P=1/2^+$ ground-state multiplet, although
their spin and parity have not been directly measured.
They must therefore be distinguished from the
$J^P=3/2^+$ reference states $\Xi_{cc}^*$ and
$\Omega_{cc}^*$ considered here, and are not included
as inputs in the present spin-$3/2$ fits.

Under the conventional assignments, differences between
the extrapolated $3/2^+$ masses and the observed
$1/2^+$ reference masses provide a conditional test
of the expected hyperfine ordering.
Defining $\Delta_\Xi=M_{\Xi_{cc}^*}^{\rm fit}
-M_{\Xi_{cc}}^{\rm exp}$ and
$\Delta_\Omega=M_{\Omega_{cc}^*}^{\rm fit}
-M_{\Omega_{cc}}^{\rm exp}$, the masses in
Table~\ref{tab:baryon-20S-leading-masses}
give, in MeV,
\begin{equation}
\begin{array}{lrr}
 & \Delta_\Xi & \Delta_\Omega \\
\hline
SU(4)_c\ {\rm Leading} & 139.95 & 175.24 \\
SU(4)_c\ {\rm Sym}     & -45.64 & -50.52 \\
SU(5)\ {\rm Leading}   & 140.59 & 174.60 \\
SU(5)\ {\rm Sym}       &  78.77 & 101.20
\end{array}
\label{eq:baryon-doubly-charm-hyperfine-benchmarks}
\end{equation}
The separate $SU(4)_c$ Sym extrapolation reverses
the conventionally expected spin-$3/2$--spin-$1/2$
ordering of both doubly charmed partners, despite
its small in-sample RMS. The joint $SU(5)$ Sym
extrapolation instead gives positive mass differences,
although their magnitudes cannot be tested without
direct spin-$3/2$ measurements. Moreover, the measured
strange-flavor increment of $105.14$~MeV is close to
the $SU(4)_c$ Sym value
$M_{\Omega_{cc}^*}-M_{\Xi_{cc}^*}=100.26$~MeV.
However, this agreement does not remedy the inverted
hyperfine ordering. These comparisons illustrate
the model dependence of the extrapolations, rather
than providing direct measurements of the spin-$3/2$
heavy--heavy response.

The nine experimental masses used in the joint fit
contain at most one heavy quark. Consequently, the
heavy--heavy curvatures $\kappa_{cc}$,
$\kappa_{bc}$, and $\kappa_{bb}$ cannot be
independently determined from these data.
Their values in the Sym model are instead induced
by the restricted common-source relations in
Eq.~(\ref{eq:baryon-35S-Sym-curvatures}).
For example, $\kappa_{bc}=-302.04$~MeV in
Table~\ref{tab:baryon-20S-leading-masses}
is a model-dependent extrapolation rather than
an experimentally constrained response.
Since these relations connect the heavy--heavy
curvatures to the fitted strange--strange and
heavy--strange responses, uncertainties in the
latter can be amplified, particularly when the
strange--strange curvature is small.
The substantial differences between the leading
and Sym mass extrapolations therefore indicate
considerable model dependence rather than
controlled predictive accuracy for the unobserved
spin-$3/2$ doubly- and triply-heavy baryons.

From the group-expansion perspective, this
sensitivity raises the question of how reliably
low-order symmetry-breaking responses can be
extended across different flavor configurations.
As also suggested by the vector-meson analysis,
simple relations among flavor responses need not
remain quantitatively accurate.
Additional heavy--heavy correlations, higher-order
interactions, and changes in the internal structure
of physical states may become important.
However, changing the number of heavy quarks at
fixed physical quark masses does not directly
determine the local curvature along a continuous
symmetry-breaking path.
The present results therefore reveal limitations
of the restricted quadratic extrapolation, rather
than establishing curvature growth or a specific
state-flow mechanism.
The observed doubly charmed baryons, conventionally
assigned to $J^P=1/2^+$, may provide complementary
constraints if their hyperfine relations to the
$J^P=3/2^+$ partners can be established independently.
No such cross-spin relation is imposed in the present fits.
Additional spin-$3/2$ experimental masses and
independent dynamical constraints are needed to
determine the heavy--heavy responses reliably.

\subsection{Breaking responses in mesons and baryons}
\label{sec:meson-baryon-comparison}

The vector-meson and fully symmetric $J^P=3/2^+$
baryon sectors exhibit related patterns of flavor
symmetry breaking. At leading order, the ratios of
the heavy- and light-flavor breaking responses in
Eqs.~(\ref{eq:meson-R158-cb}) and
(\ref{eq:F15F8-20S}) show similar hierarchies.
For $Q=c,b$, we denote the corresponding fitted ratios by
$R_{15/8}^{\mathrm{meson}}=D_{15}^{SU(4)_Q}/D_8^{SU(4)_Q}$
and $R_{15/8}^{\mathrm{baryon}}=F_{15}^{\mathbf S}/F_8^{\mathbf S}$,
where the latter is evaluated in the corresponding
$SU(4)_Q$ system. As shown in
Table~\ref{tab:meson-baryon-normalized-responses},
both ratios increase substantially from charm to bottom,
while the baryon values remain smaller than the meson
values. Thus, the leading responses share a hierarchy
but not an exact numerical relation.

\begin{table}[t]
\renewcommand{\arraystretch}{1.2}
\setlength{\tabcolsep}{6pt}
\caption{Comparison of leading flavor-breaking ratios
and normalized nonadditive responses in vector mesons
and fully symmetric $J^P=3/2^+$ baryons. The leading
ratios are taken from the independent fits in
Eqs.~(\ref{eq:meson-R158-cb}) and
(\ref{eq:F15F8-20S}), whereas the $\eta_Q$ values
are constructed from experimental mass differences
without additional fitting.}
\label{tab:meson-baryon-normalized-responses}
\begin{ruledtabular}
\begin{tabular}{lrr}
Response & $Q=c$ & $Q=b$ \\
\hline
$R_{15/8}^{\mathrm{meson}}$ (leading) & $10.20$ & $39.79$ \\
$R_{15/8}^{\mathrm{baryon}}$ (leading) & $9.15$ & $33.24$ \\
\hline
$\eta_Q^{\mathrm{meson}}$ (nonadditive) & $-0.1243$ & $-0.2277$ \\
$\eta_Q^{\mathrm{baryon}}$ (nonadditive) & $-0.1638$ & $-0.2047$
\end{tabular}
\end{ruledtabular}
\end{table}

To compare departures from the leading additive mass
structures without identifying the different quadratic
couplings, we use experimental mass finite differences.
With the meson nonadditivities ${\cal N}_Q^{\rm exp}$
in Eq.~(\ref{eq:meson-SU4-null-exp}) and the baryon
nonadditivities ${\cal N}_{Qs}^{q,\rm exp}$ in
Eq.~(\ref{eq:baryon-nonadditivity-Qs}), we define
\begin{eqnarray}
\eta_Q^{\mathrm{meson}}
&\equiv&
\frac{{\cal N}_Q^{\rm exp}}
{M_{K^*}-M_\rho},
\nonumber\\
\eta_Q^{\mathrm{baryon}}
&\equiv&
\frac{{\cal N}_{Qs}^{q,\rm exp}}
{M_{\Sigma^*}-M_\Delta},
\qquad Q=c,b.
\label{eq:meson-baryon-normalized-responses}
\end{eqnarray}
The denominators are the strange-flavor mass
increments in the corresponding light-hadron
reference configurations. Hence, each $\eta_Q$
measures the fractional change in that increment when
a light spectator is replaced by a heavy flavor.
The resulting values are included in
Table~\ref{tab:meson-baryon-normalized-responses}.
Both sectors exhibit negative nonadditivities of
comparable relative magnitude, indicating that the
strange-flavor mass increment decreases in the
presence of a heavy spectator. The normalized
responses also become more negative from charm to
bottom in both sectors. Nevertheless, their numerical
values differ between mesons and baryons, so these
patterns do not establish a universal quantitative
relation.

The normalized comparison relies on physical mass
finite differences rather than fitted quadratic
couplings. The available quadratic flavor structures
and their normalizations differ between mesons and
baryons, and neither $\eta_Q^{\mathrm{meson}}$ nor
$\eta_Q^{\mathrm{baryon}}$ represents a local derivative
along a continuous symmetry-breaking path.
Taken together, the leading response hierarchy and
the nonadditive mass patterns suggest a common
qualitative organization of flavor breaking, while
the quantitative response strengths remain sensitive
to the internal dynamics of each hadronic sector.

\section{Mixed-symmetry baryons}
\label{sec:20M}

We now turn to the ground-state $J^P=1/2^+$ baryons,
for which the mixed-symmetry flavor representations
provide natural reference configurations.
As in the fully symmetric sector studied in
Sec.~\ref{sec:20S}, these flavor representations
should not be identified with the complete physical
wave functions. Rather, they serve as reference
structures within group expansion, while the physical
states may contain contributions from different
flavor, spin, and spatial configurations compatible
with the same total angular momentum and parity.
Compared with the fully symmetric flavor representations,
the mixed-symmetry representations admit a richer
pattern of symmetry-breaking responses: whereas
the former admit only one independent adjoint response
at leading order, the latter admit two, conventionally
referred to as the $F$- and $D$-type responses.

The additional flavor structure allows symmetry
breaking to generate not only diagonal mass shifts
but also non-diagonal matrix elements between reference
states with the same conserved quantum numbers.
Consequently, symmetry breaking can modify both
the mass spectrum and the composition of the
corresponding eigenstates. This provides a natural
setting for examining the interplay between mass
flow and state flow within group expansion.
We first establish the leading flavor structures
and the resulting mass relations, and then investigate
higher-order symmetry-breaking responses, the baryon
spectra in flavor $SU(4)$ and $SU(5)$, and the extent
to which the effective mass Hamiltonian can be
determined from experimental masses.

\subsection{Leading symmetry-breaking Hamiltonian}
\label{sec:baryon-M-leading}

We first consider flavor $SU(4)_Q$, with $Q=c$ or $b$
and the ordered basis $(u,d,s,Q)$. Under its light-flavor
$SU(3)$ subgroup, the mixed representation branches as
\begin{equation}
\mathbf{20_M}\longrightarrow
\mathbf8\oplus\overline{\mathbf3}
\oplus\mathbf6\oplus\mathbf3
\quad [SU(4)\to SU(3)]\,.
\label{eq:baryon-20M-SU3-branching}
\end{equation}
These components contain zero, one, one, and two
heavy quarks, respectively. In particular, the
$\overline{\mathbf3}$ and $\mathbf6$ both contain
reference configurations with the flavor content
$qsQ$. They therefore furnish two distinct flavor
basis states in the $\Xi_Q$ sector.

As in Appendix~\ref{app:baryon-S-matrices}, we denote
the mixed flavor components by
\begin{equation}
N^I=\mathbb M^I_{ABC}\,q^Aq^Bq^C,
\qquad I=1,\ldots,20,
\label{eq:baryon-M-flavor-components}
\end{equation}
where the normalized tensors span a fixed mixed-symmetry
carrier, with $\mathbb M^I_{BAC}=-\mathbb M^I_{ABC}$.
The adjoint representation occurs twice in the
mixed-state bilinear:
\begin{equation}
\overline{\mathbf{20_M}}\otimes\mathbf{20_M}
\supset\mathbf1\oplus2\,\mathbf{15}.
\label{eq:baryon-M-two-adjoints}
\end{equation}
Consequently, the mixed representation admits two
independent adjoint flavor structures, conventionally
denoted by ${\bf F}^N$ and ${\bf D}^N$, whereas
the fully symmetric representation admits only one.
This is a statement about the operators allowed by
flavor symmetry. It does not require both reduced
responses to be nonzero in a specified dynamical
realization.

Suppressing Lorentz and spin indices as in
Sec.~\ref{sec:baryon-S-leading}, the corresponding
leading flavor-invariant interactions may be written as
\begin{eqnarray}
{\cal L}_{\rm baryon}^{\mathbf M}
&=&g_0\,\delta_{IJ}M^0\bar N^I N^J
\nonumber\\
&+&M^N\bar N^I N^J
\left(g_F{\bf F}^N_{IJ}+g_D{\bf D}^N_{IJ}\right).
\label{eq:baryon-M-leading-L}
\end{eqnarray}
Here ${\bf F}^N$ and ${\bf D}^N$ are the two
independent flavor structures defined in
Eq.~(\ref{eq:app-mixed-DF}) of Appendix~\ref{app:baryon-S-matrices}.
As in the fully symmetric sector, the interaction in
Eq.~(\ref{eq:baryon-M-leading-L}) represents schematic
flavor-invariant contractions. Replacing the singlet
and adjoint insertions by the corresponding diagonal
Lorentz-scalar backgrounds, we obtain the leading
mass term in the mixed-symmetry reference space:
\begin{eqnarray}
{\cal H}_{\rm baryon}^{\mathbf M}
&=&m_{\mathbf M}\,\delta_{IJ}\,\bar N^I N^J
\nonumber\\
&+&\left(
F_C^{\mathbf M}\,[{\bf F}]^C_{IJ}
+D_C^{\mathbf M}\,[{\bf D}]^C_{IJ}
\right)\bar N^I N^J.
\label{eq:baryon-M-leading-H}
\end{eqnarray}
Here $C$ is summed over the diagonal flavor-breaking
directions. Within the strict single-source leading
construction, the coefficients are
\begin{eqnarray}
m_{\mathbf M}&=&g_0\langle M^0\rangle,
\nonumber\\
F_C^{\mathbf M}&=&g_F\langle M^C\rangle,
\nonumber\\
D_C^{\mathbf M}&=&g_D\langle M^C\rangle.
\label{eq:baryon-M-leading-parameters}
\end{eqnarray}
More general flavor-symmetric contributions can be
absorbed into $m_{\mathbf M}$. The reduced couplings
are not fixed by flavor symmetry. Within this leading
construction, the two responses satisfy
\begin{equation}
\frac{F_C^{\mathbf M}}{F_{C'}^{\mathbf M}}
=\frac{D_C^{\mathbf M}}{D_{C'}^{\mathbf M}}
=\frac{\langle M^C\rangle}{\langle M^{C'}\rangle}
\equiv R_{C/C'},
\label{eq:baryon-M-breaking-ratio}
\end{equation}
provided that the denominators are nonzero.
This relation is analogous to
Eq.~(\ref{eq:baryon-S-breaking-ratio}) for the fully
symmetric baryons. 
It is retained for the linear
coefficients in the common-source models studied below,
but need not hold for general higher-order effective
responses.

For flavor $SU(4)_Q$, the two diagonal breaking
directions are represented by
$\lambda_8=\operatorname{diag}(1,1,-2,0)/\sqrt3$
and
$\lambda_{15}=\operatorname{diag}(1,1,1,-3)/\sqrt6$,
with $\operatorname{Tr}(\lambda_A\lambda_B)=2\delta_{AB}$.
The corresponding $\mathbf{20_M}$ mass kernel is
\begin{eqnarray}
{\cal H}_{\mathbf{20_M}}
=m_{\mathbf M}{\bf1}_{20}
&+&F_8^{\mathbf M}{\bf F}^{8}
\,+\,D_8^{\mathbf M}{\bf D}^{8}
\nonumber\\
&+&F_{15}^{\mathbf M}{\bf F}^{15}
\,+\,D_{15}^{\mathbf M}{\bf D}^{15}.
\label{eq:baryon-20M-leading-H}
\end{eqnarray}
The four adjoint coefficients in this equation
satisfy
\begin{equation}
F_{15}^{\mathbf M}D_8^{\mathbf M}
-D_{15}^{\mathbf M}F_8^{\mathbf M}=0.
\label{eq:baryon-M-leading-common-source}
\end{equation}
Accordingly, the leading Hamiltonian contains four
independent parameters, including $m_{\mathbf M}$.

Using the mixed tensor basis and matrix conventions
fixed in Appendix~\ref{app:baryon-S-matrices}, the
leading mass formulas can be organized in terms of
the four $SU(3)$-symmetric flavor-reference centers:
\begin{eqnarray}
C_8
&=&m_{\mathbf M}
+\sqrt{\frac38}\,F_{15}^{\mathbf M}
-\frac{D_{15}^{\mathbf M}}{\sqrt{24}},
\nonumber\\
C_{\bar{3}}
&=&m_{\mathbf M}
-\frac{F_{15}^{\mathbf M}}{\sqrt{24}}
-\frac{7D_{15}^{\mathbf M}}{\sqrt{216}},
\nonumber\\
C_6
&=&m_{\mathbf M}
-\frac{F_{15}^{\mathbf M}}{\sqrt{24}}
+\sqrt{\frac38}\,D_{15}^{\mathbf M},
\nonumber\\
C_3
&=&m_{\mathbf M}
-\frac{5F_{15}^{\mathbf M}}{\sqrt{24}}
-\frac{D_{15}^{\mathbf M}}{\sqrt{24}}.
\label{eq:baryon-M-SU3-centers}
\end{eqnarray}
In Eq.~(\ref{eq:baryon-20M-SU3-branching}),
$\mathbf8$ contains light baryons, $\overline{\mathbf3}$ and
$\mathbf6$ contain singly-heavy baryons, and $\mathbf3$ contains doubly-heavy baryons:
\begin{itemize}

\item For the light baryon octet, one obtains
\begin{eqnarray}
M_N
&=&C_8+\frac{\sqrt3}{2}F_8^{\mathbf M}
-\frac{D_8^{\mathbf M}}{\sqrt{12}},
\nonumber\\
M_\Lambda
&=&C_8-\frac{D_8^{\mathbf M}}{\sqrt3},
\nonumber\\
M_\Sigma
&=&C_8+\frac{D_8^{\mathbf M}}{\sqrt3},
\nonumber\\
M_\Xi
&=&C_8-\frac{\sqrt3}{2}F_8^{\mathbf M}
-\frac{D_8^{\mathbf M}}{\sqrt{12}}.
\label{eq:baryon-M-leading-light-masses}
\end{eqnarray}

\item For the singly-heavy baryons, one obtains
\begin{eqnarray}
M_{\Lambda_Q}
&=&C_{\bar{3}}
+\frac{F_8^{\mathbf M}}{\sqrt3}
-\frac{2D_8^{\mathbf M}}{\sqrt{27}},
\nonumber\\
M_{\Sigma_Q}
&=&C_6+\frac{F_8^{\mathbf M}}{\sqrt3},
\nonumber\\
M_{\Omega_Q}
&=&C_6-\frac{2F_8^{\mathbf M}}{\sqrt3},
\nonumber\\
M_{\Xi_Q}+M_{\Xi'_Q}
&=&C_{\bar{3}}+C_6
-\frac{F_8^{\mathbf M}}{\sqrt3}
+\frac{D_8^{\mathbf M}}{\sqrt{27}}.
\label{eq:baryon-M-leading-single-masses}
\end{eqnarray}

\item The doubly heavy triplet gives
\begin{eqnarray}
M_{\Xi_{QQ}}
&=&C_3+\frac{F_8^{\mathbf M}-D_8^{\mathbf M}}{\sqrt{12}},
\nonumber\\
M_{\Omega_{QQ}}
&=&C_3-\frac{F_8^{\mathbf M}-D_8^{\mathbf M}}{\sqrt3}.
\label{eq:baryon-M-leading-double-masses}
\end{eqnarray}
\end{itemize}

The two $qsQ$ configurations require a separate
$2\times2$ mass matrix. In the fixed
$(|\overline{\mathbf3}\rangle,|\mathbf6\rangle)$
flavor-reference basis, it reads
\begin{equation}
{\cal H}_{\Xi_Q}=
\begin{pmatrix}
C_{\bar{3}}-\dfrac{F_8^{\mathbf M}}{\sqrt{12}}
+\dfrac{D_8^{\mathbf M}}{\sqrt{27}}
&\dfrac{D_8^{\mathbf M}}2\\[2mm]
\dfrac{D_8^{\mathbf M}}2
&C_6-\dfrac{F_8^{\mathbf M}}{\sqrt{12}}
\end{pmatrix},
\label{eq:baryon-M-Xi-leading-matrix}
\end{equation}
where $H_{\bar{3}\bar{3}}$ and $H_{66}$ denote the
$\overline{\mathbf3}$ and $\mathbf6$ diagonal entries,
respectively, and $V_Q=D_8^{\mathbf M}/2$ is the
off-diagonal entry. The two eigenmasses of this
restricted flavor kernel are
\begin{eqnarray}
M_{\Xi_Q}
&=&\frac{H_{\bar{3}\bar{3}}+H_{66}}2
-\frac12\sqrt{(H_{\bar{3}\bar{3}}-H_{66})^2+4V_Q^2},
\nonumber\\
M_{\Xi'_Q}
&=&\frac{H_{\bar{3}\bar{3}}+H_{66}}2
+\frac12\sqrt{(H_{\bar{3}\bar{3}}-H_{66})^2+4V_Q^2}.
\nonumber\\
\label{eq:baryon-M-Xi-leading-eigenmasses}
\end{eqnarray}
The labels in the last equation follow the lower-
and higher-mass ordering within this two-state model.
The masses in the above formulas denote leading-order
model eigenvalues, which should not be identified with
the flavor-basis diagonal elements when mixing is present. Moreover, the flavor-space
matrix element $V_Q=D_8^{\mathbf M}/2$ is not, by
itself, a matched physical $j_\ell$-changing mixing
amplitude. Such an identification requires additional
spin--flavor and spatial information.

We perform two independent unweighted least-squares
fits to the observed baryon masses in the charm and
bottom systems, using the leading Hamiltonian in
Eq.~(\ref{eq:baryon-20M-leading-H}) subject to
Eq.~(\ref{eq:baryon-M-leading-common-source}).
The charm fit contains the four light-octet masses,
five singly-charmed masses, and two doubly-charmed
masses, for eleven inputs in total. The bottom fit
contains the four light-octet masses and five
singly-bottom masses, for nine inputs in total.
Arithmetic isospin averages are used where available.
In particular, $M_{\Xi_{cc}}^{\rm exp}=3620.76$~MeV~\cite{LHCb:2019epo,LHCb:2026pxn}
and $M_{\Omega_{cc}}^{\rm exp}=3725.90$~MeV~\cite{LHCb:2026lnj}
are adopted as conventional ground-state $1/2^+$
references. 
For the bottom baryons, we use
$M_{\Sigma_b}^{\rm exp}
=(M_{\Sigma_b^+}+M_{\Sigma_b^-})/2
=5813.10$~MeV.
The measured charged-state mass
$M_{\Xi_b^{\prime-}}^{\rm exp}=5934.9$~MeV
is adopted as an approximation to the
isospin-averaged $\Xi_b'$ mass. 
The fitting results are summarized in
Table~\ref{tab:baryon-M-global-fit-summary}, together
with those obtained after introducing the quadratic
responses and extending the analysis to flavor $SU(5)$
in the following subsections.

\begin{table*}[!tp]
\renewcommand{\arraystretch}{1.3}
\setlength{\tabcolsep}{2.6pt}
\footnotesize
\caption{
Leading and higher-order mass fits for the mixed-symmetry
$J^P=1/2^+$ flavor-reference baryons in $SU(4)_c$,
$SU(4)_b$, and $SU(5)_{udscb}$, in the
$\mathbf{20_M}$, $\mathbf{20_M}$, and $\mathbf{40_M}$
representations, respectively. Each Leading fit uses a
normalized common adjoint source
$X=s\lambda_8+c\lambda_{15}(+b\lambda_{24})$,
with independent linear $F$- and $D$-type couplings.
For all three flavor sectors, Sym ($-$) denotes the
negative factorized quadratic response
$-[f'F_M(X)+d'D_M(X)]^2$,
where the quadratic $F/D$ ratio is independent of the
leading ratio.
The fitted linear coefficients are
$F_a^{\mathbf M}=f x_a$ and $D_a^{\mathbf M}=d x_a$,
where $x_a=(s,c,b)$ for $a=(8,15,24)$, with
$s^2+c^2=1$ in $SU(4)$ and $s^2+c^2+b^2=1$ in $SU(5)$.
The $f',d'$ entries are in $\sqrt{\mathrm{MeV}}$,
while all other dimensional entries are in MeV.
The fits use 11, 9, and 16 mass inputs, respectively,
with the measured $\Xi_b^{\prime-}$ mass of $5934.90$ MeV
used as an isospin-reference proxy in the bottom and joint fits.
Unmeasured $bc$ and $bb$ masses are conditional model
extrapolations without controlled prediction uncertainties.
The superscripts $(L)$ and $(H)$ denote the lower and
higher eigenmasses of mixed-heavy reference blocks.
}
\label{tab:baryon-M-global-fit-summary}
\begin{ruledtabular}
\begin{tabular}{lrrrrrrr}
 & Experiment
 & \multicolumn{2}{c}{$SU(4)_c$}
 & \multicolumn{2}{c}{$SU(4)_b$}
 & \multicolumn{2}{c}{$SU(5)_{udscb}$} \\
 & & Leading & Sym ($-$) & Leading & Sym ($-$)
 & Leading & Sym ($-$) \\
\hline
$N_{\rm input}$ & -- & $11$ & $11$ & $9$ & $9$ & $16$ & $16$ \\
$N_{\rm par}$ & -- & $4$ & $6$ & $4$ & $6$ & $5$ & $7$ \\
\hline
$m_{\mathbf M}$ & -- & $2123.40$ & $2249.81$ & $4640.21$ & $4999.56$ & $4712.89$ & $4800.23$ \\
$F_8^{\mathbf M}$ & -- & $-172.10$ & $-178.03$ & $-179.35$ & $-175.33$ & $-171.25$ & $-153.68$ \\
$D_8^{\mathbf M}$ & -- & $15.05$ & $27.94$ & $4.29$ & $19.14$ & $5.15$ & $7.40$ \\
$F_{15}^{\mathbf M}$ & -- & $-1538.63$ & $-1509.67$ & $-5667.16$ & $-5233.98$ & $-1548.02$ & $-1321.32$ \\
$D_{15}^{\mathbf M}$ & -- & $134.54$ & $236.91$ & $135.51$ & $571.39$ & $46.55$ & $63.66$ \\
$F_{24}^{\mathbf M}$ & -- & -- & -- & -- & -- & $-5412.18$ & $-5042.86$ \\
$D_{24}^{\mathbf M}$ & -- & -- & -- & -- & -- & $162.75$ & $242.98$ \\
\hline
$f'\ (\sqrt{\mathrm{MeV}})$ & -- & -- & $12.65$ & -- & $28.16$ & -- & $30.17$ \\
$d'\ (\sqrt{\mathrm{MeV}})$ & -- & -- & $-16.98$ & -- & $-28.02$ & -- & $-7.67$ \\
\hline
$d/f$ & -- & $-0.087$ & $-0.157$ & $-0.024$ & $-0.109$ & $-0.030$ & $-0.048$ \\
$d'/f'$ & -- & -- & $-1.343$ & -- & $-0.995$ & -- & $-0.254$ \\
$c/s$ & -- & $8.941$ & $8.480$ & $31.599$ & $29.851$ & $9.039$ & $8.598$ \\
$b/s$ & -- & -- & -- & -- & -- & $31.604$ & $32.814$ \\
\hline
RMS & -- & $33.47$ & $5.45$ & $31.27$ & $3.94$ & $43.62$ & $21.82$ \\
\hline
$N$ & $938.920$ & $1000.34$ & $946.00$ & $985.59$ & $942.54$ & $1012.67$ & $974.49$ \\
$\Lambda$ & $1115.683$ & $1145.03$ & $1110.01$ & $1139.67$ & $1114.80$ & $1159.50$ & $1134.44$ \\
$\Sigma$ & $1193.154$ & $1162.41$ & $1193.41$ & $1144.62$ & $1186.63$ & $1165.44$ & $1153.48$ \\
$\Xi$ & $1318.285$ & $1298.41$ & $1314.41$ & $1296.22$ & $1322.08$ & $1309.29$ & $1302.76$ \\
\hline
$\Lambda_c$ & $2286.460$ & $2268.24$ & $2281.32$ & -- & -- & $2312.90$ & $2293.22$ \\
$\Sigma_c$ & $2453.457$ & $2420.50$ & $2455.39$ & -- & -- & $2365.56$ & $2426.54$ \\
$\Xi_c$ & $2469.145$ & $2425.57$ & $2478.91$ & -- & -- & $2464.05$ & $2456.38$ \\
$\Xi'_c$ & $2578.550$ & $2569.93$ & $2576.35$ & -- & -- & $2514.00$ & $2579.06$ \\
$\Omega_c$ & $2695.300$ & $2718.58$ & $2695.42$ & -- & -- & $2662.17$ & $2729.70$ \\
\hline
$\Xi_{cc}$ & $3620.760$ & $3612.27$ & $3612.68$ & -- & -- & $3639.45$ & $3595.97$ \\
$\Omega_{cc}$ & $3725.900$ & $3774.34$ & $3731.71$ & -- & -- & $3792.22$ & $3749.86$ \\
\hline
$\Lambda_b$ & $5619.570$ & -- & -- & $5627.28$ & $5614.97$ & $5610.25$ & $5640.73$ \\
$\Sigma_b$ & $5813.100$ & -- & -- & $5776.45$ & $5817.12$ & $5796.45$ & $5814.66$ \\
$\Xi_b$ & $5794.350$ & -- & -- & $5785.04$ & $5798.86$ & $5761.50$ & $5773.92$ \\
$\Xi'_b$ & $5934.900$ & -- & -- & $5931.80$ & $5931.61$ & $5944.80$ & $5931.33$ \\
$\Omega_b$ & $6045.800$ & -- & -- & $6087.09$ & $6045.15$ & $6093.07$ & $6046.79$ \\
\hline
$\Xi_{bb}$ & -- & -- & -- & $10343.56$ & $9663.97$ & $10300.93$ & $9182.28$ \\
$\Omega_{bb}$ & -- & -- & -- & $10502.59$ & $9756.54$ & $10453.70$ & $9265.14$ \\
\hline
$\Xi_{bc}^{(L)}$ & -- & -- & -- & -- & -- & $6946.78$ & $6712.48$ \\
$\Xi_{bc}^{(H)}$ & -- & -- & -- & -- & -- & $7113.03$ & $6761.85$ \\
$\Omega_{bc}^{(L)}$ & -- & -- & -- & -- & -- & $7098.69$ & $6832.54$ \\
$\Omega_{bc}^{(H)}$ & -- & -- & -- & -- & -- & $7260.72$ & $6868.38$ \\
$\Omega_{ccb}$ & -- & -- & -- & -- & -- & $8423.23$ & $7624.80$ \\
$\Omega_{cbb}$ & -- & -- & -- & -- & -- & $11653.82$ & $9823.00$ \\
\end{tabular}
\end{ruledtabular}
\end{table*}

As shown in Table~\ref{tab:baryon-M-global-fit-summary},
the leading Hamiltonian gives RMS mass deviations of
$33.47$ and $31.27$~MeV for the charm and bottom
systems, respectively. The sizeable deviations indicate
that the leading flavor responses alone do not provide
an adequate description of the observed spectra.
More explicitly, the leading mass formulas imply the
following parameter-independent relations, where $M_X$
denotes the leading-order model mass of baryon $X$:
\begin{eqnarray}
&& 2(M_N+M_\Xi)
-3M_\Lambda-M_\Sigma=0,
\nonumber\\
&& M_{\Omega_Q}-M_{\Sigma_Q}
=M_\Xi-M_N,
\nonumber\\
&& M_{\Omega_{QQ}}-M_{\Xi_{QQ}}
=M_\Sigma-M_N.
\label{eq:baryon-M-leading-mass-relations}
\end{eqnarray}
The first is the familiar Gell-Mann--Okubo mass
relation. The latter two relate strange-flavor
replacements in different heavy-flavor sectors.
None of these relations involves the mixed $\Xi_Q^{(\prime)}$
sector.

However, the experimental baryon masses exhibit
clear departures from these leading relations.
In the following numerical mass differences, $M_X$
denotes the adopted experimental mass of baryon $X$.
For the light baryons, we obtain
\begin{eqnarray}
\overline M_{qqs}-M_N
&=&196.13~{\rm MeV}\,,
\nonumber\\
M_\Xi-\overline M_{qqs}
&=&183.23~{\rm MeV}\,,
\label{eq:baryon-M-light-exp-spacing}
\end{eqnarray}
where $\overline M_{qqs}=(3M_\Lambda+M_\Sigma)/4$
is a convenient mass combination, rather than the
mass of an individual reference state.
For comparison, the average strange-flavor increments
in the light octet and the singly-heavy sextets are
\begin{eqnarray}
\frac{M_\Xi-M_N}{2}
&=&189.68~{\rm MeV}\,,
\nonumber\\
\frac{M_{\Omega_c}-M_{\Sigma_c}}{2}
&=&120.92~{\rm MeV}\,,
\nonumber\\
\frac{M_{\Omega_b}-M_{\Sigma_b}}{2}
&=&116.35~{\rm MeV}\,.
\label{eq:baryon-M-heavy-exp-spacing}
\end{eqnarray}
Moreover, comparing the light and doubly-charmed
baryons, we obtain
\begin{eqnarray}
M_\Sigma-M_N&=&254.23~{\rm MeV}\,,
\nonumber\\
M_{\Omega_{cc}}-M_{\Xi_{cc}}
&=&105.14~{\rm MeV}\,.
\label{eq:baryon-M-double-exp-spacing}
\end{eqnarray}
These comparisons reveal sizeable departures from
the leading mass-increment relations when the
flavor composition of the reference configuration
is changed.

This nonadditive behavior can be quantified by
constructing combinations of the experimental
baryon masses that vanish in the leading model.
For the light baryons, we obtain
\begin{equation}
{\cal N}_{ss}^{\mathbf8,{\rm exp}}
\equiv M_\Xi-2\overline M_{qqs}+M_N
=-12.90~{\rm MeV}\,.
\label{eq:baryon-M-nonadditivity-ss}
\end{equation}
Comparing the strange-flavor increments in the
light and singly-heavy baryons gives
\begin{eqnarray}
{\cal N}_{cs}^{\mathbf6,{\rm exp}}
&\equiv&\frac{M_{\Omega_c}-M_{\Sigma_c}-M_\Xi+M_N}{2}
=-68.76~{\rm MeV}\,,
\nonumber\\[1mm]
{\cal N}_{bs}^{\mathbf6,{\rm exp}}
&\equiv&\frac{M_{\Omega_b}-M_{\Sigma_b}-M_\Xi+M_N}{2}
=-73.33~{\rm MeV}\,.
\nonumber\\
\label{eq:baryon-M-nonadditivity-Qs}
\end{eqnarray}
For the doubly-charmed baryons, we similarly obtain
\begin{eqnarray}
{\cal N}_{s}^{cc,qq;{\rm exp}}
&\equiv&M_{\Omega_{cc}}-M_{\Xi_{cc}}-M_\Sigma+M_N
\nonumber\\
&=&-149.09~{\rm MeV}\,.
\label{eq:baryon-M-nonadditivity-ccs}
\end{eqnarray}
Here the representation superscripts distinguish the
light-octet and singly-heavy-sextet mass combinations,
while $cc,qq$ specifies the two flavor environments
being compared. Unlike in the fully symmetric sector,
the $F$- and $D$-type responses need not give the
same strange-flavor increment for all reference
configurations. The above combinations are therefore chosen to cancel
both leading responses.

For the mixed $\Xi_Q^{(\prime)}$ sector, we can further
examine the non-diagonal symmetry-breaking responses.
From Eq.~(\ref{eq:baryon-M-leading-single-masses}),
we obtain the leading relation
\begin{equation}
V_Q=\frac{D_8^{\mathbf M}}{2}
=\frac{\sqrt3}{2}
\left(M_{\Xi_Q}+M_{\Xi'_Q}
-M_{\Lambda_Q}-M_{\Omega_Q}\right).
\label{eq:baryon-M-leading-mixing-relation}
\end{equation}
Inserting the experimental singly-heavy baryon
masses into this relation, we obtain the
leading-order estimates
\begin{eqnarray}
V_c&\simeq&57.10~{\rm MeV}\,,
\nonumber\\
V_b&\simeq&55.32~{\rm MeV}\,.
\label{eq:baryon-M-leading-mixing-estimate}
\end{eqnarray}
The charm and bottom sectors give numerically
similar estimates when the experimental masses
are inserted into the leading trace relation.
These values provide conditional consistency tests
of the leading flavor-reference Hamiltonian,
rather than independent determinations of
physical mixing matrix elements.

Alternatively, we can estimate the magnitudes of
the non-diagonal matrix elements by retaining the
leading equal-spacing relation for the diagonal
entries in the singly-heavy sextet reference:
\begin{equation}
M_{\Omega_Q}-2H_{66}+M_{\Sigma_Q}=0.
\label{eq:baryon-M-leading-sextet-curvature}
\end{equation}
The corresponding experimental spectral combinations
in the charm and bottom systems are
\begin{eqnarray}
M_{\Omega_c}-2M_{\Xi'_c}+M_{\Sigma_c}
&=&-8.34~{\rm MeV}\,,
\nonumber\\
M_{\Omega_b}-2M_{\Xi'_b}+M_{\Sigma_b}
&=&-10.90~{\rm MeV}\,.
\label{eq:baryon-M-exp-sextet-curvature}
\end{eqnarray}
Here the bottom combination uses the charged-state
mass as specified above.
Within the adopted two-state mass kernel, the two
mass eigenvalues satisfy the exact relation
\begin{equation}
|V_Q|^2=
(M_{\Xi'_Q}-H_{66})(H_{66}-M_{\Xi_Q}).
\label{eq:baryon-M-mixing-spectral-relation}
\end{equation}
Combining this relation with
Eq.~(\ref{eq:baryon-M-leading-sextet-curvature}),
we obtain
\begin{eqnarray}
|V_Q|^2
&=&
\left(M_{\Xi'_Q}-\frac{M_{\Sigma_Q}+M_{\Omega_Q}}{2}\right)
\nonumber\\
&&\times
\left(\frac{M_{\Sigma_Q}+M_{\Omega_Q}}{2}-M_{\Xi_Q}\right).
\label{eq:baryon-M-mixing-exp-relation}
\end{eqnarray}
Inserting the experimental baryon masses, we obtain
\begin{eqnarray}
|V_c|&\simeq&20.95~{\rm MeV}\,,
\nonumber\\
|V_b|&\simeq&27.13~{\rm MeV}\,.
\label{eq:baryon-M-mixing-exp-estimates}
\end{eqnarray}
These values are significantly smaller than the
corresponding leading-order estimates given in
Eq.~(\ref{eq:baryon-M-leading-mixing-estimate}).
Moreover, the two-state mass kernel necessarily
satisfies
\begin{equation}
|V_Q|\leq\frac{M_{\Xi'_Q}-M_{\Xi_Q}}{2}.
\label{eq:baryon-M-mixing-upper-bound}
\end{equation}
In particular, for the adopted charm mass inputs,
the leading estimate of $|V_c|\simeq57.10$~MeV
exceeds the upper bound of $54.70$~MeV imposed
by the experimental mass splitting.
This discrepancy provides a consistency test
of the leading flavor-reference Hamiltonian
for the specified mass inputs, rather than
a determination of the physical mixing amplitude
or its experimental significance.
Together with the difference between the two
conditional estimates, it indicates that the
leading relations between the diagonal and
non-diagonal responses cannot simultaneously
account for the observed singly-heavy baryon
masses.

\begin{figure*}[!t]
\centering
\includegraphics[width=0.75\textwidth]{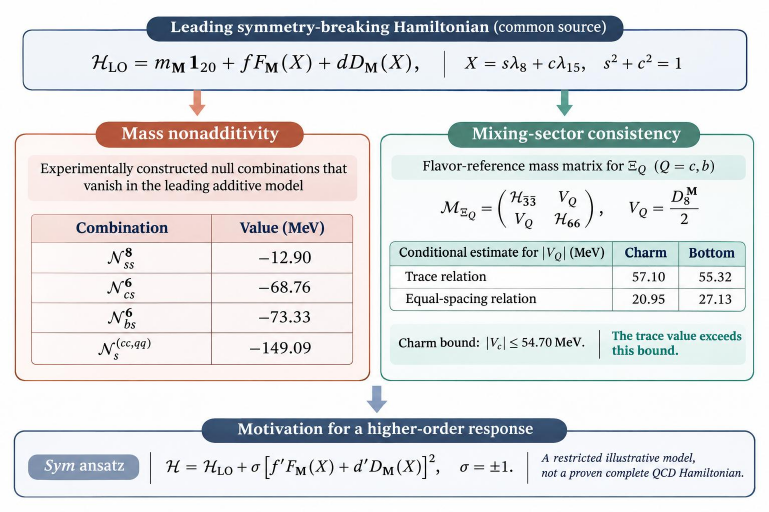}
\caption{
Schematic illustration of the limitations of the
leading flavor-symmetry-breaking Hamiltonian for
mixed-symmetry $J^P=1/2^+$ baryons.
The experimental nonadditive mass combinations and the
two conditional estimates of the $\Xi_Q$ mixing matrix
elements motivate higher-order responses. All mass
quantities are in MeV.
}
\label{fig:baryon-M-leading-limitations}
\end{figure*}

As summarized in
Fig.~\ref{fig:baryon-M-leading-limitations},
the limitations of the leading flavor Hamiltonian
manifest themselves in two related aspects.
First, the experimental mass combinations in
Eqs.~(\ref{eq:baryon-M-nonadditivity-ss})--
(\ref{eq:baryon-M-nonadditivity-ccs})
reveal nonadditive mass responses beyond the
leading description, as already encountered
in the fully symmetric baryon sector.
Second, the mixed-symmetry sector exhibits
an additional discrepancy in the relations
between the diagonal and non-diagonal entries
of the leading mass matrix.
Within group expansion, these observations
motivate the study of higher-order responses
in both mass flow and state mixing.
Although the non-diagonal matrix elements
can affect the flavor-reference state composition,
the experimental masses alone do not determine
the physical state flow.
Moreover, the two discrepancies need not
originate from independent dynamical mechanisms.
In the following subsections, we examine how
higher-order symmetry-breaking responses modify the
mass relations and constrain reference-state mixing.

\subsection{Higher-order symmetry-breaking responses}
\label{sec:baryon-M-higher}

The leading $\mathbf{20_M}$ Hamiltonian contains two
independent adjoint responses, of $F$ and $D$ type.
The nonadditive mass combinations discussed in
Sec.~\ref{sec:baryon-M-leading} show that a purely
linear response is insufficient. As in the vector-meson
analysis, we consider flavor-invariant four-field
interactions involving two adjoint symmetry-breaking
sources. This subsection focuses on their consequences
for the mass spectrum, while flavor admixtures are
discussed separately.

For flavor $SU(4)$, the relevant representation
products are
\begin{eqnarray}
\operatorname{Sym}^2(\mathbf{15})
&=&\mathbf1\oplus\mathbf{15}\oplus\mathbf{20'}
\oplus\mathbf{84},
\nonumber\\
\overline{\mathbf{20_M}}\otimes\mathbf{20_M}
&=&\mathbf1\oplus2\,\mathbf{15}\oplus\mathbf{20'}
\nonumber\\
&\oplus&\mathbf{45}\oplus\overline{\mathbf{45}}
\oplus\mathbf{84}\oplus\mathbf{175}.
\label{eq:baryon-M-quadratic-irreps}
\end{eqnarray}
For a single adjoint scalar source and a specified
$\mathbf{20_M}$ baryon field, the shared representations
allow five independent quadratic covariants, consisting
of one singlet, two adjoints, one $\mathbf{20'}$, and
one $\mathbf{84}$. Their four-field reduced coefficients
are not required by flavor symmetry to inherit the
leading $F/D$ ratio. We compare selected restricted
quadratic responses rather than the most general
four-field Hamiltonian.

We use the normalized diagonal source of the leading
analysis,
\begin{equation}
X=c\lambda_{15}+s\lambda_8,
\label{eq:baryon-M-quadratic-source}
\end{equation}
with $s^2+c^2=1$, and define
\begin{eqnarray}
F_M(X)&=&s{\bf F}^8+c{\bf F}^{15},
\nonumber\\
D_M(X)&=&s{\bf D}^8+c{\bf D}^{15}.
\label{eq:baryon-M-FD-source-operators}
\end{eqnarray}
The leading response is
\begin{equation}
{\cal X}_{\mathbf M}=fF_M(X)+dD_M(X),
\label{eq:baryon-M-leading-response-operator}
\end{equation}
with $F_8^{\mathbf M}=fs$, $F_{15}^{\mathbf M}=fc$,
$D_8^{\mathbf M}=ds$, and $D_{15}^{\mathbf M}=dc$.
Thus,
\begin{equation}
{\cal H}_{\mathbf{20_M}}^{\rm leading}=m_{\mathbf M}{\bf1}_{20}
+{\cal X}_{\mathbf M}
\label{eq:baryon-M-linear-operator}
\end{equation}
contains four independent parameters, including $c/s$.

To examine the roles of the two adjoint structures at
quadratic order, we first consider their separate squares,
\begin{eqnarray}
{\cal H}_{\mathbf{20_M}}^{F^2}&=&{\cal H}_{\mathbf{20_M}}^{\rm leading}
+\kappa_{F^2}F_M(X)^2,
\nonumber\\
{\cal H}_{\mathbf{20_M}}^{D^2}&=&{\cal H}_{\mathbf{20_M}}^{\rm leading}
+\kappa_{D^2}D_M(X)^2.
\label{eq:baryon-M-pure-square-models}
\end{eqnarray}
Both contain five continuous parameters. The labels
$F^2$ and $D^2$ denote squares of the leading
adjoint operators. The leading-aligned square (AS)
ansatz instead uses the complete leading response,
\begin{equation}
{\cal H}_{\mathbf{20_M}}^{\rm AS}
={\cal H}_{\mathbf{20_M}}^{\rm leading}
+\kappa_{\mathbf M}{\cal X}_{\mathbf M}^2.
\label{eq:baryon-M-Sym-fit-H}
\end{equation}
It also contains five parameters. In addition to
assuming a factorized quadratic vertex, AS requires
the quadratic $F/D$ ratio to coincide with that of
the leading response. This alignment is a model
assumption rather than a consequence of $SU(4)$ symmetry.

We next allow the quadratic interaction to have its own
reduced $F$- and $D$-type couplings, independently of
those entering the leading term. Introducing
\begin{equation}
{\cal X}'_{\mathbf M}=f'F_M(X)+d'D_M(X),
\label{eq:baryon-M-factorized-vertex}
\end{equation}
and the source-independent vertex matrices
${\cal V}^a=f'{\bf F}^a+d'{\bf D}^a$, a schematic
four-field interaction takes the form
\begin{equation}
{\cal L}_{\rm baryon}^{\mathbf M,(4)}
\propto\frac12 M^aM^b\bar N^I N^J
\big\{{\cal V}^a,{\cal V}^b\big\}_{IJ}.
\label{eq:baryon-M-factorized-L4}
\end{equation}
Here $M^a$ denotes an adjoint scalar source field.
The overall coupling and source normalization are
absorbed into the effective quadratic amplitudes.
Allowing either sign of the induced quadratic response,
we obtain the symmetrized-square (Sym) Hamiltonian
\begin{equation}
{\cal H}_{\mathbf{20_M}}^{\rm Sym}
=m_{\mathbf M}{\bf1}_{20}
+{\cal X}_{\mathbf M}
+\sigma\big({\cal X}'_{\mathbf M}\big)^2,
\quad\sigma=\pm1.
\label{eq:baryon-M-factorized-H}
\end{equation}
The six continuous fit parameters are
$m_{\mathbf M},f,d,f',d'$, and $c/s$, while the
discrete sign $\sigma$ is not counted as a parameter.
We choose $f'\geq0$ to remove the simultaneous sign
redundancy of $f'$ and $d'$. For $f'\ne0$, the
quadratic term can equivalently be written as
$\kappa[F_M(X)+rD_M(X)]^2$, with
$\kappa=\sigma(f')^2$ and $r=d'/f'$.
The designation Sym follows the square response
used for fully symmetric baryons in Sec.~\ref{sec:20S},
but here the quadratic vertex need not be aligned
with the leading response. AS imposes $r=d/f$,
whereas the separate $F^2$ and $D^2$ models are
obtained by setting $d'=0$ and $f'=0$, respectively,
with the appropriate sign branch. The square
factorization is a restricted four-field ansatz,
not a general requirement of flavor symmetry.
It assumes both a rank-one quadratic coupling in the
$F/D$ operator space and the same flavor-source
direction for the two insertions. These are restrictions
on operator coefficients and do not require the
curvatures of individual mass eigenvalues to factorize.

\begin{table}[t]
\renewcommand{\arraystretch}{1.16}
\setlength{\tabcolsep}{3pt}
\caption{Unweighted mass-fit RMS deviations and
heavy--strange nonadditivity combinations for
mixed-symmetry $J^P=1/2^+$ flavor-reference baryons.
All mass quantities are in MeV, and $N_p$ counts
continuous fit parameters. AS denotes the square
of the leading response, while Sym allows an independent
quadratic $F/D$ ratio. The signs in Sym ($+$) and
Sym ($-$) indicate the sign $\sigma$ of the quadratic
term. The charm and bottom samples contain 11 and 9
masses, respectively. The $F^2$ and $D^2$ labels denote squares
of the leading adjoint operators.}
\label{tab:baryon-M-higher-nested}
\begin{ruledtabular}
\begin{tabular}{lrrrrr}
Model & $N_p$ & ${\rm RMS}_c$ & ${\rm RMS}_b$
& ${\cal N}_{cs}^{\mathbf6}$ & ${\cal N}_{bs}^{\mathbf6}$ \\
\hline
Leading & 4 & 33.47 & 31.27 & $0$ & $0$ \\
$F^2$ only & 5 & 25.53 & 21.96 & $-11.81$ & $-63.64$ \\
$D^2$ only & 5 & 30.05 & 26.33 & $0$ & $0$ \\
AS & 5 & 25.13 & 21.05 & $-12.90$ & $-67.38$ \\
Sym ($+$) & 6 & 4.84 & 3.66 & $-65.59$ & $-75.73$ \\
Sym ($-$) & 6 & 5.45 & 3.94 & $-64.19$ & $-75.75$ \\
Experiment & --- & --- & --- & $-68.76$ & $-73.33$ \\
\end{tabular}
\end{ruledtabular}
\end{table}

For a like-for-like comparison, we refit all models
to the same eleven charm and nine bottom mass inputs
by unweighted least squares. The results are summarized
in Table~\ref{tab:baryon-M-higher-nested}.
All quadratic models improve the RMS relative to the
leading reference fits. The $F^2$ and AS responses
yield comparable moderate improvements, whereas the
isolated $D^2$ response is less effective and leaves
the chosen heavy--strange nonadditivity combinations
identically zero. The Sym model provides a substantially
larger improvement in both sectors. The positive branch
gives RMS deviations of $4.84$~MeV for charm and
$3.66$~MeV for bottom, while the negative branch
gives $5.45$ and $3.94$~MeV, respectively.
Both branches reproduce the heavy--strange nonadditivity
combinations reasonably well. Although the positive
branch gives slightly smaller RMS deviations, the
mass fits alone do not establish the preferred sign
of the quadratic response.

For the positive Sym branches, the fitted leading
ratios are $d/f=-0.1481$ for charm and $-0.0915$
for bottom, while the corresponding quadratic ratios
are $d'/f'=1.9368$ and $2.0300$, respectively.
For the negative branches, the leading ratios are
$d/f=-0.1569$ and $-0.1092$, while the quadratic
ratios are $d'/f'=-1.3430$ and $-0.9950$.
Thus, in both branches, the quadratic $F/D$
combinations differ substantially from their
leading counterparts. The near equality of the
two positive-branch quadratic ratios, and their
proximity to two, is an empirical feature of
the restricted Sym ansatz rather than a universal
relation.

The two sign branches also differ in their effects
on the fitted flavor-breaking source directions.
For the charm and bottom sectors, the leading fits
give $c/s=8.941$ and $31.599$, respectively.
These ratios become $5.477$ and $18.008$ in the
positive branches, but $8.480$ and $29.851$ in the
negative branches. Relative to the leading fits,
the negative branches change these ratios by
approximately $5\%$, whereas the positive branches
change them by approximately $40\%$.
The negative quadratic responses therefore
substantially improve the description of the mass
spectrum while largely preserving the effective
source directions extracted at leading order.

It is instructive to compare these results with
the quadratic responses of the fully symmetric
baryons discussed in Sec.~\ref{sec:20S}, for which
the fitted quadratic coefficients are negative.
Moreover, as will be discussed in
Sec.~\ref{sec:shared-spin-flavor}, the negative
Sym branches are more compatible with the
approximate relation between the linear flavor
response coefficients of the two spin families
suggested by common spin--flavor dynamics.
We therefore adopt the negative Sym branches
as our reference quadratic fits in
Table~\ref{tab:baryon-M-global-fit-summary}
and in the subsequent dynamical analysis.
This choice allows us to investigate the two
baryon families within a common negative
quadratic-response framework, without assuming
that their quadratic interactions have the same
microscopic origin.

All the quadratic models above are polynomials in
$F_M(X)$ and $D_M(X)$, which commute for a fixed
common source,
\begin{equation}
[F_M(X),D_M(X)]=0.
\label{eq:baryon-M-FD-commutator}
\end{equation}
They therefore generate nonlinear mass responses
without introducing an additional noncommuting
structure into the fixed-source Hamiltonian.
However, refitting the source direction can modify
the corresponding reference-state compositions.
The mass fits alone do not determine physical
$\overline{\mathbf3}_F$--$\mathbf6_F$ admixture
probabilities or their heavy-quark limits, which
are addressed separately. Furthermore, the fitted
response coefficients and source directions remain
dependent on the effective Hamiltonian employed.
The small in-sample RMS deviations do not establish
a unique microscopic four-field mechanism or
convergence of the symmetry-breaking expansion.

\subsection{Joint charm--bottom analysis in flavor $SU(5)$}
\label{sec:baryon-M-SU5}

We now consider the joint flavor-$SU(5)_{udscb}$ system,
where the mixed-symmetry baryons belong to the
$\mathbf{40_M}$ flavor-reference representation.
The relevant representation products are
\begin{eqnarray}
\operatorname{Sym}^2(\mathbf{24})
&=&\mathbf1\oplus\mathbf{24}\oplus\mathbf{75}
\oplus\mathbf{200},
\nonumber\\
\overline{\mathbf{40_M}}\otimes\mathbf{40_M}
&=&\mathbf1\oplus2\,\mathbf{24}\oplus\mathbf{75}
\nonumber\\
&\oplus&\mathbf{126}\oplus\overline{\mathbf{126}}
\oplus\mathbf{200}\oplus\mathbf{1024}.
\label{eq:baryon-M-SU5-quadratic-irreps}
\end{eqnarray}
As in flavor $SU(4)$, the shared representations
allow five independent homogeneous quadratic
covariants for a single adjoint scalar source
and a specified $\mathbf{40_M}$ baryon field,
consisting of one singlet, two adjoints,
one $\mathbf{75}$, and one $\mathbf{200}$.
We examine whether the restricted factorized
quadratic response introduced in the preceding
subsection can also provide a common description
of the charm and bottom sectors.

We retain one normalized diagonal source,
\begin{equation}
X=b\lambda_{24}+c\lambda_{15}+s\lambda_8,
\label{eq:baryon-M-SU5-source}
\end{equation}
where $s^2+c^2+b^2=1$. The operators
$F_M(X)$ and $D_M(X)$ are the corresponding
$40\times40$ representation matrices. The leading
Hamiltonian is
\begin{equation}
{\cal H}_{\mathbf{40_M}}^{\rm leading}
=m_{\mathbf M}{\bf1}_{40}+fF_M(X)+dD_M(X),
\label{eq:baryon-M-SU5-leading}
\end{equation}
which contains five continuous parameters,
including the two source ratios $c/s$ and $b/s$.
Following the Sym construction in the preceding
subsection, we consider
\begin{equation}
{\cal H}_{\mathbf{40_M}}^{\rm Sym}
={\cal H}_{\mathbf{40_M}}^{\rm leading}
+\sigma\big[f'F_M(X)+d'D_M(X)\big]^2,
\quad\sigma=\pm1.
\label{eq:baryon-M-SU5-Sym}
\end{equation}
The quadratic couplings $f'$ and $d'$ are independent
of the leading couplings $f$ and $d$.
As before, we choose $f'\geq0$ to remove the
simultaneous sign redundancy of $f'$ and $d'$.
The Sym model contains seven continuous parameters,
$m_{\mathbf M},f,d,f',d',c/s$, and $b/s$,
and represents a restricted realization of the
symmetry-allowed quadratic response space.

\begin{table}[t]
\renewcommand{\arraystretch}{1.16}
\setlength{\tabcolsep}{3pt}
\caption{
Unweighted mass-fit RMS deviations and
heavy--strange nonadditivity combinations for the
joint flavor-$SU(5)_{udscb}$ mixed-symmetry baryons.
All mass quantities are in MeV, and $N_p$ counts
continuous fit parameters.
Sym denotes the factorized quadratic response
with an independent quadratic $F/D$ ratio.
The signs in Sym ($+$) and Sym ($-$) indicate
the sign $\sigma$ of the quadratic term.
}
\label{tab:baryon-M-SU5-quadratic-comparison}
\begin{ruledtabular}
\begin{tabular}{lrrrr}
Model & $N_p$ & RMS
& ${\cal N}_{cs}^{\mathbf6}$
& ${\cal N}_{bs}^{\mathbf6}$ \\
\hline
Leading & 5 & 43.62 & $0$ & $0$ \\
Sym ($+$) & 7 & 20.45 & $-16.89$ & $-49.49$ \\
Sym ($-$) & 7 & 21.82 & $-12.55$ & $-48.07$ \\
Experiment & --- & --- & $-68.76$ & $-73.33$ \\
\end{tabular}
\end{ruledtabular}
\end{table}

We fit these models by unweighted least squares to
the same sixteen baryon masses, including four light,
five singly charmed, two doubly charmed, and five
singly bottom states. The results are summarized
in Table~\ref{tab:baryon-M-SU5-quadratic-comparison}.
Both Sym branches substantially improve the
description of the mass spectrum, reducing the
leading RMS of $43.62$~MeV to $20.45$ and
$21.82$~MeV for the positive and negative
branches, respectively.
Nevertheless, the joint flavor-$SU(5)$ fits
are considerably less accurate than the separate
$SU(4)_c$ and $SU(4)_b$ fits.
The joint analysis imposes common response
coefficients across the charm and bottom sectors
and uses a different fitting sample, thus
providing a more restrictive test of the
factorized quadratic ansatz.
In particular, neither branch satisfactorily
reproduces the heavy--strange nonadditivity
combinations in both heavy-flavor sectors.

For the positive Sym branch, the fitted
linear and quadratic ratios are
$d/f=-0.0467$ and $d'/f'=2.5720$,
respectively, whereas the negative branch gives
$d/f=-0.0482$ and $d'/f'=-0.2543$.
These quadratic ratios differ from those
obtained in the separate $SU(4)$ fits.
In particular, the approximate equality of the
quadratic $F/D$ ratios in the positive $SU(4)$
branches does not by itself ensure a successful
factorized description of the combined
heavy-flavor system.

As in the preceding subsection, the two sign
branches differ in their readjustments of the
fitted flavor-breaking source.
The leading fit gives
\begin{equation}
(c/s,b/s)=(9.039,\,31.604),
\end{equation}
whereas the two Sym branches give
\begin{eqnarray}
(c/s,b/s)_{\rm Sym(+)}
&=&(7.558,\,21.632),
\nonumber\\
(c/s,b/s)_{\rm Sym(-)}
&=&(8.598,\,32.814).
\end{eqnarray}
Thus, the negative branch again preserves
the effective source ratios extracted at
leading order more closely, with changes
of less than $5\%$ in both ratios.
We therefore adopt the negative Sym branch
as our reference quadratic fit in
Table~\ref{tab:baryon-M-global-fit-summary},
consistent with the choice made in the
separate $SU(4)$ analyses and the negative
quadratic responses of the fully symmetric
baryons.

The limitations of the common factorized
response are further illustrated by the
doubly-charmed nonadditivity combination
${\cal N}_{s}^{cc,qq}$ defined in
Eq.~(\ref{eq:baryon-M-nonadditivity-ccs}).
Its experimental reference value is
$-149.09$~MeV, whereas the positive and
negative Sym branches give $-33.77$ and
$-25.11$~MeV, respectively.
Within the present common-source quadratic
ansatz, the relation
\begin{equation}
{\cal N}_{s}^{cc,qq}
=2{\cal N}_{cs}^{\mathbf6}
\label{eq:baryon-M-SU5-nonadditivity-relation}
\end{equation}
holds exactly.
For the adopted experimental inputs,
the difference
${\cal N}_{s}^{cc,qq}-2{\cal N}_{cs}^{\mathbf6}$
is $-11.57$~MeV.
The discrepancies between the fitted and
experimental nonadditivity combinations
are, however, substantially larger than
this violation alone.
The limitations of the joint fits therefore
cannot be attributed solely to the above
mass relation.

These results show that the factorized
quadratic response, while effective in the
separate charm and bottom systems, is less
successful when extended to a common
flavor-$SU(5)$ Hamiltonian.
Although the negative branch largely preserves
the effective source ratios extracted at
leading order, its quadratic response cannot
account for all the observed mass relations.
The fitted response coefficients remain
effective parameters of a restricted Hamiltonian,
and the improvement in the mass-fit RMS does
not establish convergence of the expansion
or a unique microscopic four-field interaction.
In particular, extrapolations to unmeasured
$bc$ and $bb$ baryons remain sensitive to
the assumed quadratic response.

These limitations motivate a closer examination
of the internal correlations underlying the
flavor responses. Their dynamical origin and
the additional constraints relating different
baryon families will be discussed in
Sec.~\ref{sec:shared-dynamics}.

\subsection{State mixing and identifiability}
\label{sec:baryon-M-identifiability}

The preceding subsections show that higher-order flavor
responses substantially improve the description of
mixed-symmetry baryon masses. However, mass eigenvalues
alone do not uniquely determine the mixing of their
reference-state components.

Consider a real symmetric two-state Hamiltonian,
such as the $\Xi_Q$ flavor-reference block,
\begin{equation}
H_\alpha=
\begin{pmatrix}
A_\alpha&V_\alpha\\
V_\alpha&B_\alpha
\end{pmatrix},
\qquad
\Delta_\alpha=B_\alpha-A_\alpha.
\label{eq:baryon-M-two-state-H}
\end{equation}
Its eigenvalues and mixing angle satisfy
\begin{eqnarray}
M_{\alpha\pm}
&=&\frac{A_\alpha+B_\alpha}{2}
\pm\frac12\sqrt{\Delta_\alpha^2+4V_\alpha^2},
\label{eq:baryon-M-two-state-eigenvalues}\\
\tan 2\theta_\alpha
&=&\frac{2V_\alpha}{\Delta_\alpha}.
\label{eq:baryon-M-two-state-angle}
\end{eqnarray}
For a fixed reference-state phase convention, the two
eigenvalues determine the trace and determinant of the
Hamiltonian, but not its three independent matrix
elements. Different mixing angles can therefore
correspond to the same isolated two-state spectrum.

The common-source Hamiltonians considered above impose
additional restrictions. At a fixed flavor-breaking
source $X$, the adjoint operators satisfy
\begin{equation}
[F_M(X),D_M(X)]=0.
\label{eq:baryon-M-fixed-source-commutator}
\end{equation}
All the leading and quadratic Hamiltonians considered
above are polynomials in these commuting operators
and thus admit a common eigenbasis at fixed $X$.
Higher-order coefficients can modify the mass eigenvalues
without inducing an additional continuous eigenvector
rotation, although the ordering of the eigenstates
may change. Different fitted source directions,
however, may lead to different reference-state
compositions.

The $\Xi_Q$ sector illustrates this distinction.
In the fixed $(|\overline{\mathbf3}\rangle,
|\mathbf6\rangle)$ reference basis, $F_M(X)$ is
proportional to the identity, while for
$X=s\lambda_8+c\lambda_{15}$ with $s^2+c^2=1$,
\begin{equation}
D_Q\equiv D_M(X)\big|_{\Xi_Q}=
\begin{pmatrix}
-\dfrac{7c}{\sqrt{216}}+\dfrac{s}{\sqrt{27}}
&\dfrac{s}{2}\\[2mm]
\dfrac{s}{2}&\sqrt{\dfrac38}\,c
\end{pmatrix}.
\label{eq:baryon-M-Xi-common-source-D}
\end{equation}
By the Cayley--Hamilton identity for a two-state
matrix, the Sym Hamiltonian in this block reduces to
$H_{\Xi_Q}^{\rm Sym}=\alpha_Q{\bf1}_2+\beta_QD_Q$.
Consequently,
\begin{eqnarray}
V_Q&=&\frac{\beta_Qs}{2},
\nonumber\\
\Delta_Q&=&\beta_Q\left(
\frac{16c}{\sqrt{216}}-\frac{s}{\sqrt{27}}
\right),
\label{eq:baryon-M-Xi-Sym-reduced}
\end{eqnarray}
and hence, for a nondegenerate branch with fixed
eigenstate ordering,
\begin{equation}
\tan 2\theta_Q=
\frac{s}{16c/\sqrt{216}-s/\sqrt{27}}.
\label{eq:baryon-M-Sym-mixing-angle}
\end{equation}
Thus, the reference-state mixing angle depends on the
source ratio $c/s$, rather than directly on the leading
or quadratic coupling strengths. The latter affect
$V_Q$ and the diagonal splitting through the common
factor $\beta_Q$.

\begin{table}[t]
\renewcommand{\arraystretch}{1.15}
\setlength{\tabcolsep}{4pt}
\caption{
Source ratios, off-diagonal matrix elements,
and reference-state mixing angles from the separate
$SU(4)_c$ and $SU(4)_b$ Leading and Sym fits.
The $+$ and $-$ signs label the sign of the Sym
quadratic response. The $|V_Q|$ values are in MeV.
The angles refer to the specified flavor-reference
basis, not to experimentally determined physical
mixing angles.
}
\label{tab:baryon-M-Sym-mixing-angles}
\begin{ruledtabular}
\begin{tabular}{llrrr}
Sector & Model & $c/s$ & $|V_Q|$ & $|\theta_Q|$ \\
\hline
$SU(4)_c$ & Leading & 8.9406 & 7.52 & $2.99^\circ$ \\
$SU(4)_c$ & Sym ($+$) & 5.4771 & 8.45 & $4.92^\circ$ \\
$SU(4)_c$ & Sym ($-$) & 8.4798 & 5.36 & $3.16^\circ$ \\
$SU(4)_b$ & Leading & 31.5991 & 2.14 & $0.84^\circ$ \\
$SU(4)_b$ & Sym ($+$) & 18.0077 & 3.43 & $1.47^\circ$ \\
$SU(4)_b$ & Sym ($-$) & 29.8515 & 2.05 & $0.89^\circ$ \\
\end{tabular}
\end{ruledtabular}
\end{table}

As shown in Table~\ref{tab:baryon-M-Sym-mixing-angles},
both Sym branches improve the mass spectra without
requiring large reference-state mixing.
In particular, the negative branches adopted as
our reference quadratic fits yield mixing angles
of $3.16^\circ$ for charm and $0.89^\circ$ for
bottom, which remain close to the corresponding
leading values of $2.99^\circ$ and $0.84^\circ$.
This reflects the relatively small readjustments
of the fitted source directions discussed in
Sec.~\ref{sec:baryon-M-higher}.
Both Sym branches yield smaller reference-state
mixing angles for bottom than for charm.
The fitted $V_Q$ values should not be identified
with the conditional estimates obtained from
leading sextet equal spacing in
Eq.~(\ref{eq:baryon-M-mixing-exp-estimates}),
since quadratic responses modify that leading
relation.

More generally, the flavor-breaking source may evolve
along a path in the parameter space $\boldsymbol\xi$.
In a fixed reference basis, the mixing angle of a
smooth nondegenerate two-state branch satisfies
\begin{equation}
\partial_a\theta_\alpha=
\frac{\Delta_\alpha\,\partial_a V_\alpha
-V_\alpha\,\partial_a\Delta_\alpha}
{\Delta_\alpha^2+4V_\alpha^2}.
\label{eq:baryon-M-mixing-angle-flow}
\end{equation}
This is a two-state realization of the state-flow
relation in Eq.~(\ref{eq:gsb-state-flow}), complementing
the mass-flow relation in Eq.~(\ref{eq:gsb-mass-flow}).
Diagonal response matrix elements in the instantaneous
eigenbasis determine local mass shifts, while
off-diagonal response matrix elements, relative to
the level separations, govern eigenvector evolution.
Thus, a small mass shift need not imply a small
state rearrangement. In the present common-source
models, changing coupling strengths at fixed $X$
does not continuously rotate the eigenvectors,
whereas a changing source direction can generate
state flow. The formula does not apply at an exact
degeneracy, and a fit at the physical flavor point
does not uniquely reconstruct the full breaking
trajectory.

The local identifiability of a specified model can be
tested through the mass-sensitivity matrix
\begin{equation}
J_{ia}=\frac{\partial M_i}{\partial p_a}
=\left\langle\psi_i\left|
\frac{\partial H(\boldsymbol p)}{\partial p_a}
\right|\psi_i\right\rangle,
\label{eq:baryon-M-observable-Jacobian}
\end{equation}
for nondegenerate eigenvalues, where $\boldsymbol p$
denotes the continuous fit parameters.
The joint $SU(5)$ Leading fit and both Sym branches
have full local column ranks of five and seven,
respectively. Nevertheless, local parameter
identifiability does not guarantee global uniqueness,
numerical stability, or a unique microscopic
interpretation of the fitted interactions.

Finally, reference-state mixing must be distinguished
from physical baryon mixing.
A physical interpretation of
$\overline{\mathbf3}_F$--$\mathbf6_F$ mixing requires
a Pauli-compatible spin--flavor and spatial embedding.
In the lowest spatially symmetric singly-heavy reference,
these configurations are associated with $j_\ell=0$
and $j_\ell=1$, respectively.
In the heavy-quark limit, heavy-quark spin symmetry
(HQSS) forbids mixing between states with different
light-cloud angular momenta.
However, in an enlarged Pauli-compatible Hilbert space,
both flavor representations can occur within the same
$j_\ell$ sector and may mix when light-flavor symmetry
is broken.
The constraints imposed by HQSS on the matched
physical Hamiltonian, rather than directly on the
flavor-reference matrix elements, will be discussed
in Sec.~\ref{sec:shared-HQSS}.
Transition observables, heavy-quark spin partners,
and lattice-QCD state overlaps could provide
complementary constraints on physical state compositions.

The present analysis therefore distinguishes mass
responses from state responses within group expansion.
In particular, the negative Sym branches substantially
improve the baryon mass fits while largely preserving
the reference-state compositions obtained at leading
order. This illustrates how sizable higher-order
mass corrections can arise without substantial
eigenvector rearrangements.
Nevertheless, the resulting mixing angles remain
model-dependent, and mass spectra alone cannot
establish a unique microscopic mechanism or
a complete trajectory of physical state evolution.

\section{Dynamics and constraints}
\label{sec:shared-dynamics}

The preceding sections have classified the flavor responses
allowed in the vector-meson, fully symmetric baryon, and
mixed-symmetry baryon reference spaces. In particular, we
have identified one independent leading-order flavor structure
in the fully symmetric baryon representation and two in the
mixed representation. Although these structures are independent
from a group-theoretical perspective, their reduced coefficients
may be related through common internal dynamics. Moreover,
the corresponding physical responses may be further constrained
by symmetries other than flavor symmetry.

In this section, we investigate these issues within the lowest
spatially symmetric three-quark spin--flavor reference space.
We first examine how common spin--flavor operators generate
and relate the flavor responses of the $J^P=1/2^+$ and
$3/2^+$ baryons. We then investigate how the flavor of a
spectator quark modifies the effective correlations between
the other two quarks, thereby introducing additional response
directions. Finally, we discuss how heavy-quark spin symmetry
(HQSS) constrains the physical responses after the appropriate
spin--flavor matching. These studies illustrate how the
symmetry-allowed response spaces are realized and constrained
by the internal dynamics and other physical symmetries,
without assuming a unique microscopic origin for their
effective coefficients.

\subsection{Common spin--flavor dynamics}
\label{sec:shared-spin-flavor}

For three quarks in the lowest spatially symmetric reference,
the antisymmetric color wave function requires a symmetric
spin--flavor configuration. The corresponding spin--flavor
decompositions are
\begin{eqnarray}
\operatorname{Sym}^{3}(\mathbf4\otimes\mathbf2)
&=&(\mathbf{20_S},\mathbf4)
\oplus(\mathbf{20_M},\mathbf2),
\nonumber\\
\operatorname{Sym}^{3}(\mathbf5\otimes\mathbf2)
&=&(\mathbf{35_S},\mathbf4)
\oplus(\mathbf{40_M},\mathbf2),
\label{eq:shared-spin-flavor-branching}
\end{eqnarray}
where the second factor denotes the spin representation.
Although the $J^P=1/2^+$ and $3/2^+$ baryons belong to
different flavor representations, they are thus embedded
in a common spin--flavor reference space. This allows us
to investigate their flavor responses in terms of common
internal operators, without assuming an exact spin--flavor
symmetry of the Hamiltonian.

In particular, the Pauli-compatible spin--flavor reference
in the mixed sector involves correlated flavor and spin
permutation structures, schematically
\begin{equation}
W_M=\frac{1}{\sqrt2}
\left(V_\rho\chi_\rho+V_\lambda\chi_\lambda\right),
\label{eq:shared-Pauli-reference}
\end{equation}
where $V_{\rho,\lambda}$ and $\chi_{\rho,\lambda}$
denote the corresponding mixed-symmetry flavor and spin
components, with consistent normalization and phase
conventions. This combined reference is distinct from
a single flavor carrier used to define the matrices
${\bf F}$ and ${\bf D}$.

\begin{figure*}[t]
\centering
\includegraphics[width=0.8\textwidth]{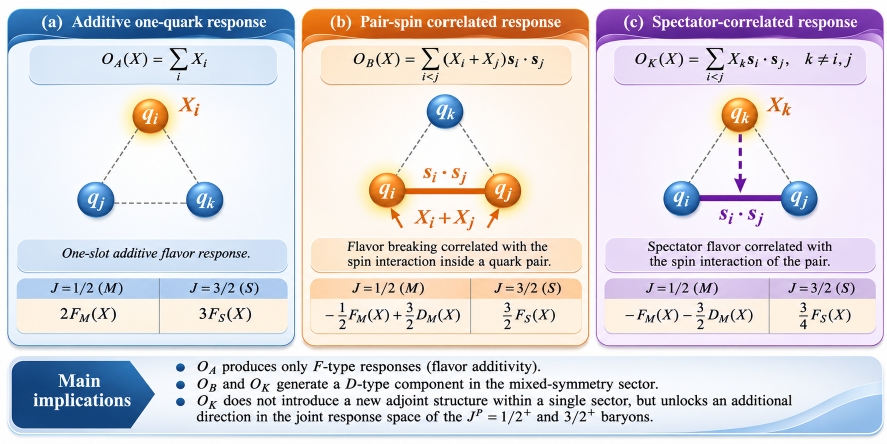}
\caption{
Schematic illustration of the three first-order spin--flavor
operators defined in Eq.~(\ref{eq:shared-first-order-operators})
and their projections onto the $J^P=1/2^+$ ($M$) and
$3/2^+$ ($S$) baryon reference spaces.
(a) Additive one-quark response.
(b) Pair-spin correlated response.
(c) Spectator-correlated response.
Only representative terms of the permutation-symmetric
operators are illustrated.
}
\label{fig:shared-spin-flavor}
\end{figure*}

Let $X$ denote a traceless diagonal flavor-breaking source
and $X_i$ its action on quark slot $i$. With
$\mathbf s_i=\boldsymbol\sigma_i/2$, we consider three
permutation-symmetric operators linear in $X$:
\begin{eqnarray}
O_A(X)&=&\sum_iX_i,
\nonumber\\
O_B(X)&=&\sum_{i<j}(X_i+X_j)
\mathbf s_i\cdot\mathbf s_j,
\nonumber\\
O_K(X)&=&\sum_{i<j}X_k
\mathbf s_i\cdot\mathbf s_j,
\quad k\ne i,j.
\label{eq:shared-first-order-operators}
\end{eqnarray}
Their physical interpretations are schematically illustrated
in Fig.~\ref{fig:shared-spin-flavor}.
The first operator describes an additive one-quark
response, independent of the spin correlations among
the quarks. The second describes the change in the
effective spin interaction between two quarks induced
by their own flavor breaking. The third describes
a different type of correlation, in which the flavor
of the remaining spectator quark affects the effective
spin interaction within the pair. Although all three
operators are of first order in the flavor-breaking
source, they involve different internal correlations.

The projections of these operators can be understood
from the spin configurations of the two baryon families.
In the fully symmetric $J=3/2$ sector, every quark pair
is in a spin-triplet configuration, for which
$\mathbf s_i\cdot\mathbf s_j=1/4$. Consequently,
\begin{eqnarray}
O_B(X)\big|_S
&=&\frac12 O_A(X)\big|_S,
\nonumber\\
O_K(X)\big|_S
&=&\frac14 O_A(X)\big|_S.
\end{eqnarray}
Thus, despite their different internal correlations,
all three operators generate the same flavor structure
in the fully symmetric representation.

The situation is different in the mixed-symmetry
$J=1/2$ sector, where the spin-singlet and spin-triplet
pair configurations, with
$\mathbf s_i\cdot\mathbf s_j=-3/4$ and $1/4$,
respectively, are correlated with different flavor
permutation structures. Projecting the three operators
onto the normalized $M$ and $S$ spin--flavor references,
we obtain
\begin{equation}
\begin{array}{c|cc}
 &J=1/2~(M)&J=3/2~(S)\\ \hline
O_A&2F_M(X)&3F_S(X)\\[2pt]
O_B&-\frac12F_M(X)+\frac32D_M(X)&\frac32F_S(X)\\[2pt]
O_K&-F_M(X)-\frac32D_M(X)&\frac34F_S(X)
\end{array}
\label{eq:shared-first-order-dictionary}
\end{equation}
where $F_M$, $D_M$, and $F_S$ have the normalizations
adopted in the preceding baryon sections.
The additive operator $O_A$, being the total flavor
generator, produces only an $F$-type response.
In contrast, the spin-correlated operators
$O_B$ and $O_K$ also generate a $D$-type response in
the mixed representation. This provides a concrete
dynamical realization of the additional adjoint
structure identified in Sec.~\ref{sec:20M},
although other internal correlations may generate
the same effective flavor structure.

We now consider a minimal common spin--flavor
Hamiltonian at first order in $X$:
\begin{equation}
H_{\rm sf}^{(1)}(X)=aO_A(X)+bO_B(X)+dO_K(X),
\label{eq:shared-first-order-model}
\end{equation}
where $a$, $b$, and $d$ are assumed to be common to
the two spin families for the same flavor-breaking
source and reference convention.
The corresponding reduced response coefficients are
\begin{eqnarray}
F^M&=&2a-\frac b2-d,
\nonumber\\
D^M&=&\frac32(b-d),
\nonumber\\
F^S&=&3a+\frac32b+\frac34d.
\label{eq:shared-reduced-couplings}
\end{eqnarray}
These coefficients consequently satisfy
\begin{equation}
F^S-\frac32(F^M+D^M)=\frac92d.
\label{eq:shared-one-spectator-relation}
\end{equation}
If the common dynamics is restricted to the additive
and pair-spin operators $O_A$ and $O_B$, the right-hand
side vanishes, leading to
\begin{equation}
F^S=\frac32(F^M+D^M).
\label{eq:shared-restricted-relation}
\end{equation}
Although the three reduced coefficients are independent
from the viewpoint of flavor group theory, they are
determined by only two independent dynamical coefficients
in this restricted realization. The spectator-correlated
operator $O_K$ lifts this additional constraint,
allowing all three flavor response coefficients to
vary independently within the common operator basis.

To clarify the origin of this additional freedom,
we define
\begin{equation}
J_{\rm pair}=\sum_{i<j}\mathbf s_i\cdot\mathbf s_j
=\frac12\left(\mathbf J^2-\frac94\right),
\end{equation}
where $\mathbf J=\sum_i\mathbf s_i$ in the present
lowest spatial reference. Using
$X_k=O_A(X)-(X_i+X_j)$ for $k\ne i,j$, we obtain
the operator identity
\begin{equation}
O_K(X)=O_A(X)J_{\rm pair}-O_B(X).
\label{eq:shared-spectator-identity}
\end{equation}
Since $J_{\rm pair}$ takes the values $-3/4$ and $3/4$
in the $J=1/2$ and $3/2$ sectors, respectively,
the operator $O_K$ can be expressed as a linear
combination of $O_A$ and $O_B$ within either sector:
\begin{eqnarray}
O_K(X)\big|_M
&=&-\frac34 O_A(X)\big|_M-O_B(X)\big|_M,
\nonumber\\
O_K(X)\big|_S
&=&+\frac34 O_A(X)\big|_S-O_B(X)\big|_S.
\end{eqnarray}
However, these linear combinations are different
in the two sectors, so the contribution of $O_K$
cannot be absorbed simultaneously into the same
redefinition of the common coefficients $a$ and $b$.
In the language of Sec.~\ref{sec:group-expansion-picture},
the spectator-correlated operator therefore unlocks
an additional direction in the joint response space,
increasing the dimension of the dynamically accessible
first-order response space from two to three.
This unlocking results from relaxing a dynamical
restriction rather than from lowering the flavor
symmetry or introducing a new independent adjoint
structure within either representation.

It is interesting to examine whether the restricted
common-dynamics relation in
Eq.~(\ref{eq:shared-restricted-relation}) is approximately
satisfied by the independently fitted response coefficients
obtained in Secs.~\ref{sec:20S} and \ref{sec:20M}.
For each diagonal flavor-breaking direction $\lambda_A$,
we define the deviation
\begin{equation}
\Delta_A^{\rm sf}
=
F_A^{\mathbf S}
-\frac32\left(F_A^{\mathbf M}+D_A^{\mathbf M}\right),
\label{eq:shared-response-deviation}
\end{equation}
and its relative magnitude
\begin{equation}
R_A^{\rm sf}
=
\frac{|\Delta_A^{\rm sf}|}{|F_A^{\mathbf S}|}.
\label{eq:shared-relative-deviation}
\end{equation}
By applying Eq.~(\ref{eq:shared-one-spectator-relation})
to each generator direction separately, the deviation
can be parametrized in terms of an effective
spectator coefficient,
\begin{equation}
d_A^{\rm eff}=\frac29\Delta_A^{\rm sf}.
\label{eq:shared-effective-spectator}
\end{equation}
Here we compare the coefficients multiplying the same
normalized flavor generator in the two baryon families,
without imposing a common normalized source direction
in their preceding mass fits. Accordingly, this
comparison constitutes a direction-by-direction
diagnostic of the common-dynamics relation, rather
than a simultaneous fit of the common spin--flavor
Hamiltonian in Eq.~(\ref{eq:shared-first-order-model}).

\begin{table*}[t]
\renewcommand{\arraystretch}{1.25}
\setlength{\tabcolsep}{4pt}
\small
\caption{
Comparison of the independently fitted linear flavor
response coefficients of the fully symmetric and
mixed-symmetry baryons, taken from
Tables~\ref{tab:baryon-20S-leading-masses}
and \ref{tab:baryon-M-global-fit-summary}.
The coefficients $F_A^{\mathbf S}$,
$F_A^{\mathbf M}$, $D_A^{\mathbf M}$,
$\Delta_A^{\rm sf}$, and $d_A^{\rm eff}$ are
evaluated using the Leading fits and are given in MeV.
The last two columns give the relative deviations
$R_A^{\rm sf}$ obtained from the Leading and
Sym fits, respectively.
The Sym column uses the Sym fits for the
fully symmetric baryons and the reference Sym ($-$)
fits for the mixed-symmetry baryons in all three
flavor systems.
}
\label{tab:shared-first-order-comparison}
\begin{ruledtabular}
\begin{tabular}{lcrrrrrrr}
Flavor group
& $A$
& $F_A^{\mathbf S}$
& $F_A^{\mathbf M}$
& $D_A^{\mathbf M}$
& $\Delta_A^{\rm sf}$
& $d_A^{\rm eff}$
& \multicolumn{2}{c}{$R_A^{\rm sf}$ (\%)} \\
& & & & & & & Leading & Sym \\
\hline
$SU(4)_c$
& 8
& $-243.22$
& $-172.10$
& $15.05$
& $-7.65$
& $-1.70$
& $3.15$
& $0.44$ \\
& 15
& $-2224.86$
& $-1538.63$
& $134.54$
& $-118.71$
& $-26.38$
& $5.34$
& $10.88$ \\
\hline
$SU(4)_b$
& 8
& $-250.60$
& $-179.35$
& $4.29$
& $11.99$
& $2.66$
& $4.78$
& $6.05$ \\
& 15
& $-8329.59$
& $-5667.16$
& $135.51$
& $-32.12$
& $-7.14$
& $0.39$
& $11.61$ \\
\hline
$SU(5)_{udscb}$
& 8
& $-241.02$
& $-171.25$
& $5.15$
& $8.13$
& $1.81$
& $3.37$
& $6.32$ \\
& 15
& $-2224.47$
& $-1548.02$
& $46.55$
& $27.74$
& $6.16$
& $1.25$
& $0.70$ \\
& 24
& $-8021.42$
& $-5412.18$
& $162.75$
& $-147.26$
& $-32.73$
& $1.84$
& $4.61$ \\
\end{tabular}
\end{ruledtabular}
\end{table*}

The numerical results are summarized in
Table~\ref{tab:shared-first-order-comparison}.
Remarkably, the relative deviations obtained from
the Leading fits range from $0.39\%$ to $5.34\%$
across the seven comparisons, although the relation
in Eq.~(\ref{eq:shared-restricted-relation}) was not
imposed in the preceding mass fits.
This approximate agreement is consistent with a
common spin--flavor organization of the dominant
linear flavor responses of the two baryon families.
In particular, it is compatible with the dominance
of the additive one-quark response, supplemented
by pair-spin correlations.
However, the small relative deviations partly
reflect the large additive contributions, and the
different flavor-group fits share some experimental
mass inputs. They should therefore not be interpreted
as independent determinations of the underlying
dynamical mechanism.

When negative quadratic responses are included in
both baryon families, the relative deviations range
from $0.44\%$ to $11.61\%$, as shown in the last
column of Table~\ref{tab:shared-first-order-comparison}.
In particular, all three deviations in the joint
flavor-$SU(5)$ system remain below $7\%$.
Although the deviations increase in five of the
seven comparisons, the independently refitted
linear coefficients still approximately satisfy
Eq.~(\ref{eq:shared-restricted-relation}).
Thus, the approximate common-dynamics relation
is compatible with the improved description of
the baryon masses provided by the selected
quadratic responses.
Since this compatibility is one of the motivations
for choosing the negative Sym branches in
Sec.~\ref{sec:20M}, the quadratic comparison
should not be regarded as an independent
verification of the common-dynamics relation.

The remaining deviations cannot, however, be attributed
uniquely to the spectator-correlated operator $O_K$.
The preceding fits employ different quadratic
parametrizations in the two baryon families,
with independently refitted source directions
and linear coefficients.
Moreover, these finite-point effective coefficients
need not coincide with the local first-order
response coefficients at the exact symmetry point.
The values of $d_A^{\rm eff}$ should therefore be
regarded as effective measures of the departure
from the restricted common-dynamics relation,
rather than direct determinations of a microscopic
spectator interaction.

The above analysis illustrates how common internal
dynamics can relate flavor responses belonging to
different representations, and how additional internal
correlations can release such dynamical constraints.
It also motivates a closer examination of how the
flavor of a spectator quark affects the effective
correlations between the other two quarks.

\subsection{Spectator-dependent dynamics}
\label{sec:shared-spectator}

The preceding subsection illustrates how spectator
correlations can release dynamical constraints between
different baryon representations. We now investigate
how the flavor of a spectator quark affects the
effective interaction between the other two quarks.

Within the lowest spatially symmetric three-quark
reference, we consider an effective pair-spin Hamiltonian,
\begin{equation}
H_{\rm pair}
=E_{123}{\bf1}
+\sum_{i<j}b_{ij}^{(k)}
\mathbf s_i\cdot\mathbf s_j,
\quad k\ne i,j,
\label{eq:shared-effective-pair-H}
\end{equation}
where $E_{123}$ denotes a common spin-independent
energy for the specified quark flavor composition,
and $b_{ij}^{(k)}$ characterizes the effective spin
interaction between quarks $i$ and $j$ in the
presence of the spectator quark $k$.
For baryons with the flavor composition $qqQ$,
where $q=u,d$ in the isospin limit and $Q=s,c,b$,
the above Hamiltonian gives
\begin{eqnarray}
M_{\Lambda_Q}
&=&E_{qqQ}-\frac34b_{qq}^{(Q)},
\nonumber\\
M_{\Sigma_Q}
&=&E_{qqQ}+\frac14b_{qq}^{(Q)}
-b_{qQ}^{(q)},
\nonumber\\
M_{\Sigma_Q^*}
&=&E_{qqQ}+\frac14b_{qq}^{(Q)}
+\frac12b_{qQ}^{(q)},
\label{eq:shared-qqQ-pair-masses}
\end{eqnarray}
where $\Lambda_s$, $\Sigma_s$, and $\Sigma_s^*$
denote the ordinary strange baryons
$\Lambda$, $\Sigma$, and $\Sigma^*$, respectively.
The spin-weighted average of the $\Sigma_Q$
and $\Sigma_Q^*$ masses eliminates the spin
interaction between the $q$ and $Q$ quarks.
Consequently, the effective pair-spin coefficients
can be extracted as
\begin{eqnarray}
b_{qq}^{(Q)}
&=&\frac{M_{\Sigma_Q}+2M_{\Sigma_Q^*}}{3}
-M_{\Lambda_Q},
\nonumber\\
b_{qQ}^{(q)}
&=&\frac23(M_{\Sigma_Q^*}-M_{\Sigma_Q}),
\nonumber\\
b_{qq}^{(q)}
&=&\frac23(M_\Delta-M_N).
\label{eq:shared-qqQ-pair-extraction}
\end{eqnarray}
Using the experimental masses adopted in the
preceding baryon sections, we obtain
\begin{equation}\nonumber
\begin{array}{c|cccc}
k&q&s&c&b\\ \hline
b_{qq}^{(k)}~({\rm MeV})
&195.39&205.11&210.09&206.48
\end{array}
\end{equation}
Although the interacting pair has the same light
flavors in all four cases, the presence of a
strange, charm, or bottom spectator increases
its extracted effective spin-interaction coefficient
by approximately $5$--$8\%$ relative to
$b_{qq}^{(q)}$.
This motivates the spectator-dependent response
$O_K(X)$ introduced in
Eq.~(\ref{eq:shared-first-order-operators}).
Such an environmental dependence may arise from
changes in the internal spatial distributions
and correlations of the quarks.
The extracted coefficients are conditional on
the effective Hamiltonian in
Eq.~(\ref{eq:shared-effective-pair-H}) and should
not be identified with fundamental QCD couplings.

A further question is whether the same change of
spectator flavor affects different interacting pairs
in the same way.
For the $qs$ pair, we similarly obtain
\begin{eqnarray}
b_{qs}^{(q)}
&=&\frac23(M_{\Sigma^*}-M_\Sigma)
=127.64~{\rm MeV},
\nonumber\\
b_{qs}^{(s)}
&=&\frac23(M_{\Xi^*}-M_\Xi)
=143.41~{\rm MeV}.
\label{eq:shared-qs-pair-extraction}
\end{eqnarray}
Replacing the spectator quark $q$ with $s$
increases the effective $qq$ and $qs$ interaction
coefficients by $9.72$ and $15.77$~MeV,
respectively.
We therefore define the environmental nonadditivity
combination
\begin{equation}
L_{\rm env}
=\big[b_{qs}^{(s)}-b_{qs}^{(q)}\big]
-\big[b_{qq}^{(s)}-b_{qq}^{(q)}\big]
=6.05~{\rm MeV}.
\label{eq:shared-environment-null}
\end{equation}
A universal additive spectator correction,
independent of the flavors of the interacting pair,
would give $L_{\rm env}=0$.
Its nonzero central value motivates a correlated
environmental response that depends on both the
spectator and the interacting pair.

Let $x_i$ denote the eigenvalue of the diagonal
flavor-breaking source $X$ corresponding to
the flavor of quark $i$.
Through second order in $X$, we consider
the expansion
\begin{eqnarray}
b_{ij}^{(k)}
&=&b_0+b_1(x_i+x_j)+d\,x_k
\nonumber\\
&&+b_2(x_i^2+x_j^2)+b_3x_ix_j
\nonumber\\
&&+d_E x_k(x_i+x_j)+d_Kx_k^2
+{\cal O}(X^3).
\label{eq:shared-pair-response-expansion}
\end{eqnarray}
A common quadratic singlet contribution proportional
to $\operatorname{Tr}(X^2)$ is omitted here, since
it does not contribute to $L_{\rm env}$.
The terms proportional to $b_1$, $b_2$, and $b_3$
describe the responses associated with the flavors
of the interacting pair, whereas those proportional
to $d$, $d_E$, and $d_K$ describe the responses
associated with the spectator flavor.

The spectator-dependent terms generate the
first-order operator $O_K(X)$ and the two
second-order operators
\begin{eqnarray}
O_{2E}(X)
&=&\sum_{i<j}X_k(X_i+X_j)
\mathbf s_i\cdot\mathbf s_j,
\nonumber\\
O_{2K}(X)
&=&\sum_{i<j}X_k^2
\mathbf s_i\cdot\mathbf s_j,
\quad k\ne i,j.
\label{eq:shared-second-order-environment}
\end{eqnarray}
The former quadratic operator correlates the
flavor breaking of the spectator with that of
the interacting pair, whereas the latter describes
a quadratic response to the spectator flavor
itself.
Within the quadratic truncation of
Eq.~(\ref{eq:shared-pair-response-expansion}),
one obtains
\begin{equation}
L_{\rm env}=d_E(x_s-x_q)^2.
\label{eq:shared-environment-response}
\end{equation}
Thus, all spectator-independent contributions
and all contributions depending only on the
spectator flavor cancel in $L_{\rm env}$.
This mass combination provides a diagnostic
of the correlated environmental response,
although it cannot distinguish the quadratic
contribution from possible higher-order responses.

We now investigate how these environmental
operators affect the common quadratic responses
of the two baryon families.
Introducing the traceless adjoint source
\begin{equation}
Y=X^2-\frac{\operatorname{Tr}(X^2)}{N_f}
{\bf1}_{N_f},
\label{eq:shared-quadratic-adjoint-source}
\end{equation}
where $N_f$ is the number of quark flavors,
the complete homogeneous quadratic covariants
for the $SU(4)$ and $SU(5)$ baryon representations
considered here can be chosen as
\begin{eqnarray}
Q_0^R&=&\operatorname{Tr}(X^2){\bf1}_R,
\nonumber\\
Q_F^R&=&F_R(Y),
\nonumber\\
Q_{FF}^R&=&F_R(X)^2,
\quad R=M,S,
\label{eq:shared-quadratic-S-basis}
\end{eqnarray}
together with two additional covariants in the
mixed representation,
\begin{eqnarray}
Q_D^M&=&D_M(Y),
\nonumber\\
Q_{FD}^M&=&\frac12\{F_M(X),D_M(X)\}.
\label{eq:shared-quadratic-M-basis}
\end{eqnarray}
In particular, the operator $D_M(X)^2$ is not
independent of these covariants, since
\begin{equation}
D_M(X)^2
=\frac{Q_0^M}{2N_f}
+\frac13Q_F^M
-\frac13Q_{FF}^M
-\frac23Q_{FD}^M.
\label{eq:shared-quadratic-D2-identity}
\end{equation}
Writing the quadratic Hamiltonians as
$H_R^{(2)}=\sum_i c_i^R Q_i^R$,
we thus have three independent coefficients
in the symmetric representation and five in
the mixed representation.
These complete homogeneous response spaces
should be distinguished from the restricted
factorized quadratic models employed in the
preceding baryon mass fits.

In analogy with the first-order analysis,
we first consider six common quadratic operators
without spectator-dependent pair-spin correlations:
\begin{eqnarray}
O_{2A}&=&\sum_iX_i^2,
\nonumber\\
O_{2B}&=&\sum_{i<j}(X_i^2+X_j^2)
\mathbf s_i\cdot\mathbf s_j,
\nonumber\\
O_{2C}&=&\sum_{i<j}X_iX_j,
\nonumber\\
O_{2D}&=&\sum_{i<j}X_iX_j
\mathbf s_i\cdot\mathbf s_j,
\nonumber\\
O_{2I}&=&\operatorname{Tr}(X^2){\bf1},
\nonumber\\
O_{2J}&=&\operatorname{Tr}(X^2)J_{\rm pair}.
\label{eq:shared-common-quadratic-operators}
\end{eqnarray}
Their projections generate six independent directions
in the combined quadratic response space.
The eight reduced coefficients consequently satisfy
two independent dynamical constraints:
\begin{eqnarray}
{\cal N}_1
&\equiv&c_F^S-\frac32(c_F^M+c_D^M)
-\frac34c_{FD}^M=0,
\nonumber\\
{\cal N}_2
&\equiv&c_{FF}^S-\frac94c_{FF}^M=0.
\label{eq:shared-second-order-null}
\end{eqnarray}
These relations are the quadratic counterparts
of the restricted common-dynamics relation
in Eq.~(\ref{eq:shared-restricted-relation}).

The projections of the two environmental operators
can be obtained from the identities
\begin{eqnarray}
O_{2K}(X)
&=&O_K(Y)
+\frac{\operatorname{Tr}(X^2)}{N_f}J_{\rm pair},
\nonumber\\
O_{2E}(X)
&=&O_A(X)O_K(X)-O_{2K}(X),
\label{eq:shared-environment-operator-identities}
\end{eqnarray}
together with the first-order projection dictionary
in Eq.~(\ref{eq:shared-first-order-dictionary}).
Adding $d_EO_{2E}+d_KO_{2K}$ to the restricted
common quadratic Hamiltonian gives
\begin{equation}
\begin{pmatrix}
{\cal N}_1\\
{\cal N}_2
\end{pmatrix}
=
\begin{pmatrix}
-\dfrac94&\dfrac92\\[2mm]
\dfrac{27}{4}&0
\end{pmatrix}
\begin{pmatrix}
d_E\\
d_K
\end{pmatrix}.
\label{eq:shared-quadratic-quotient}
\end{equation}
Since the transformation matrix is nonsingular,
the two environmental operators release both
quadratic dynamical constraints.
Together with the first-order result in
Sec.~\ref{sec:shared-spin-flavor}, we obtain
\begin{equation}\nonumber
\begin{array}{c|cc}
&\text{First order}&\text{Second order}\\ \hline
\text{Common pair dynamics}&2&6\\
\text{Environmental completion}&3&8
\end{array}
\end{equation}
The spectator-dependent correlations therefore
complete the joint homogeneous response spaces
at first and second order within the specified
spin--flavor reference.
This completion results from relaxing dynamical
constraints rather than introducing additional
flavor covariants beyond those allowed by
group theory.

These dimensions should be distinguished from
the finite-point effective mass-kernel ranks.
In the isospin limit of the joint $SU(5)$ system,
we assign independent effective coefficients to the
operators associated with the three Cartan directions
$\lambda_8$, $\lambda_{15}$, and $\lambda_{24}$
and their quadratic products. The combined mass-kernel
space through second order then has the rank sequence
\begin{equation}
17\longrightarrow20\longrightarrow26
=16_M+10_S,
\label{eq:shared-finite-rank-completion}
\end{equation}
upon including the first-order and then the
quadratic environmental correlations.
In this finite-point description, the contributions
of $O_{2K}$ can be absorbed into the first-order
spectator operators and $J_{\rm pair}$,
whereas $O_{2E}$ supplies the remaining
independent directions.
These ranks characterize the effective operator
space, rather than the number of identifiable
parameters in a mass fit.
The environmental completion removes the additional
constraints between the two representations, but
not the intrinsic quadratic relations within
either representation.

The above analysis illustrates how the effective
interaction between two quarks can depend on
the flavor environment provided by the third.
The extracted hyperfine mass combinations
motivate such environmental responses, while
the spin--flavor projections demonstrate how
they can unlock additional response directions
without further lowering the flavor symmetry.
Neither the environmental dependence nor the
completion of the response space establishes
a unique microscopic three-body interaction
or the convergence of the symmetry-breaking
expansion.

\subsection{Heavy-quark spin symmetry}
\label{sec:shared-HQSS}

The preceding two subsections demonstrate that
common spin--flavor operators can generate the
complete flavor response spaces through second order.
Including the two zeroth-order singlet structures,
there are $2+3+8=13$ independent structures in the
combined $M$ and $S$ representations.
We now evaluate these structures at a fixed source,
project onto a specific flavor subspace, and examine
the constraints imposed by heavy-quark spin symmetry
(HQSS).

Consider the singly-heavy baryons containing one
heavy quark $Q=c$ or $b$ and two light quarks.
For a fixed heavy flavor $Q$, their flavor compositions
in the isospin limit can be characterized by the
number $n=0,1,2$ of strange quarks.
For a fixed diagonal flavor-breaking source $X$,
we have
\begin{eqnarray}
F_S(X)\big|_n
&=&\frac13[(2-n)x_q+nx_s+x_Q],
\nonumber\\
F_M(X)\big|_n
&=&\frac12[(2-n)x_q+nx_s+x_Q],
\label{eq:shared-HQSS-F-projection}
\end{eqnarray}
where $x_q$, $x_s$, and $x_Q$ are the corresponding
eigenvalues of $X$.
At fixed $X$, the quadratic singlet covariants
$Q_0^M$ and $Q_0^S$ are already proportional to the
respective zeroth-order identities on the full
representations. Restriction to the singly-heavy
subspace introduces a further redundancy. For the
quadratic source $Y$ defined in
Eq.~(\ref{eq:shared-quadratic-adjoint-source}), we obtain
\begin{equation}
F_R(Y)=(x_q+x_s)F_R(X)+K_R{\bf1}_R,
\quad R=M,S,
\label{eq:shared-HQSS-quadratic-redundancy}
\end{equation}
where $K_R$ is independent of $n$.
Hence, the quadratic covariants $Q_F^M$ and $Q_F^S$
can also be absorbed into the corresponding
zeroth- and first-order structures at the specified
source.

The remaining operator structures can be chosen as
\begin{eqnarray}
S:&&{\bf1}_S,\ F_S(X),\ F_S(X)^2,
\nonumber\\
M:&&{\bf1}_M,\ F_M(X),\ D_M(X),\ D_M(Y),\ 
\nonumber\\
&&F_M(X)^2,\
\frac12\{F_M(X),D_M(X)\}.
\label{eq:shared-HQSS-projected-basis}
\end{eqnarray}
For generic distinct values of $x_q$, $x_s$, and
$x_Q$, these projected operators are linearly
independent within their respective representations.
The dimension of the combined operator space
is therefore reduced as
\begin{equation}
\underbrace{5_S+8_M}_{13}
\longrightarrow
\underbrace{4_S+7_M}_{11}
\longrightarrow
\underbrace{3_S+6_M}_{9}.
\label{eq:shared-HQSS-projection-counting}
\end{equation}
The intermediate dimension is the rank on the full
representations at a generic fixed source. The last
dimension is the rank on the singly-heavy subspace.
The original homogeneous covariants are independent
as functions of the source on the full representations.

In the lowest spatially symmetric $S$-wave reference,
the relevant singly-heavy configurations can be
labeled by
\begin{eqnarray}
|\bar3\rangle
&:&(\overline{\mathbf3}_F,j_\ell=0,J=1/2),
\nonumber\\
|6\rangle
&:&(\mathbf6_F,j_\ell=1,J=1/2),
\nonumber\\
|6^*\rangle
&:&(\mathbf6_F,j_\ell=1,J=3/2),
\label{eq:shared-lowest-light-cloud}
\end{eqnarray}
where $j_\ell$ denotes the total angular momentum
of the light degrees of freedom.
The last two configurations are heavy-quark spin
partners built from the same $j_\ell=1$ light cloud.
In the heavy-quark limit, the heavy-quark spin
decouples from the light degrees of freedom.
The rotationally invariant static Hamiltonian
$H_\infty$ commutes with both the total angular
momentum $\mathbf J$ and the heavy-quark spin
$\mathbf s_Q$, and therefore satisfies
\begin{equation}
[H_\infty,\mathbf j_\ell]=0,
\qquad
\mathbf j_\ell=\mathbf J-\mathbf s_Q.
\label{eq:shared-HQSS-light-angular-momentum}
\end{equation}
Consequently,
\begin{equation}
P_{j'_\ell}H_\infty P_{j_\ell}=0,
\qquad j'_\ell\ne j_\ell,
\label{eq:shared-HQSS-static-selection}
\end{equation}
where $P_{j_\ell}$ projects onto the corresponding
light-cloud angular-momentum sector.
The same selection rule applies to the static
response operators associated with light-quark
mass variations when defined in a common fixed
reference space.
It should be emphasized that HQSS protects the
light-cloud angular momentum rather than the
light-flavor representations.
In an enlarged Pauli-compatible Hilbert space,
both $\overline{\mathbf3}_F$ and $\mathbf6_F$
can occur within the same $j_\ell$ sector and
may mix when light-flavor symmetry is broken.
In the following, we restrict our discussion
to the effective mass kernels in the lowest
matched reference space.

The nine independent operator directions in
Eq.~(\ref{eq:shared-HQSS-projection-counting})
span the real mass-kernel space of this reference.
The diagonal elements of the antitriplet and
the two sextets can be parametrized as
\begin{eqnarray}
H_{\bar3}(n)
&=&C_{\bar3}+s_{\bar3}n,
\nonumber\\
H_6^M(n)
&=&C_M+s_Mn+\frac12\epsilon_Mn(n-1),
\nonumber\\
H_6^S(n)
&=&C_S+s_Sn+\frac12\epsilon_Sn(n-1),
\label{eq:shared-HQSS-mass-shapes}
\end{eqnarray}
where $n=0,1$ for the antitriplet and
$n=0,1,2$ for the sextets.
Here $C_R$, $s_R$, and $\epsilon_R$ characterize
the reference energy, the first strange-quark
mass increment, and the second finite difference
of the corresponding mass kernel, respectively.
The two $J^P=1/2^+$ reference configurations
with the flavor composition $qsQ$ may also mix
through an effective real matrix element $V$,
leading to the mass kernel
\begin{equation}
H_{\Xi_Q}=
\begin{pmatrix}
C_{\bar3}+s_{\bar3} & V\\
V & C_M+s_M
\end{pmatrix}.
\label{eq:shared-HQSS-Xi-kernel}
\end{equation}
The nine real kernel coordinates can thus be
chosen as
\begin{equation}
\boldsymbol z=
(C_{\bar3},s_{\bar3},
C_M,s_M,\epsilon_M,V,C_S,s_S,\epsilon_S).
\label{eq:shared-HQSS-coordinates}
\end{equation}

In the heavy-quark limit, the two spin partners
with $j_\ell=1$ share the same static light-cloud
Hamiltonian, including its responses to light-flavor
symmetry breaking.
Within the same matched reference and after
subtracting the common heavy-quark mass,
HQSS therefore imposes the following constraints
on the static kernel coordinates:
\begin{equation}
\begin{array}{ll}
C_S=C_M, & s_S=s_M,\\[1mm]
\epsilon_S=\epsilon_M, & V=0.
\end{array}
\label{eq:shared-HQSS-kernel-constraints}
\end{equation}
The first three relations identify the static
mass kernels of the two spin partners, whereas
the last excludes transitions between the
lowest $j_\ell=0$ and $j_\ell=1$ branches.
These four independent constraints reduce the
nine-dimensional real mass-kernel space to
five dimensions, parametrized by
$(C_{\bar3},s_{\bar3},C,s,\epsilon)$.
Combining this result with
Eq.~(\ref{eq:shared-HQSS-projection-counting}),
we obtain
\begin{equation}
13
\xrightarrow{\rm evaluation}
11
\xrightarrow{\rm projection}
9
\xrightarrow{\rm HQSS}
5.
\label{eq:shared-HQSS-response-counting}
\end{equation}
The two quadratic singlets become redundant at fixed $X$.
The singly-heavy projection makes each $F_R(Y)$ a linear
combination of ${\bf1}_R$ and $F_R(X)$. HQSS then imposes
four constraints on the matched static Hamiltonian.
This counting concerns the effective mass-kernel
space rather than the dimension of the complete
light-cloud Hilbert space.
In particular, the static condition $V=0$
does not require the flavor response coefficient
$D_8^M$ to vanish.
A flavor transition matrix element such as
$D_8^M/2$ does not by itself determine the
physical cross-$j_\ell$ transition, which must
be evaluated after the appropriate spin--flavor
and spatial matching.

For finite heavy-quark masses, the two spin
partners receive both spin-independent and
chromomagnetic corrections.
Their matched sextet kernels may be organized as
\begin{equation}
H_{6,J}
=m_Q+H_\ell+
\frac{K_\ell+C_{\rm mag}d_JG_\ell}{2m_Q}
+{\cal O}(m_Q^{-2}),
\label{eq:shared-HQET-kernels}
\end{equation}
with $d_{1/2}=-2$ and $d_{3/2}=1$.
Here $H_\ell$, $K_\ell$, and $G_\ell$ denote
the static, kinetic, and chromomagnetic
light-cloud operators, respectively, and $m_Q$
is defined in the same heavy-quark mass scheme
for both partners.
The spin-weighted average
\begin{equation}
\overline H_6
=\frac{H_{6,1/2}+2H_{6,3/2}}{3}
=m_Q+H_\ell+\frac{K_\ell}{2m_Q}
+{\cal O}(m_Q^{-2})
\label{eq:shared-HQET-spin-average}
\end{equation}
eliminates the leading diagonal chromomagnetic
contribution.
This relation applies to the matched sextet
kernels rather than directly to mass eigenvalues
affected by cross-$j_\ell$ mixing.
At finite $m_Q$, transitions between different
$j_\ell$ sectors may also arise from subleading
interactions that break HQSS.
Their effects on the physical mixing angles
depend additionally on the corresponding level
separations, as discussed in
Sec.~\ref{sec:baryon-M-identifiability}.
For nondegenerate static levels, the corresponding
cross-$j_\ell$ mixing angles vanish in the
heavy-quark limit.

The above analysis illustrates how the physical
response space can be constrained by a symmetry
other than flavor symmetry.
While additional internal correlations can unlock
response directions, as demonstrated in the
preceding subsections, additional physical
symmetries can restrict their realization.
These two effects characterize complementary
aspects of the dynamics of symmetry breaking.

\subsection{Dynamical response spaces}
\label{sec:shared-limitations}

The preceding subsections illustrate how the
realization of flavor responses depends on the
internal dynamics, the chosen reference space,
and additional physical symmetries.
Within the common spin--flavor reference,
the inclusion of spectator-dependent correlations
completes the joint response spaces at first and
second order, while leaving the zeroth-order
singlet structures unchanged:
\begin{equation}
2\longrightarrow2,\qquad
2\longrightarrow3,\qquad
6\longrightarrow8,
\end{equation}
at zeroth, first, and second order, respectively.
The complete joint response space through second
order thus contains $2+3+8=13$ independent structures.
Conversely, evaluating these structures at a generic
fixed source, projecting onto the singly-heavy
subspace, and imposing HQSS on the matched static
mass kernels gives
\begin{equation}
13\xrightarrow{\rm evaluation}11
\xrightarrow{\rm projection}9\xrightarrow{\rm HQSS}5.
\end{equation}
Although these dimensions refer to different
operator or mass-kernel spaces, they illustrate
three distinct mechanisms through which response
directions can be unlocked by internal correlations,
become redundant upon evaluation and projection, or
be constrained by additional physical symmetries.

These results distinguish the symmetry-allowed
response structures from their dynamical
realization.
Flavor group theory determines the allowed
covariants, whereas the underlying dynamics
determines their coefficients and correlations,
subject to additional physical constraints.
Within the group-expansion framework introduced
in Sec.~\ref{sec:gsb-differential}, the evolution
of a hadronic system involves both the variation
of its effective Hamiltonian and the associated
evolution of its eigenstates.
The present spin--flavor analysis provides a
concrete illustration of how internal correlations
can modify the former and thereby influence
the latter.

The operator identities and rank relations
obtained in this section are established within
the specified spin--flavor reference.
In physical QCD, additional spatial configurations,
light-cloud excitations, and higher Fock components
may modify the effective response coefficients
and the matching between reference spaces.
The completeness of an effective operator basis
therefore does not imply a unique microscopic
interpretation of its coefficients, just as a
successful description of the observed mass
spectrum does not necessarily ensure reliable
extrapolations to unexplored flavor sectors.

The group expansion thus provides a systematic
organization of the allowed response structures,
their dynamical correlations, and their physical
constraints.
While the present analysis establishes such
relations through second order within the
specified reference spaces, it does not determine
a universal evolution law for the response
coefficients.
Deriving their evolution from the underlying
QCD dynamics remains an important step toward
a more complete understanding of symmetry
breaking in hadronic systems.

\section{Summary and Discussions}
\label{sec:summary}

In this work, we have investigated flavor-symmetry
breaking in vector mesons and baryons within a
framework that we refer to as the \emph{group expansion}.
Starting from the observed nonadditivity of
vector-meson masses, we have examined how the
response to one symmetry-breaking direction can
depend on the presence of other breaking backgrounds.
Higher-order responses may not only modify the
coefficients of existing flavor structures but also
unlock additional independent response directions
when a parent symmetry is reduced to a daughter
symmetry.
The vector-meson analyses provide concrete examples
of these two mechanisms, which we characterize as
\emph{coefficient evolution} and \emph{response
unlocking}, respectively.

The group expansion describes the continuous
evolution of a system from a specified symmetric
reference through its effective Hamiltonian
${\cal H}$ and the associated response operators,
\begin{equation}
B_a=\partial_a{\cal H},
\qquad
K_{ab}=\partial_b B_a \, .
\end{equation}
The former characterizes the local response to
a symmetry-breaking direction, while the latter
describes how this response varies along another
breaking direction.
We refer to the evolution of the effective
Hamiltonian and its response operators as the
\emph{interaction flow}, which in turn generates
the \emph{mass flow} and \emph{state flow}
within the effective description.
In particular, states belonging to different
flavor representations at the symmetric point
may become mixed as the symmetry is broken,
provided that the corresponding transition
channels are allowed and dynamically activated.
The evolution of effective interactions, the
variation of mass eigenvalues, and the reorganization
of eigenstates thus constitute three interconnected
aspects of symmetry breaking.

The baryon analyses further demonstrate that the
response space depends on the internal representation
of the system.
For an adjoint flavor-breaking source,
the fully symmetric three-quark flavor representation
admits one independent first-order $F$-type response,
whereas the mixed-symmetry representation admits
two independent $F$- and $D$-type responses.
Their complete homogeneous quadratic response
spaces contain three and five independent
structures, respectively.
The inclusion of selected quadratic responses
substantially improves the description of the
baryon mass spectra.
However, the limited success of extending the
separate flavor-$SU(4)$ fits to a common
flavor-$SU(5)$ description, particularly for the
mixed-symmetry baryons, illustrates the limitations
of a simple factorized quadratic response.
In the separate $SU(4)$ fits for the mixed-symmetry
baryons, the negative-sign quadratic solutions
substantially improve the mass fits while largely
preserving the reference-state compositions obtained
at leading order.
For example, in the flavor-$SU(4)_c$ system,
the Sym ($-$) solution reduces the RMS mass residual
from $33.47$ to $5.45~{\rm MeV}$, while the
magnitude of the corresponding reference-state
mixing angle in the $\Xi_c$--$\Xi_c^\prime$ system
changes only from $2.99^\circ$ to $3.16^\circ$.
This comparison illustrates that sizable mass
corrections need not be accompanied by substantial
eigenvector rearrangements, although the physical
state compositions cannot be uniquely determined
from mass spectra alone.

We have further investigated how the allowed
flavor responses are realized through common
internal dynamics.
Within the lowest spatially symmetric spin--flavor
reference, the inclusion of spectator-dependent
correlations completes the joint response spaces
of the two baryon families, with their dimensions
changing as
\begin{equation}
2\longrightarrow2,\qquad
2\longrightarrow3,\qquad
6\longrightarrow8,
\end{equation}
at zeroth, first, and second order, respectively.
Starting from the resulting $2+3+8=13$
independent operator structures through second order,
we evaluate them at a generic fixed source and project
them onto the singly-heavy subspace.
Heavy-quark spin symmetry (HQSS) further restricts
the matched static mass kernels, leading to
the successive reductions
\begin{equation}
13\xrightarrow{\rm evaluation}11
\xrightarrow{\rm projection}9
\xrightarrow{\rm HQSS}5 \, .
\end{equation}
Although these dimensions refer to different
operator or mass-kernel spaces, they illustrate
how internal correlations can unlock response
directions, how operator structures can become
redundant upon evaluation and projection, and
how additional physical symmetries can constrain
their realization.
In particular, HQSS protects the light-cloud
angular momentum rather than the light-flavor
representations themselves.

More generally, the basic idea of the group expansion
is to distinguish the system under consideration,
represented by a chosen symmetry-adapted reference
space, from the degrees of freedom outside this space.
Rather than describing all these degrees of freedom
explicitly, we may incorporate their effects into
the effective Hamiltonian and organize their influence
on symmetry breaking in terms of effective responses.
The analyses in Secs.~\ref{sec:group-expansion}
and \ref{sec:shared-dynamics} illustrate two
complementary mechanisms of response unlocking.
In the former, a symmetry-breaking background
allows higher-order parent-group structures to
generate independent daughter-group responses
outside the inherited response space.
In the latter, additional internal correlations
relax the restrictions of a common dynamical model
and allow it to realize response directions already
permitted by flavor symmetry.
These mechanisms illustrate how the response
to symmetry breaking can itself depend on the
breaking background and the internal dynamics
of the system.
Different symmetry-breaking processes need not
be independent, but may form a hierarchy of
interconnected responses.
Flavor symmetry organizes the allowed response
structures, while the underlying dynamics and
additional physical symmetries govern their
realization.

Several limitations of the present study should
be emphasized.
First, our meson and baryon analyses are based
primarily on quark--antiquark and three-quark
flavor reference spaces, respectively.
Higher Fock components, gluonic degrees of
freedom, and additional spatial configurations
may enlarge the physical configuration space
and give rise to further effective response channels.
Their effects are not explicitly resolved in
the present analysis.
Second, the QCD vacuum itself responds to
flavor-symmetry breaking.
While a quark--antiquark condensate can contain
both singlet and adjoint components under
vector flavor symmetry, more general vacuum
correlations may involve richer flavor structures.
The coupled evolution of the vacuum and hadronic
responses has not been investigated here.
Third, our analysis is largely restricted to
the isospin limit and to selected responses
through second order.
Higher-order responses, electromagnetic effects,
and couplings to hadronic decay channels may
further modify the effective Hamiltonians.
Given the strong breaking of flavor $SU(4)$ and
$SU(5)$, the effective coefficients extracted
from mass fits at finite symmetry breaking need
not coincide with local Taylor coefficients at
the exact symmetry point.
An improved mass fit does not establish the
convergence of the expansion, a unique microscopic
interpretation of its coefficients, or the
reliability of extrapolations to unmeasured
flavor sectors.

From a methodological perspective, the group
expansion involves extensive group-theoretical,
algebraic, and numerical calculations, whose
complexity grows rapidly with the dimensions
of the reference spaces and the orders of the
symmetry-breaking responses.
Recent advances in artificial intelligence,
together with symbolic and numerical computation,
make such systematic investigations increasingly
tractable.
AI-assisted calculations facilitate the exploration
of alternative response structures, the verification
of operator relations, and the comparison of
different dynamical models, while mathematical
consistency and independent computational checks
remain essential.
These developments open further possibilities
for extending the group expansion beyond the
hadronic systems considered here and investigating
symmetry-breaking dynamics in more complex
physical systems.

Further progress requires information beyond
the mass spectrum at the physical flavor point.
Measurements of additional heavy-quark spin
partners and mixed-heavy baryons may provide
further constraints on the effective responses.
Transition observables and appropriately matched
lattice-QCD state overlaps may provide complementary
information on representation mixing and the
evolution of physical state compositions.
Lattice-QCD calculations along different quark-mass
trajectories, with consistent reference-state
matching, would be particularly valuable for
distinguishing the evolution of the effective
response operators from the associated rearrangement
of the eigenstates.
More generally, the group expansion provides
a unified framework for organizing the evolution,
correlations, and constraints of symmetry-breaking
responses in physical systems with approximate
symmetries.
Whether the evolution of these responses is
governed by more general dynamical principles,
and how such principles depend on the internal
structure of a physical system, remain important
questions for future investigations.

\section*{Acknowledgments}

This work is dedicated to the memory of
Veljko Dmitra\v sinovi\' c and Keitaro Nagata.
The author acknowledges the extensive assistance of ChatGPT (OpenAI) in 
theoretical discussions,
the development of the theoretical framework,
analytical derivations,
symbolic and numerical computations,
the exploration of alternative theoretical models,
cross-checks of analytical and numerical results,
and manuscript preparation.
The author takes full responsibility for the scientific content of this work.
This work is supported by
the National Natural Science Foundation of China under Grant No.~12075019,
the Jiangsu Provincial Double-Innovation Program under Grant No.~JSSCRC2021488,
and 
the Fundamental Research Funds for the Central Universities.

\appendix

\section{Matrices for the flavor $SU(4)$ group}
\label{app:baryon-S-matrices}

In this appendix, we summarize the flavor tensors, transition
matrices, and invariant contractions used to construct meson and
baryon interactions in flavor $SU(4)$. The explicit tensor bases
refer to the ordered flavors $(u,d,s,c)$. The construction extends
to flavor $SU(5)$ by enlarging the fundamental flavor space and
constructing the corresponding representation bases. In particular,
the fully symmetric three-quark representation $\mathbf{20_S}$
becomes $\mathbf{35_S}$, whereas the fully antisymmetric
three-quark representation $\overline{\mathbf 4}_A$ becomes
$\overline{\mathbf{10}}_A$, not $\overline{\mathbf 5}_A$.
Related flavor-$SU(3)$ constructions are discussed in
Refs.~\cite{Chen:2008qv,Chen:2009sf,Chen:2010ba,Chen:2011rh,Chen:2013aga,Dmitrasinovic:2016hup}.

\subsection{Matrices for mesons and baryons}

The quark--antiquark flavor components in $SU(4)$ are
\begin{equation}
M^N=\lambda^N_{AB}\,\bar q^A q^B,
\end{equation}
where $A,B=1,\ldots,4$ and $N=0,\ldots,15$. Here
$\lambda^N_{AB}\equiv[\lambda^N_{(4)}]_{AB}$, with the singlet
and adjoint generators normalized as in
Eq.~(\ref{eq:meson-generator-normalization}).

The three-quark baryon flavor components are written as
\begin{eqnarray}
\Lambda^D &=& \epsilon^{ABCD} \times q_A q_B q_C \, ,
\label{def:Lambda}
\\
N^I &=& \mathbb{M}^I_{ABC} \times q^A q^B q^C \, ,
\label{def:N}
\\
\Delta^P &=& \mathbb{S}^{P}_{ABC} \times q^A q^B q^C \, ,
\label{eq:app-Delta-def}
\end{eqnarray}
where color, spin, and Lorentz structures are suppressed.
These tensors specify reference flavor configurations rather
than complete physical baryon wave functions at broken
flavor symmetry. We adopt the following conventions:
\begin{itemize}

\item The components $\Lambda^D$ form the antisymmetric
three-quark $SU(4)$ anti-quartet ($\mathbf{\bar{4}_A}$), with
$A,B,C,D=1,\ldots,4$. In the chosen flavor basis,
\begin{equation}
\Lambda^1 \sim - \Xi_{c \bf A}^0 , \, \Lambda^2 \sim \Xi_{c \bf A}^+ , \, \Lambda^3 \sim - \Lambda_{c \bf A}^+ , \, \Lambda^4 \sim \Lambda_{\bf A}^0 .
\end{equation}
The tensor $\epsilon^{ABCD}$ is the totally antisymmetric
Levi-Civita tensor. The state identifications indicate the reference flavor
assignments, with the phases displayed explicitly.

\item The components $N^I$, with $I=1,\ldots,20$,
span the mixed-symmetry flavor representation $\mathbf{20_M}$.
The tensor basis is chosen so that
$\mathbb{M}^I_{BAC}=-\mathbb{M}^I_{ABC}$.
Its independent nonzero components are listed in
Table~\ref{tab:app-SU4-mixed-basis}.

\item The components $\Delta^P$, with $P=1,\ldots,20$,
span the fully symmetric flavor representation $\mathbf{20_S}$.
The tensors $\mathbb{S}^{P}_{ABC}$ are symmetric under all
permutations of $A,B,C$, and their nonzero components are given in
Table~\ref{tab:app-SU4-symmetric-basis}.

\end{itemize}
The corresponding flavor representations and Young diagrams are
\begin{gather}
\nonumber
M^{N=0}\sim\mathbf{1},\qquad
M^{N=1\cdots15}\sim
\mathbf{15}\left(\tiny\yng(2,1,1)\right),
\\\nonumber
\Lambda^{D=1\cdots4}\sim
\mathbf{\bar{4}_A}\left(\tiny\yng(1,1,1)\right),
\\\nonumber
N^{I=1\cdots20}\sim
\mathbf{20_M}\left(\tiny\yng(2,1)\right),
\\\nonumber
\Delta^{P=1\cdots20}\sim
\mathbf{20_S}\left(\tiny\yng(3)\right).
\end{gather}

\begin{table*}[t]
\caption{Orthonormal fully symmetric $SU(4)_c$
flavor-tensor basis. In each row the displayed
component $\mathbb S^P_{ABC}$ has the given value
for every distinct permutation of the listed
triple $ABC$. All other components vanish.}
\label{tab:app-SU4-symmetric-basis}
\begin{ruledtabular}
\begin{tabular}{rllc@{\hspace{1.0cm}}rllc}
$P$ & State & $ABC$ & $\mathbb S^P_{ABC}$
& $P$ & State & $ABC$ & $\mathbb S^P_{ABC}$ \\
\hline
1  & $\Delta^{++}$         & $111$ & $1$
& 11 & $\Sigma_c^{*++}$    & $114$ & $1/\sqrt3$ \\
2  & $\Delta^+$            & $112$ & $1/\sqrt3$
& 12 & $\Sigma_c^{*+}$     & $124$ & $1/\sqrt6$ \\
3  & $\Delta^0$            & $122$ & $1/\sqrt3$
& 13 & $\Sigma_c^{*0}$     & $224$ & $1/\sqrt3$ \\
4  & $\Delta^-$            & $222$ & $1$
& 14 & $\Xi_c^{*+}$        & $134$ & $1/\sqrt6$ \\
5  & $\Sigma^{*+}$         & $113$ & $1/\sqrt3$
& 15 & $\Xi_c^{*0}$        & $234$ & $1/\sqrt6$ \\
6  & $\Sigma^{*0}$         & $123$ & $1/\sqrt6$
& 16 & $\Omega_c^{*0}$     & $334$ & $1/\sqrt3$ \\
7  & $\Sigma^{*-}$         & $223$ & $1/\sqrt3$
& 17 & $\Xi_{cc}^{*++}$     & $144$ & $1/\sqrt3$ \\
8  & $\Xi^{*0}$            & $133$ & $1/\sqrt3$
& 18 & $\Xi_{cc}^{*+}$      & $244$ & $1/\sqrt3$ \\
9  & $\Xi^{*-}$            & $233$ & $1/\sqrt3$
& 19 & $\Omega_{cc}^{*+}$  & $344$ & $1/\sqrt3$ \\
10 & $\Omega^-$            & $333$ & $1$
& 20 & $\Omega_{ccc}^{++}$  & $444$ & $1$ \\
\end{tabular}
\end{ruledtabular}
\end{table*}

\begin{table*}[t]
\caption{Independent nonzero components of a
normalized mixed-symmetry $\mathbf{20_M}$ tensor
basis in flavor $SU(4)_c$. The entries with
the first two indices interchanged have opposite
signs. All unspecified components vanish.}
\label{tab:app-SU4-mixed-basis}
\begin{ruledtabular}
\begin{tabular}{rll@{\hspace{1cm}}rll}
$I$ & State & $ABC:\mathbb M^I_{ABC}$
& $I$ & State & $ABC:\mathbb M^I_{ABC}$ \\
\hline
1 & $\Sigma^+$ & $131:1/\sqrt2$
& 11 & $\Sigma_c^0$ & $242:1/\sqrt2$ \\
2 & $\Sigma^0$ & $231:1/2,\;132:1/2$
& 12 & $\Xi_c^{\prime+}$ & $341:1/2,\;143:1/2$ \\
3 & $\Sigma^-$ & $232:1/\sqrt2$
& 13 & $\Xi_c^{\prime0}$ & $342:1/2,\;243:1/2$ \\
4 & $p$ & $121:1/\sqrt2$
& 14 & $\Omega_c^0$ & $343:1/\sqrt2$ \\
5 & $n$ & $122:1/\sqrt2$
& 15 & $\Lambda_c^+$ & $241:1/\sqrt{12},\;412:1/\sqrt{12},\;214:1/\sqrt3$ \\
6 & $\Xi^0$ & $133:1/\sqrt2$
& 16 & $\Xi_c^+$ & $341:1/\sqrt{12},\;413:1/\sqrt{12},\;314:1/\sqrt3$ \\
7 & $\Xi^-$ & $233:1/\sqrt2$
& 17 & $\Xi_c^0$ & $342:1/\sqrt{12},\;423:1/\sqrt{12},\;324:1/\sqrt3$ \\
8 & $\Lambda$ & $231:1/\sqrt{12},\;312:1/\sqrt{12},\;213:1/\sqrt3$
& 18 & $\Xi_{cc}^{++}$ & $144:1/\sqrt2$ \\
9 & $\Sigma_c^{++}$ & $141:1/\sqrt2$
& 19 & $\Xi_{cc}^{+}$ & $244:1/\sqrt2$ \\
10 & $\Sigma_c^{+}$ & $241:1/2,\;142:1/2$
& 20 & $\Omega_{cc}^{+}$ & $344:1/\sqrt2$ \\
\end{tabular}
\end{ruledtabular}
\end{table*}

\subsection{Transition matrices}

The relevant flavor-$SU(4)$ tensor-product decompositions include
\begin{eqnarray}
\nonumber \mathbf{\bar{4}_A} \otimes \mathbf{15} &=& \mathbf{\bar{4}_A} \oplus \mathbf{20_M} \oplus \mathbf{36} \, ,
\\
\mathbf{20_M} \otimes \mathbf{15} &=& \mathbf{\bar{4}_A} \oplus \mathbf{20_M} \oplus \mathbf{20_M} \oplus \mathbf{20_S} \oplus \mathbf{36}
\\ \nonumber && ~~~~~~~~~~~~~~~~~~~~~~~~\, \oplus \mathbf{60} \oplus \mathbf{140} \, ,
\\ \nonumber
\mathbf{20_S} \otimes \mathbf{15} &=& \mathbf{20_M} \oplus \mathbf{20_S} \oplus \mathbf{120} \oplus \mathbf{140} \, .
\end{eqnarray}
Using the tensor bases specified above, we introduce the
transition matrices through the following decomposition identities
in the adopted conventions~\cite{Chen:2008qv,Chen:2009sf}:
\begin{eqnarray}
\label{eq:con1}
             && \epsilon_{ABDE} \times \lambda^N_{DC}
\\ \nonumber && ~~~~~~~~~ = -{1\over3} \lambda^N_{FE} \epsilon_{ABCF} + \left[{\bf T}_{\Lambda}\right]^N_{EI} \mathbb{M}^I_{ABC} \, ,
\\ \label{eq:con2}
             && \mathbb{M}^I_{ABD} \times \lambda^N_{DC}
\\ \nonumber && ~~~~~~~~~ = + {1\over6} \left[{\bf T}_{\Lambda}\right]^{N\dagger}_{ID} \epsilon_{ABCD} + \left[{\bf D} + {\bf F}\right]^{N}_{IJ} \mathbb{M}^J_{ABC} \, ,
\\ \label{eq:con3}
             && \mathbb{M}^I_{ADC} \times \lambda^N_{DB}
\\ \nonumber && ~~~~~~~~~ = - {1\over12} \left[{\bf T}_{\Lambda}\right]^{N\dagger}_{ID} \epsilon_{ABCD} + \left[{\bf T}_{\Delta}\right]^{N\dagger}_{IP} \mathbb{S}^P_{ABC}
\\ \nonumber && ~~~~~~~~~~~~\, + \frac23{\bf F}^{N}_{IJ} \mathbb{M}^J_{ABC} + \left[{\bf D} + \frac13 {\bf F}\right]^{N}_{IJ} \mathbb{M}^J_{BCA} \, ,
\\ \label{eq:con4}
             && \mathbb{S}^P_{ABD} \times \lambda^N_{DC}
\\ \nonumber && ~~~~~~~~~ =  + {2\over3}\left[{\bf T}_{\Delta}\right]^N_{PI} \mathbb{M}^I_{ABC} + {4\over3}\left[{\bf T}_{\Delta}\right]^N_{PI} \mathbb{M}^I_{BCA}
\\ \nonumber && ~~~~~~~~~~~~\, + \left[{\bf F}_{\Delta}\right]^{N}_{PQ} \mathbb{S}^Q_{ABC} \, .
\end{eqnarray}
The two independent mixed-symmetry adjoint matrices
${\bf D}^N$ and ${\bf F}^N$ are fixed by these identities.
For clarity, define the projection coefficients
\begin{eqnarray}
{\cal A}^N_{IJ}
&=&
(\mathbb{M}^{J}_{ABC})^*\,
\mathbb{M}^I_{ABD}\lambda^N_{DC},
\nonumber\\
{\cal B}^N_{IJ}
&=&
(\mathbb{M}^{J}_{ABC})^*\,
\mathbb{M}^I_{ADC}\lambda^N_{DB},
\label{eq:app-mixed-projections}
\end{eqnarray}
where repeated flavor indices are summed. In the
conventions used above, these coefficients give
\begin{equation}
{\bf F}^N=\frac12{\cal A}^N+{\cal B}^N,
\qquad
{\bf D}^N=\frac12{\cal A}^N-{\cal B}^N.
\label{eq:app-mixed-DF}
\end{equation}
In this convention, ${\bf F}^N$ is the generator matrix
obtained from one half of the sum of the three fundamental
flavor generators acting on the $\mathbf{20_M}$ tensor
basis, whereas ${\bf D}^N$ is a second independent
adjoint response structure.
The remaining transition matrices ${\bf F}_{\Delta}$,
${\bf T}_{\Lambda}$, and ${\bf T}_{\Delta}$ are likewise
obtained by projection onto the normalized tensor bases.
In particular, the transition matrix
${\bf T}_{\Delta}$ is defined by
\begin{equation}
\left[{\bf T}_{\Delta}\right]^N_{PI}
=
(\mathbb{M}^{I}_{BCA})^*
\mathbb{S}^{P}_{ABD}\lambda^N_{DC},
\label{eq:app-TDelta-projection}
\end{equation}
which fixes its relative phase in the
transition identities above.

The matrix elements can be obtained by symbolic
tensor contractions and independently verified
through the decomposition identities.
For reproducibility, the explicit matrix
representations of
$\mathbf{D}^N$, $\mathbf{F}^N$,
$\mathbf{F}_\Delta^N$,
$\mathbf{T}_\Lambda^N$, and
$\mathbf{T}_\Delta^N$
are provided in the accompanying Mathematica
notebook \texttt{matrix.nb},
rather than reproduced here as lengthy
basis-dependent tables.

\subsection{Flavor-singlet combinations}

The transition matrices determine the allowed flavor-singlet
contractions of a meson component with a baryon bilinear.
The following expressions display the schematic flavor-invariant
interactions, with Lorentz and spin structures suppressed.
Distinct contractions generally carry independent reduced
couplings.
\begin{itemize}

\item A flavor-singlet meson component $M^0$ gives the
following three baryon--antibaryon couplings:
\begin{eqnarray}
   g_{\bar \Lambda \Lambda} \times \delta_{EF} \times M^0 \times \bar \Lambda^E \Lambda^F \, ,~~&
\label{eq:M0LL}
\\ g_{\bar N N}             \times \delta_{IJ} \times M^0 \times \bar N^I N^J             \, ,~~&
\label{eq:M0NN}
\\ g_{\bar \Delta \Delta}   \times \delta_{PQ} \times M^0 \times \bar \Delta^P \Delta^Q   \, ,~~&
\label{eq:M0DD}
\end{eqnarray}

\item An adjoint meson component $M^N$, with
$N=1,\ldots,15$, admits the following six independent
flavor contractions involving baryon--antibaryon fields:
\begin{eqnarray}
   g_{M \bar \Lambda \Lambda} \times \lambda^N_{FE}                            \times M^N \times \bar \Lambda^E \Lambda^F \, ,~~&
\label{eq:M15LL}
\\ g_{M \bar N N}             \times {\bf D}^{N}_{IJ}                          \times M^N \times \bar N^I N^J \, ,~~&
\label{eq:M15NND}
\\ g^\prime_{M \bar N N}      \times {\bf F}^{N}_{IJ}                          \times M^N \times \bar N^I N^J \, ,~~&
\label{eq:M15NNF}
\\ g_{M \bar \Delta \Delta}   \times \left[{\bf F}_{\Delta}\right]^{N}_{PQ}    \times M^N \times \bar \Delta^P \Delta^Q \, ,~~&
\label{eq:M15DD}
\\ g_{M \bar \Lambda N}       \times \left[{\bf T}_{\Lambda}\right]^{N}_{DI}   \times M^N \times \bar \Lambda^D N^I ~+& c.\,c. \, ,
\label{eq:M15LN}
\\ g_{M \bar \Delta N}        \times \left[{\bf T}_{\Delta}\right]^{N}_{PI}    \times M^N \times \bar \Delta^P N^I  ~+& c.\,c. \, .
\label{eq:M15DN}
\end{eqnarray}

\item For completeness, the symmetric flavor-invariant
cubic interactions among three meson components are
\begin{eqnarray}
{\cal L}_{\rm meson}
&=& {g_{SSS}\over3!}\,M^0M^0M^0
\nonumber\\
&+& {g_{SOO}\over2}\,M^0M^NM^N
\nonumber\\
&+& {g_{OOO}\over3!}\,d_{NPQ}M^NM^PM^Q,
\label{eq:app-meson-invariant-L3}
\end{eqnarray}
where the adjoint indices are summed and
$ d_{NPQ}={1\over4}{\rm Tr}\!\left(
\{\lambda^N,\lambda^P\}\lambda^Q\right)$.
These are the same symmetric flavor structures used
in Eq.~(\ref{eq:meson-general-L3}). An antisymmetric
$f_{NPQ}$ contraction may also occur for suitably
distinguished field insertions, but it does not
contribute to the real symmetric mass kernel
considered here.

\end{itemize}
Note that the $SU(4)$ coupling in Eq.~(\ref{eq:M15LL}) involves
the antisymmetric three-quark anti-quartet. For $SU(3)$,
the fully antisymmetric three-quark flavor representation
is a singlet, and the corresponding adjoint coupling
vanishes. Adjoint couplings involving other $SU(3)$
representations are, of course, still allowed.

These expressions classify the allowed flavor structures,
but do not determine their reduced couplings or microscopic
dynamics. As discussed in Sec.~\ref{sec:meson-leading},
Lorentz-scalar singlet and diagonal adjoint backgrounds
provide a leading effective description of flavor-symmetric
and symmetry-breaking mass contributions. Such backgrounds
are not vacuum expectation values of the physical vector-meson
fields. Their possible connections to QCD condensates and
current-quark masses depend on the dynamical realization.
The distinction is important when comparing effective
symmetry-breaking responses in the meson and baryon sectors.

\section{Young tableaux and state flow}
\label{sec:young-breaking}

In Sec.~\ref{sec:group-expansion}, we formulated
the group expansion in terms of the effective
Hamiltonian and its response operators.
The resulting interaction flow gives rise to
the mass flow and the state flow, with the
latter describing the evolution of the eigenstates
in the chosen reference space.
In this appendix, we develop a complementary
graphical description of the state flow using
Young tableaux, which provides an intuitive
visualization of possible symmetry-breaking
directions in representation space.

To initialize the graphical notation, we employ
a white box to denote the reference flavors
and a marked box to denote the flavor direction
distinguished by a symmetry-breaking coordinate.
Schematically,
\begin{equation}
\young({\bb})
-
\young({~})
\quad\Longleftrightarrow\quad
\delta_B \, ,
\label{eq:young-breaking-coordinate}
\end{equation}
where $\delta_B$ denotes a symmetry-breaking
coordinate, for example, $m_d-m_u$ or
$m_c-m_{u,d}$.
The marked box does not represent an additional
constituent, and the subtraction in
Eq.~(\ref{eq:young-breaking-coordinate})
does not denote a literal subtraction of
two flavor states.
Rather, it indicates the flavor direction
distinguished from the reference flavors
by the symmetry-breaking Hamiltonian.

Taking the white box as a fixed reference,
we introduce the graphical differential
\begin{equation}
{\cal D}
\left(
\young({~})
\right)
=
0 \, ,
\end{equation}
and identify
\begin{equation}
{\cal D}
\left(
\young({\bb})
\right)
\quad\Longleftrightarrow\quad
d\delta_B \, .
\label{eq:young-box-differential}
\end{equation}
Here ${\cal D}$ is a graphical symbol rather
than an independently defined differential
operator acting on Young tableaux.
Its physical interpretation is provided by
the differential formulation of the state flow
developed in Sec.~\ref{sec:gsb-interaction-state-flow}.

According to Eq.~(\ref{eq:gsb-state-flow}),
along a smooth, isolated eigenstate branch
and with the corresponding local phase
convention, we have
\begin{equation}
\frac{\partial|\psi_i\rangle}{\partial\delta_B}
=
\sum_{j\neq i}
|\psi_j\rangle
\frac{
\langle\psi_j|B_B|\psi_i\rangle
}{
M_i-M_j
} \, ,
\label{eq:young-state-flow}
\end{equation}
where
\begin{equation}
B_B
=
\frac{\partial{\cal H}}{\partial\delta_B}
\end{equation}
is the response operator associated with
the breaking coordinate $\delta_B$.
At degeneracies, the corresponding eigenspaces
must instead be treated using degenerate
perturbation theory.

Let $P_{R'}$ denote the projector onto a
flavor representation $R'$ in the fixed
reference space.
The projected state-flow response
\begin{equation}
P_{R'}
\frac{\partial|\psi_i\rangle}{\partial\delta_B}
\label{eq:young-projected-state-flow}
\end{equation}
then describes the component of the local
state flow along this representation direction.
While flavor group theory organizes the allowed
transition channels, their actual amplitudes
are determined by the response matrix elements
and the corresponding mass separations.
The Young-tableau construction below visualizes
such possible representation directions without
specifying their amplitudes.

As a simple illustration, consider the
$\Delta^+$ baryon at the level of its
flavor wave function.
Its fully symmetric reference configuration
can be represented schematically as
\begin{equation}
\,\young(uud)\,
\quad\widehat{=}\quad
\,\young({~}{~}{\bb})\, ,
\end{equation}
where the marked box identifies the $d$-flavor
direction relative to the reference $u$ flavors.
When the up--down flavor-breaking source
is switched off, the marking disappears,
\begin{equation}
\,\young({~}{~}{\bb})\,
\longrightarrow
\,\young({~}{~}{~})\, .
\end{equation}
This operation removes the graphical distinction
associated with the breaking source without
changing the flavor content of the baryon.

Starting from a fully symmetric flavor-decuplet
reference configuration, an isovector breaking
source may induce a response along a
mixed-symmetry flavor-octet direction,
as allowed by
\begin{equation}
\mathbf{10}\otimes\mathbf{8}
\supset
\mathbf{8} \, .
\end{equation}
The $SU(3)$ representation labels used here
refer to the parent flavor reference and do
not require exact flavor $SU(3)$ symmetry
along the isospin-breaking trajectory.
Restricting attention to these two
representation channels, we may write
schematically
\begin{equation}
|\psi_\Delta(\delta_B)\rangle
\quad\widehat{=}\quad
\,\young({~}{~}{~})\,
\oplus
\,\young({~}{~},{~})\, ,
\label{eq:young-Delta}
\end{equation}
where
$\mathbf{10}
\left(\scalebox{\sn}{\young({~}{~}{~})}\right)$
represents the leading reference component,
while
$\mathbf{8}
\left(\scalebox{\sn}{\young({~}{~},{~})}\right)$
represents a possible mixed-symmetry response
direction.
Here and below, $\widehat{=}$ denotes a
schematic graphical correspondence, while
$\oplus$ in Eq.~(\ref{eq:young-Delta})
indicates the representation channels
under consideration rather than a complete
decomposition of the physical state.

To clarify the physical meaning of this
graphical correspondence, consider a state
that is purely decuplet-like at the chosen
reference point,
\begin{equation}
|\psi_\Delta(\delta_B)\rangle
=
a(\delta_B)|\mathbf{10},\Delta\rangle
+\sum_\beta b_\beta(\delta_B)
|\mathbf{8},\beta\rangle
+\cdots \, ,
\end{equation}
where $\beta$ distinguishes different
compatible octet configurations.
At the reference point,
$a(0)=1$ and $b_\beta(0)=0$.
Along a smooth normalized eigenstate branch,
with the phase convention adopted above,
we have
\begin{eqnarray}
a(\delta_B)
&=&
1+{\cal O}(\delta_B^2) \, ,
\nonumber\\
b_\beta(\delta_B)
&=&
\kappa_\beta\delta_B
+{\cal O}(\delta_B^2) \, .
\end{eqnarray}
Consequently, the local state-flow response
projected onto the octet reference space is
\begin{equation}
\left.
P_{\mathbf{8}}
\frac{\partial|\psi_\Delta\rangle}
{\partial\delta_B}
\right|_{\delta_B=0}
=
\sum_\beta
\kappa_\beta|\mathbf{8},\beta\rangle \, .
\label{eq:young-Delta-state-flow}
\end{equation}
The coefficients $\kappa_\beta$ are determined
by the corresponding Hamiltonian response
matrix elements and mass separations,
as described by Eq.~(\ref{eq:young-state-flow}).

The possible octet direction of this projected
state-flow response can therefore be represented
graphically as
\begin{equation}
\left.
\frac{
{\cal D}
\left(
\,\young({~}{~}{\bb})\,
\right)
}{
{\cal D}
\left(
\,\young({\bb})\,
\right)
}
\right|_{\scalebox{\sn}{\young({~}{~}{~})}}
\quad\widehat{=}\quad
\,\young({~}{~},{~})\, .
\label{eq:young-Delta-differential}
\end{equation}
The Young tableau on the right-hand side
identifies a possible representation component
of the local tangent to the state-flow
trajectory.
It does not represent the complete state
derivative or imply that the corresponding
amplitude must be nonzero.
In particular, the allowed flavor transition
channels are subject to additional constraints
from spin, parity, and the Pauli principle.
The ordinary ground-state spin-$1/2$ octet
cannot mix directly with the spin-$3/2$
$\Delta$ through a rotationally invariant
mass Hamiltonian.
Eq.~(\ref{eq:young-Delta-differential})
thus refers to compatible flavor-octet
configurations within an appropriately enlarged
reference space.

A similar graphical construction can be
applied to the neutral vector-meson sector.
The flavor-$SU(4)$ adjoint representation
$\mathbf{15}$ has the Young shape $[2,1,1]$,
while a column of four boxes represents
a singlet of $SU(4)$.
Since
\begin{equation}
\mathbf{15}\otimes\mathbf{15}
\supset
\mathbf{1} \, ,
\end{equation}
an adjoint breaking source may induce
a singlet response from a compatible
adjoint reference configuration.
For example, the charm-breaking source
can connect the light-flavor singlet component
of the adjoint representation to the
$SU(4)$ singlet.
This response direction may be represented
schematically as
\begin{equation}
\left.
\frac{
{\cal D}
\left(
\,\young({~}{\bb},{~},{~})\,
\right)
}{
{\cal D}
\left(
\,\young({\bb})\,
\right)
}
\right|_{\scalebox{\sn}{\young({~}{~},{~},{~})}}
\quad\widehat{=}\quad
\,\young({~},{~},{~},{~})\, ,
\label{eq:young-meson-differential}
\end{equation}
corresponding to the
$\mathbf{15}$--$\mathbf{1}$ mixing channel.
Here the four-box Young tableaux represent
the corresponding $SU(4)$ flavor
representations and do not count the
constituents of the meson.
The actual mixing is governed by the
vector-meson mass Hamiltonian discussed
in Sec.~\ref{sec:meson-ideal-mixing}.

A more instructive example is provided by
singly charmed baryons in flavor $SU(4)_{udsc}$,
with the charm flavor progressively distinguished
from three degenerate light flavors.
At the level of the parent flavor group,
the relevant transition channels are allowed by
\begin{equation}
\mathbf{20_M}\otimes\mathbf{15}
\supset
\mathbf{20_S}
\oplus
\overline{\mathbf{4}}_A \, .
\end{equation}
The corresponding flavor transition structures
are given in Appendix~\ref{app:baryon-S-matrices}.
When the charm-breaking source preserves
the light-flavor $SU(3)$ symmetry, the
corresponding response connects configurations
with the same light-flavor representation.
In the singly charmed sector, the
$\mathbf{20_M}$ representation contains
both $\mathbf{6}_F$ and
$\overline{\mathbf{3}}_F$ components,
whereas the $\mathbf{20_S}$ and
$\overline{\mathbf{4}}_A$ representations
contain the corresponding $\mathbf{6}_F$
and $\overline{\mathbf{3}}_F$ components,
respectively.

For the $\Sigma_c$-type flavor configuration,
the two light quarks belong to the symmetric
light-flavor representation $\mathbf{6}_F$.
The corresponding fully symmetric parent-group
response direction may be represented
schematically as
\begin{equation}
\left.
\frac{
{\cal D}
\left(
\,\young({~}{~},{\bb})\,
\right)
}{
{\cal D}
\left(
\,\young({\bb})\,
\right)
}
\right|_{\scalebox{\sn}{\young({~}{~},{~})}}
\quad\widehat{=}\quad
\,\young({~}{~}{~})\, ,
\label{eq:young-Sigma-differential}
\end{equation}
corresponding to the
$\mathbf{20_M}$--$\mathbf{20_S}$
flavor transition channel.

Similarly, for the $\Lambda_c$-type flavor
configuration, the two light quarks belong
to the antisymmetric light-flavor
representation $\overline{\mathbf{3}}_F$.
The corresponding fully antisymmetric
parent-group response direction may be
represented as
\begin{equation}
\left.
\frac{
{\cal D}
\left(
\,\young({~}{\bb},{~})\,
\right)
}{
{\cal D}
\left(
\,\young({\bb})\,
\right)
}
\right|_{\scalebox{\sn}{\young({~}{~},{~})}}
\quad\widehat{=}\quad
\,\young({~},{~},{~})\, ,
\label{eq:young-Lambda-differential}
\end{equation}
corresponding to the
$\mathbf{20_M}$--$\overline{\mathbf{4}}_A$
flavor transition channel.
These two examples illustrate how different
light-flavor configurations within the same
parent representation may access different
representation directions under the same
symmetry-breaking source.
The actual realization of these channels
depends on the complete spin, spatial,
and color configurations of the baryons,
as well as on the additional dynamical
and symmetry constraints discussed in
Sec.~\ref{sec:shared-dynamics}.

Young tableaux are therefore used here as
probes of symmetry-breaking directions.
The marked boxes indicate which
symmetry-breaking coordinate is varied,
while the resulting Young shapes visualize
possible representation components of the
local state-flow response.
The graphical differential relations are
thus understood as schematic descriptions
of possible tangent directions along a
symmetry-breaking trajectory in the space
of physical states.
The actual evolution along these directions
is governed by the response matrix elements
and the corresponding mass separations.
Moreover, as the symmetry-breaking background
changes, the response operators themselves
may evolve and unlock additional independent
response structures, as discussed in
Sec.~\ref{sec:group-expansion}.
The Young-tableau picture therefore provides
a complementary visualization of the state
flow, while the effective Hamiltonian and
its response operators remain the primary
dynamical objects of the group expansion.

\bibliographystyle{elsarticle-num}
\bibliography{ref}

\begin{thebibliography}{100}
\expandafter\ifx\csname url\endcsname\relax
  \def\url#1{\texttt{#1}}\fi
\expandafter\ifx\csname urlprefix\endcsname\relax\def\urlprefix{URL }\fi
\expandafter\ifx\csname href\endcsname\relax
  \def\href#1#2{#2} \def\path#1{#1}\fi

\bibitem{Gell-Mann:1964ewy}
M.~Gell-Mann, {A schematic model of baryons and mesons}, Phys. Lett. 8 (1964)
  214--215.
\newblock \href {https://doi.org/10.1016/S0031-9163(64)92001-3}
  {\path{doi:10.1016/S0031-9163(64)92001-3}}.

\bibitem{Zweig:1964ruk}
G.~Zweig, {An $SU(3)$ model for strong interaction symmetry and its breaking.
  Version 1} (1 1964).

\bibitem{pdg}
S.~Navas, et~al., {Review of particle physics}, Phys. Rev. D 110~(3) (2024)
  030001.
\newblock \href {https://doi.org/10.1103/PhysRevD.110.030001}
  {\path{doi:10.1103/PhysRevD.110.030001}}.

\bibitem{Cheng:1996jr}
H.-Y. Cheng, {Status of the proton spin problem}, Int. J. Mod. Phys. A 11
  (1996) 5109--5182.
\newblock \href {http://arxiv.org/abs/hep-ph/9607254}
  {\path{arXiv:hep-ph/9607254}}, \href
  {https://doi.org/10.1142/S0217751X96002364}
  {\path{doi:10.1142/S0217751X96002364}}.

\bibitem{Filippone:2001ux}
B.~W. Filippone, X.-D. Ji, {The Spin Structure of the Nucleon}, Adv. Nucl.
  Phys. 26 (2001) 1.
\newblock \href {http://arxiv.org/abs/hep-ph/0101224}
  {\path{arXiv:hep-ph/0101224}}, \href
  {https://doi.org/10.1007/0-306-47915-X_1}
  {\path{doi:10.1007/0-306-47915-X_1}}.

\bibitem{Bass:2004xa}
S.~D. Bass, {The spin structure of the proton}, Rev. Mod. Phys. 77 (2005)
  1257--1302.
\newblock \href {http://arxiv.org/abs/hep-ph/0411005}
  {\path{arXiv:hep-ph/0411005}}, \href
  {https://doi.org/10.1103/RevModPhys.77.1257}
  {\path{doi:10.1103/RevModPhys.77.1257}}.

\bibitem{Aidala:2012mv}
C.~A. Aidala, S.~D. Bass, D.~Hasch, G.~K. Mallot, {The spin structure of the
  nucleon}, Rev. Mod. Phys. 85 (2013) 655--691.
\newblock \href {http://arxiv.org/abs/1209.2803} {\path{arXiv:1209.2803}},
  \href {https://doi.org/10.1103/RevModPhys.85.655}
  {\path{doi:10.1103/RevModPhys.85.655}}.

\bibitem{Leader:2013jra}
E.~Leader, C.~Lorc\'e, {The angular momentum controversy:
  What\textquoteright{}s it all about and does it matter?}, Phys. Rept. 541~(3)
  (2014) 163--248.
\newblock \href {http://arxiv.org/abs/1309.4235} {\path{arXiv:1309.4235}},
  \href {https://doi.org/10.1016/j.physrep.2014.02.010}
  {\path{doi:10.1016/j.physrep.2014.02.010}}.

\bibitem{Ji:2016djn}
X.~Ji, {Proton Tomography Through Deeply Virtual Compton Scattering}, Natl.
  Sci. Rev. 4~(2) (2017) 213--223.
\newblock \href {http://arxiv.org/abs/1605.01114} {\path{arXiv:1605.01114}},
  \href {https://doi.org/10.1093/nsr/nwx024} {\path{doi:10.1093/nsr/nwx024}}.

\bibitem{Deur:2018roz}
A.~Deur, S.~J. Brodsky, G.~F. De~T\'eramond, {The spin structure of the
  nucleon}, Rept. Prog. Phys. 82~(7) (2019) 076201.
\newblock \href {http://arxiv.org/abs/1807.05250} {\path{arXiv:1807.05250}},
  \href {https://doi.org/10.1088/1361-6633/ab0b8f}
  {\path{doi:10.1088/1361-6633/ab0b8f}}.

\bibitem{Ji:2020ena}
X.~Ji, F.~Yuan, Y.~Zhao, {What we know and what we don\textquoteright{}t know
  about the proton spin after 30 years}, Nature Rev. Phys. 3~(1) (2021) 27--38.
\newblock \href {http://arxiv.org/abs/2009.01291} {\path{arXiv:2009.01291}},
  \href {https://doi.org/10.1038/s42254-020-00248-4}
  {\path{doi:10.1038/s42254-020-00248-4}}.

\bibitem{Liu:2021lke}
K.-F. Liu, {Status on lattice calculations of the proton spin decomposition},
  AAPPS Bull. 32~(1) (2022) 8.
\newblock \href {http://arxiv.org/abs/2112.08416} {\path{arXiv:2112.08416}},
  \href {https://doi.org/10.1007/s43673-022-00037-4}
  {\path{doi:10.1007/s43673-022-00037-4}}.

\bibitem{EuropeanMuon:1987isl}
J.~Ashman, et~al., {A measurement of the spin asymmetry and determination of
  the structure function $g_1$ in deep inelastic muon-proton scattering}, Phys.
  Lett. B 206 (1988) 364.
\newblock \href {https://doi.org/10.1016/0370-2693(88)91523-7}
  {\path{doi:10.1016/0370-2693(88)91523-7}}.

\bibitem{EuropeanMuon:1989yki}
J.~Ashman, et~al., {An investigation of the spin structure of the proton in
  deep inelastic scattering of polarised muons on polarised protons}, Nucl.
  Phys. B 328 (1989) 1.
\newblock \href {https://doi.org/10.1016/0550-3213(89)90089-8}
  {\path{doi:10.1016/0550-3213(89)90089-8}}.

\bibitem{Hughes:1983kf}
V.~W. Hughes, J.~Kuti, {Internal Spin Structure of the Nucleon}, Ann. Rev.
  Nucl. Part. Sci. 33 (1983) 611--644.
\newblock \href {https://doi.org/10.1146/annurev.ns.33.120183.003143}
  {\path{doi:10.1146/annurev.ns.33.120183.003143}}.

\bibitem{Zeppenfeld:1980ex}
D.~Zeppenfeld, {$SU(3)$ Relations for $B$ Meson Decays}, Z. Phys. C 8 (1981)
  77.
\newblock \href {https://doi.org/10.1007/BF01429835}
  {\path{doi:10.1007/BF01429835}}.

\bibitem{Gasser:1982ap}
J.~Gasser, H.~Leutwyler, {Quark Masses}, Phys. Rept. 87 (1982) 77--169.
\newblock \href {https://doi.org/10.1016/0370-1573(82)90035-7}
  {\path{doi:10.1016/0370-1573(82)90035-7}}.

\bibitem{Chau:1986jb}
L.~L. Chau, H.~Y. Cheng, {Quark Diagram Analysis of Two-body Charm Decays},
  Phys. Rev. Lett. 56 (1986) 1655--1658.
\newblock \href {https://doi.org/10.1103/PhysRevLett.56.1655}
  {\path{doi:10.1103/PhysRevLett.56.1655}}.

\bibitem{Li:1986iya}
B.-A. Li, M.-L. Yan, K.-F. Liu, {Electromagnetic mass differences of $N$ and
  $\Delta$ in the Skyrme model}, Phys. Lett. B 177 (1986) 409.
\newblock \href {https://doi.org/10.1016/0370-2693(86)90779-3}
  {\path{doi:10.1016/0370-2693(86)90779-3}}.

\bibitem{Chau:1987tk}
L.-L. Chau, H.-Y. Cheng, {Analysis of exclusive two-body decays of charm mesons
  using the quark diagram scheme}, Phys. Rev. D 36 (1987) 137, [Addendum: Phys.
  Rev. D 39, 2788--2791 (1989)].
\newblock \href {https://doi.org/10.1103/PhysRevD.39.2788}
  {\path{doi:10.1103/PhysRevD.39.2788}}.

\bibitem{Savage:1989ub}
M.~J. Savage, M.~B. Wise, {$SU(3)$ predictions for nonleptonic $B$-meson
  decays}, Phys. Rev. D 39 (1989) 3346, [Erratum: Phys. Rev. D 40, 3127
  (1989)].
\newblock \href {https://doi.org/10.1103/PhysRevD.39.3346}
  {\path{doi:10.1103/PhysRevD.39.3346}}.

\bibitem{Jain:1989kn}
P.~Jain, R.~Johnson, N.~W. Park, J.~Schechter, H.~Weigel, {Neutron-proton
  mass-splitting puzzle in Skyrme and chiral quark models}, Phys. Rev. D 40
  (1989) 855.
\newblock \href {https://doi.org/10.1103/PhysRevD.40.855}
  {\path{doi:10.1103/PhysRevD.40.855}}.

\bibitem{Morpurgo:1989ti}
G.~Morpurgo, {Field theory and the nonrelativistic quark model: A
  parametrization of the meson masses}, Phys. Rev. D 41 (1990) 2865.
\newblock \href {https://doi.org/10.1103/PhysRevD.41.2865}
  {\path{doi:10.1103/PhysRevD.41.2865}}.

\bibitem{Chau:1990ay}
L.-L. Chau, H.-Y. Cheng, W.~K. Sze, H.~Yao, B.~Tseng, {Charmless nonleptonic
  rare decays of $B$ mesons}, Phys. Rev. D 43 (1991) 2176--2192, [Erratum:
  Phys. Rev. D 58, 019902 (1998)].
\newblock \href {https://doi.org/10.1103/PhysRevD.43.2176}
  {\path{doi:10.1103/PhysRevD.43.2176}}.

\bibitem{Grozin:1992td}
A.~G. Grozin, O.~I. Yakovlev, {Baryonic currents and their correlators in the
  heavy quark effective theory}, Phys. Lett. B 285 (1992) 254--262.
\newblock \href {http://arxiv.org/abs/hep-ph/9908364}
  {\path{arXiv:hep-ph/9908364}}, \href
  {https://doi.org/10.1016/0370-2693(92)91462-I}
  {\path{doi:10.1016/0370-2693(92)91462-I}}.

\bibitem{Praszalowicz:1992gn}
M.~Praszalowicz, A.~Blotz, K.~Goeke, {Isospin baryon mass differences in
  $SU(3)$ Nambu-Jona-Lasinio model}, Phys. Rev. D 47 (1993) 1127--1133.
\newblock \href {https://doi.org/10.1103/PhysRevD.47.1127}
  {\path{doi:10.1103/PhysRevD.47.1127}}.

\bibitem{Adami:1993xz}
C.~Adami, E.~G. Drukarev, B.~L. Ioffe, {Isospin violation in QCD sum rules for
  baryons}, Phys. Rev. D 48 (1993) 2304, [Erratum: Phys. Rev. D 52, 4254
  (1995)].
\newblock \href {https://doi.org/10.1103/PhysRevD.48.2304}
  {\path{doi:10.1103/PhysRevD.48.2304}}.

\bibitem{Deshpande:1994ii}
N.~G. Deshpande, X.-G. He, {$CP$ Asymmetry Relations between $\bar B^0 \to \pi
  \pi$ and $\bar B^0 \to \pi K$ Rates}, Phys. Rev. Lett. 75 (1995) 1703--1706.
\newblock \href {http://arxiv.org/abs/hep-ph/9412393}
  {\path{arXiv:hep-ph/9412393}}, \href
  {https://doi.org/10.1103/PhysRevLett.75.1703}
  {\path{doi:10.1103/PhysRevLett.75.1703}}.

\bibitem{Gronau:1995hm}
M.~Gronau, O.~F. Hernandez, D.~London, J.~L. Rosner, {Broken $SU(3)$ symmetry
  in two-body $B$ decays}, Phys. Rev. D 52 (1995) 6356--6373.
\newblock \href {http://arxiv.org/abs/hep-ph/9504326}
  {\path{arXiv:hep-ph/9504326}}, \href
  {https://doi.org/10.1103/PhysRevD.52.6356}
  {\path{doi:10.1103/PhysRevD.52.6356}}.

\bibitem{Forkel:1996ty}
H.~Forkel, M.~Nielsen, {Isospin breaking and instantons in QCD nucleon sum
  rules}, Phys. Rev. D 55 (1997) 1471--1480.
\newblock \href {http://arxiv.org/abs/hep-ph/9611311}
  {\path{arXiv:hep-ph/9611311}}, \href
  {https://doi.org/10.1103/PhysRevD.55.1471}
  {\path{doi:10.1103/PhysRevD.55.1471}}.

\bibitem{Kim:1997ip}
H.-C. Kim, M.~Praszalowicz, K.~Goeke, {Magnetic moments of the $SU(3)$ decuplet
  baryons in the chiral quark-soliton model}, Phys. Rev. D 57 (1998)
  2859--2870.
\newblock \href {http://arxiv.org/abs/hep-ph/9706531}
  {\path{arXiv:hep-ph/9706531}}, \href
  {https://doi.org/10.1103/PhysRevD.57.2859}
  {\path{doi:10.1103/PhysRevD.57.2859}}.

\bibitem{Zhu:1998ai}
S.-L. Zhu, W.~Y.~P. Hwang, Z.-S. Yang, {Possible $\Sigma^0-\Lambda$ mixing in
  QCD sum rule}, Phys. Rev. D 57 (1998) 1524--1526.
\newblock \href {http://arxiv.org/abs/hep-ph/9802321}
  {\path{arXiv:hep-ph/9802321}}, \href
  {https://doi.org/10.1103/PhysRevD.57.1524}
  {\path{doi:10.1103/PhysRevD.57.1524}}.

\bibitem{He:1998rq}
X.-G. He, {$SU(3)$ analysis of annihilation contributions and $CP$ violating
  relations in $B \to P P$ decays}, Eur. Phys. J. C 9 (1999) 443--448.
\newblock \href {http://arxiv.org/abs/hep-ph/9810397}
  {\path{arXiv:hep-ph/9810397}}, \href {https://doi.org/10.1007/s100529900064}
  {\path{doi:10.1007/s100529900064}}.

\bibitem{Yagisawa:2001gz}
N.~Yagisawa, T.~Hatsuda, A.~Hayashigaki, {In-medium $\Sigma^0-\Lambda$ mixing
  in QCD sum rules}, Nucl. Phys. A 699 (2002) 665--689.
\newblock \href {http://arxiv.org/abs/hep-ph/0107023}
  {\path{arXiv:hep-ph/0107023}}, \href
  {https://doi.org/10.1016/S0375-9474(01)01293-3}
  {\path{doi:10.1016/S0375-9474(01)01293-3}}.

\bibitem{Durand:2005tb}
L.~Durand, P.~Ha, {Electromagnetic corrections to baryon masses}, Phys. Rev. D
  71 (2005) 073015, [Erratum: Phys. Rev. D 75, 039903 (2007)].
\newblock \href {http://arxiv.org/abs/hep-ph/0502090}
  {\path{arXiv:hep-ph/0502090}}, \href
  {https://doi.org/10.1103/PhysRevD.75.039903}
  {\path{doi:10.1103/PhysRevD.75.039903}}.

\bibitem{Aliev:2010ra}
T.~M. Aliev, A.~Ozpineci, V.~Zamiralov, {Mixing angle of hadrons in QCD: A new
  view}, Phys. Rev. D 83 (2011) 016008.
\newblock \href {http://arxiv.org/abs/1007.0814} {\path{arXiv:1007.0814}},
  \href {https://doi.org/10.1103/PhysRevD.83.016008}
  {\path{doi:10.1103/PhysRevD.83.016008}}.

\bibitem{Yang:2010fm}
G.-S. Yang, H.-C. Kim, {Mass splittings of $SU(3)$ baryons within a chiral
  soliton model}, Prog. Theor. Phys. 128 (2012) 397--413.
\newblock \href {http://arxiv.org/abs/1010.3792} {\path{arXiv:1010.3792}},
  \href {https://doi.org/10.1143/PTP.128.397} {\path{doi:10.1143/PTP.128.397}}.

\bibitem{Yang:2010id}
G.-S. Yang, H.-C. Kim, M.~V. Polyakov, {Electromagnetic mass differences of
  $SU(3)$ baryons within a chiral soliton model}, Phys. Lett. B 695 (2011)
  214--218.
\newblock \href {http://arxiv.org/abs/1009.5250} {\path{arXiv:1009.5250}},
  \href {https://doi.org/10.1016/j.physletb.2010.11.018}
  {\path{doi:10.1016/j.physletb.2010.11.018}}.

\bibitem{Shanahan:2012wa}
P.~E. Shanahan, A.~W. Thomas, R.~D. Young, {Strong contribution to octet baryon
  mass splittings}, Phys. Lett. B 718 (2013) 1148--1153.
\newblock \href {http://arxiv.org/abs/1209.1892} {\path{arXiv:1209.1892}},
  \href {https://doi.org/10.1016/j.physletb.2012.11.072}
  {\path{doi:10.1016/j.physletb.2012.11.072}}.

\bibitem{Walker-Loud:2012ift}
A.~Walker-Loud, C.~E. Carlson, G.~A. Miller, {The Electromagnetic Self-Energy
  Contribution to $M_p - M_n$ and the Isovector Nucleon
  MagneticPolarizability}, Phys. Rev. Lett. 108 (2012) 232301.
\newblock \href {http://arxiv.org/abs/1203.0254} {\path{arXiv:1203.0254}},
  \href {https://doi.org/10.1103/PhysRevLett.108.232301}
  {\path{doi:10.1103/PhysRevLett.108.232301}}.

\bibitem{Shanahan:2013xw}
P.~E. Shanahan, A.~W. Thomas, R.~D. Young, {Chiral expansion of moments of
  quark distributions}, Phys. Rev. D 87~(11) (2013) 114515.
\newblock \href {http://arxiv.org/abs/1301.6861} {\path{arXiv:1301.6861}},
  \href {https://doi.org/10.1103/PhysRevD.87.114515}
  {\path{doi:10.1103/PhysRevD.87.114515}}.

\bibitem{Cheng:2014rfa}
H.-Y. Cheng, C.-W. Chiang, A.-L. Kuo, {Updating $B \to PP,VP$ decays in the
  framework of flavor symmetry}, Phys. Rev. D 91~(1) (2015) 014011.
\newblock \href {http://arxiv.org/abs/1409.5026} {\path{arXiv:1409.5026}},
  \href {https://doi.org/10.1103/PhysRevD.91.014011}
  {\path{doi:10.1103/PhysRevD.91.014011}}.

\bibitem{Horsley:2014koa}
R.~Horsley, J.~Najjar, Y.~Nakamura, H.~Perlt, D.~Pleiter, P.~E.~L. Rakow,
  G.~Schierholz, A.~Schiller, H.~St\"uben, J.~M. Zanotti, {Lattice
  determination of $\Sigma-\Lambda$ mixing}, Phys. Rev. D 91~(7) (2015) 074512.
\newblock \href {http://arxiv.org/abs/1411.7665} {\path{arXiv:1411.7665}},
  \href {https://doi.org/10.1103/PhysRevD.91.074512}
  {\path{doi:10.1103/PhysRevD.91.074512}}.

\bibitem{Muller:2015lua}
S.~M\"uller, U.~Nierste, S.~Schacht, {Topological amplitudes in $D$ decays to
  two pseudoscalars: A global analysis with linear $SU(3)_F$ breaking}, Phys.
  Rev. D 92~(1) (2015) 014004.
\newblock \href {http://arxiv.org/abs/1503.06759} {\path{arXiv:1503.06759}},
  \href {https://doi.org/10.1103/PhysRevD.92.014004}
  {\path{doi:10.1103/PhysRevD.92.014004}}.

\bibitem{Lu:2016ogy}
C.-D. L\"u, W.~Wang, F.-S. Yu, {Test flavor $SU(3)$ symmetry in exclusive
  $\Lambda_c$ decays}, Phys. Rev. D 93~(5) (2016) 056008.
\newblock \href {http://arxiv.org/abs/1601.04241} {\path{arXiv:1601.04241}},
  \href {https://doi.org/10.1103/PhysRevD.93.056008}
  {\path{doi:10.1103/PhysRevD.93.056008}}.

\bibitem{Wang:2017azm}
W.~Wang, Z.-P. Xing, J.~Xu, {Weak decays of doubly heavy baryons: SU(3)
  analysis}, Eur. Phys. J. C 77~(11) (2017) 800.
\newblock \href {http://arxiv.org/abs/1707.06570} {\path{arXiv:1707.06570}},
  \href {https://doi.org/10.1140/epjc/s10052-017-5363-y}
  {\path{doi:10.1140/epjc/s10052-017-5363-y}}.

\bibitem{Shi:2017dto}
Y.-J. Shi, W.~Wang, Y.~Xing, J.~Xu, {Weak decays of doubly heavy baryons:
  multi-body decay channels}, Eur. Phys. J. C 78~(1) (2018) 56.
\newblock \href {http://arxiv.org/abs/1712.03830} {\path{arXiv:1712.03830}},
  \href {https://doi.org/10.1140/epjc/s10052-018-5532-7}
  {\path{doi:10.1140/epjc/s10052-018-5532-7}}.

\bibitem{He:2018php}
X.-G. He, W.~Wang, {Flavor $SU(3)$ topological diagram and irreducible
  representation amplitudes for heavy meson charmless hadronic decays: mismatch
  and equivalence}, Chin. Phys. C 42~(10) (2018) 103108.
\newblock \href {http://arxiv.org/abs/1803.04227} {\path{arXiv:1803.04227}},
  \href {https://doi.org/10.1088/1674-1137/42/10/103108}
  {\path{doi:10.1088/1674-1137/42/10/103108}}.

\bibitem{Geng:2020tlx}
C.~Q. Geng, C.-W. Liu, T.-H. Tsai, {Mixing effects of $\Sigma^0-\Lambda^0$ in
  $\Lambda_c^+$ decays}, Phys. Rev. D 101~(5) (2020) 054005.
\newblock \href {http://arxiv.org/abs/2002.09583} {\path{arXiv:2002.09583}},
  \href {https://doi.org/10.1103/PhysRevD.101.054005}
  {\path{doi:10.1103/PhysRevD.101.054005}}.

\bibitem{Matsui:2020wcc}
Y.~Matsui, {Mixing angle of $\Xi_Q-\Xi_Q^\prime$ in heavy quark effective
  theory}, Nucl. Phys. A 1008 (2021) 122139.
\newblock \href {http://arxiv.org/abs/2011.09653} {\path{arXiv:2011.09653}},
  \href {https://doi.org/10.1016/j.nuclphysa.2021.122139}
  {\path{doi:10.1016/j.nuclphysa.2021.122139}}.

\bibitem{Braghin:2020yri}
F.~L. Braghin, {Flavor-dependent $U(3)$ Nambu-Jona-Lasinio coupling constant},
  Phys. Rev. D 103~(9) (2021) 094028.
\newblock \href {http://arxiv.org/abs/2008.00346} {\path{arXiv:2008.00346}},
  \href {https://doi.org/10.1103/PhysRevD.103.094028}
  {\path{doi:10.1103/PhysRevD.103.094028}}.

\bibitem{Braghin:2021hmr}
F.~L. Braghin, {Strangeness content of the pion in the $U(3)$
  Nambu-Jona-Lasinio model}, J. Phys. G 49~(5) (2022) 055101, [Erratum:
  J.Phys.G 50, 119501 (2023)].
\newblock \href {http://arxiv.org/abs/2108.02748} {\path{arXiv:2108.02748}},
  \href {https://doi.org/10.1088/1361-6471/ac4d79}
  {\path{doi:10.1088/1361-6471/ac4d79}}.

\bibitem{He:2021qnc}
X.-G. He, F.~Huang, W.~Wang, Z.-P. Xing, {$SU(3)$ symmetry and its breaking
  effects in semileptonic heavy baryon decays}, Phys. Lett. B 823 (2021)
  136765.
\newblock \href {http://arxiv.org/abs/2110.04179} {\path{arXiv:2110.04179}},
  \href {https://doi.org/10.1016/j.physletb.2021.136765}
  {\path{doi:10.1016/j.physletb.2021.136765}}.

\bibitem{Geng:2022yxb}
C.-Q. Geng, X.-N. Jin, C.-W. Liu, {Resolving puzzle in $\Xi_c^0 \to \Xi^- e^+
  \nu_e$ with $\Xi_c-\Xi_c^\prime$ mixing spectrum within the quark model},
  Phys. Lett. B 838 (2023) 137736.
\newblock \href {http://arxiv.org/abs/2210.07211} {\path{arXiv:2210.07211}},
  \href {https://doi.org/10.1016/j.physletb.2023.137736}
  {\path{doi:10.1016/j.physletb.2023.137736}}.

\bibitem{Geng:2022xfz}
C.-Q. Geng, X.-N. Jin, C.-W. Liu, X.~Yu, A.-W. Zhou, {Semileptonic decays of
  doubly charmed baryons with $\Xi_c-\Xi_c^\prime$ mixing}, Phys. Lett. B 839
  (2023) 137831.
\newblock \href {http://arxiv.org/abs/2212.02971} {\path{arXiv:2212.02971}},
  \href {https://doi.org/10.1016/j.physletb.2023.137831}
  {\path{doi:10.1016/j.physletb.2023.137831}}.

\bibitem{Ke:2022gxm}
H.-W. Ke, X.-Q. Li, {Revisiting the transition $\Xi_{cc}^{++} \to \Xi_c^{(')+}$
  to understand the data from LHCb}, Phys. Rev. D 105~(9) (2022) 096011.
\newblock \href {http://arxiv.org/abs/2203.10352} {\path{arXiv:2203.10352}},
  \href {https://doi.org/10.1103/PhysRevD.105.096011}
  {\path{doi:10.1103/PhysRevD.105.096011}}.

\bibitem{Xing:2022phq}
Z.~P. Xing, Y.~j. Shi, {Novel method for searching for the
  $\Xi_c^{0/+}-\Xi_c^{'0/+}$ mixing effect in the angular distribution analysis
  of a four-body $\Xi_c^{0/+}$ decay}, Phys. Rev. D 107~(7) (2023) 074024.
\newblock \href {http://arxiv.org/abs/2212.09003} {\path{arXiv:2212.09003}},
  \href {https://doi.org/10.1103/PhysRevD.107.074024}
  {\path{doi:10.1103/PhysRevD.107.074024}}.

\bibitem{Deng:2023qaf}
Z.-F. Deng, Y.-J. Shi, W.~Wang, J.~Zeng, {QED contributions to
  $\Xi_c^+-\Xi_c^{'+}$ mixing}, Phys. Rev. D 109~(3) (2024) 036014.
\newblock \href {http://arxiv.org/abs/2309.16386} {\path{arXiv:2309.16386}},
  \href {https://doi.org/10.1103/PhysRevD.109.036014}
  {\path{doi:10.1103/PhysRevD.109.036014}}.

\bibitem{Sun:2023noo}
X.-Y. Sun, F.-W. Zhang, Y.-J. Shi, Z.-X. Zhao, {Revisiting
  $\Xi_{Q}-\Xi_{Q}^{\prime}$ mixing in QCD sum rules}, Eur. Phys. J. C 83~(10)
  (2023) 961.
\newblock \href {http://arxiv.org/abs/2305.08050} {\path{arXiv:2305.08050}},
  \href {https://doi.org/10.1140/epjc/s10052-023-12042-4}
  {\path{doi:10.1140/epjc/s10052-023-12042-4}}.

\bibitem{Liu:2023pwr}
H.~Liu, W.~Wang, Q.-A. Zhang, {Improved method to determine the
  $\Xi_c-\Xi_c^\prime$ mixing}, Phys. Rev. D 109~(3) (2024) 036037.
\newblock \href {http://arxiv.org/abs/2309.05432} {\path{arXiv:2309.05432}},
  \href {https://doi.org/10.1103/PhysRevD.109.036037}
  {\path{doi:10.1103/PhysRevD.109.036037}}.

\bibitem{Liu:2023feb}
H.~Liu, L.~Liu, P.~Sun, W.~Sun, J.-X. Tan, W.~Wang, Y.-B. Yang, Q.-A. Zhang,
  {$\Xi_c-\Xi_c^\prime$ mixing from lattice QCD}, Phys. Lett. B 841 (2023)
  137941.
\newblock \href {http://arxiv.org/abs/2303.17865} {\path{arXiv:2303.17865}},
  \href {https://doi.org/10.1016/j.physletb.2023.137941}
  {\path{doi:10.1016/j.physletb.2023.137941}}.

\bibitem{Ebrahim:1977iy}
A.~Ebrahim, {Nonlinear Lagrangian Model of Chiral $SU(4) \times SU(4)$ Symmetry
  Breaking for the Pseudoscalar Mesons}, Lett. Nuovo Cim. 19 (1977) 225--228.
\newblock \href {https://doi.org/10.1007/BF02746779}
  {\path{doi:10.1007/BF02746779}}.

\bibitem{Ebrahim:1977hb}
A.~Ebrahim, {Chiral $SU(4) \times SU(4)$ symmetry breaking and the Cabibbo
  angle}, Phys. Lett. B 69 (1977) 229--230.
\newblock \href {https://doi.org/10.1016/0370-2693(77)90650-5}
  {\path{doi:10.1016/0370-2693(77)90650-5}}.

\bibitem{Inose:1978qw}
M.~Inose, R.~Sugano, {Symmetry Breaking Pattern of $SU(4)$ and Baryon Masses},
  Prog. Theor. Phys. 59 (1978) 2171.
\newblock \href {https://doi.org/10.1143/PTP.59.2171}
  {\path{doi:10.1143/PTP.59.2171}}.

\bibitem{Hallock:1979ar}
H.~L. Hallock, S.~Oneda, {Remarks on Flavor Symmetry Breaking and $SU(4)$
  [$SU(3)$] 16-plet [Nonet] Boson Mass Formulas}, Phys. Rev. D 20 (1979) 2932.
\newblock \href {https://doi.org/10.1103/PhysRevD.20.2932}
  {\path{doi:10.1103/PhysRevD.20.2932}}.

\bibitem{Ebert:1979ba}
D.~Ebert, M.~K. Volkov, {$SU(4) \times SU(4)$ symmetry breaking with the
  cabibbo angle for a nonlinear hadron lagrangian (in Russian)}, Sov. J. Nucl.
  Phys. 31 (1980) 271.

\bibitem{Burkitt:1983ca}
A.~N. Burkitt, A.~Kenway, R.~D. Kenway, {A practical scheme for $SU(4)$ flavour
  symmetry breaking in lattice gauge theories}, Phys. Lett. B 128 (1983)
  83--85.
\newblock \href {https://doi.org/10.1016/0370-2693(83)90078-3}
  {\path{doi:10.1016/0370-2693(83)90078-3}}.

\bibitem{Suzuki:1992pa}
K.~Suzuki, H.~Toki, {Flavor $SU(4)$ baryon and meson masses in diquark quark
  model using the Pauli-Gursey symmetry}, Mod. Phys. Lett. A 7 (1992)
  2867--2875.
\newblock \href {https://doi.org/10.1142/S0217732392002238}
  {\path{doi:10.1142/S0217732392002238}}.

\bibitem{El-Bennich:2011tme}
B.~El-Bennich, G.~Krein, L.~Chang, C.~D. Roberts, D.~J. Wilson, {Flavor $SU(4)$
  breaking between effective couplings}, Phys. Rev. D 85 (2012) 031502.
\newblock \href {http://arxiv.org/abs/1111.3647} {\path{arXiv:1111.3647}},
  \href {https://doi.org/10.1103/PhysRevD.85.031502}
  {\path{doi:10.1103/PhysRevD.85.031502}}.

\bibitem{Fontoura:2017ujf}
C.~E. Fontoura, J.~Haidenbauer, G.~Krein, {$SU(4)$ flavor symmetry breaking in
  D-meson couplings to light hadrons}, Eur. Phys. J. A 53~(5) (2017) 92.
\newblock \href {http://arxiv.org/abs/1705.09408} {\path{arXiv:1705.09408}},
  \href {https://doi.org/10.1140/epja/i2017-12289-2}
  {\path{doi:10.1140/epja/i2017-12289-2}}.

\bibitem{Dillon:2009pf}
G.~Dillon, G.~Morpurgo, {The General QCD parametrization and the hierarchy of
  its parameters: Why some simple models of hadrons work so well}, Riv. Nuovo
  Cim. 33~(1) (2010) 1--55.
\newblock \href {http://arxiv.org/abs/0910.5326} {\path{arXiv:0910.5326}},
  \href {https://doi.org/10.1393/ncr/i2010-10051-4}
  {\path{doi:10.1393/ncr/i2010-10051-4}}.

\bibitem{Franklin:1975yu}
J.~Franklin, {Mass Relations for Charmed Baryons}, Phys. Rev. D 12 (1975) 2077.
\newblock \href {https://doi.org/10.1103/PhysRevD.12.2077}
  {\path{doi:10.1103/PhysRevD.12.2077}}.

\bibitem{Dashen:1994qi}
R.~F. Dashen, E.~E. Jenkins, A.~V. Manohar, {Spin flavor structure of large
  $N_c$ baryons}, Phys. Rev. D 51 (1995) 3697--3727.
\newblock \href {http://arxiv.org/abs/hep-ph/9411234}
  {\path{arXiv:hep-ph/9411234}}, \href
  {https://doi.org/10.1103/PhysRevD.51.3697}
  {\path{doi:10.1103/PhysRevD.51.3697}}.

\bibitem{Jenkins:1995td}
E.~E. Jenkins, R.~F. Lebed, {Baryon mass splittings in the $1/N_c$ expansion},
  Phys. Rev. D 52 (1995) 282--294.
\newblock \href {http://arxiv.org/abs/hep-ph/9502227}
  {\path{arXiv:hep-ph/9502227}}, \href
  {https://doi.org/10.1103/PhysRevD.52.282}
  {\path{doi:10.1103/PhysRevD.52.282}}.

\bibitem{Jenkins:1998wy}
E.~E. Jenkins, {Large $N_c$ baryons}, Ann. Rev. Nucl. Part. Sci. 48 (1998)
  81--119.
\newblock \href {http://arxiv.org/abs/hep-ph/9803349}
  {\path{arXiv:hep-ph/9803349}}, \href
  {https://doi.org/10.1146/annurev.nucl.48.1.81}
  {\path{doi:10.1146/annurev.nucl.48.1.81}}.

\bibitem{Weinberg:1969hw}
S.~Weinberg, {Algebraic realizations of chiral symmetry}, Phys. Rev. 177 (1969)
  2604--2620.
\newblock \href {https://doi.org/10.1103/PhysRev.177.2604}
  {\path{doi:10.1103/PhysRev.177.2604}}.

\bibitem{Ioffe:1981kw}
B.~L. Ioffe, {Calculation of baryon masses in quantum chromodynamics}, Nucl.
  Phys. B 188 (1981) 317--341, [Erratum: Nucl. Phys. B 191, 591--592 (1981)].
\newblock \href {https://doi.org/10.1016/0550-3213(81)90259-5}
  {\path{doi:10.1016/0550-3213(81)90259-5}}.

\bibitem{Chung:1981cc}
Y.~Chung, H.~G. Dosch, M.~Kremer, D.~Schall, {Baryon sum rules and chiral
  symmetry breaking}, Nucl. Phys. B 197 (1982) 55--75.
\newblock \href {https://doi.org/10.1016/0550-3213(82)90154-7}
  {\path{doi:10.1016/0550-3213(82)90154-7}}.

\bibitem{Espriu:1983hu}
D.~Espriu, P.~Pascual, R.~Tarrach, {Baryon masses and chiral symmetry
  breaking}, Nucl. Phys. B 214 (1983) 285--298.
\newblock \href {https://doi.org/10.1016/0550-3213(83)90663-6}
  {\path{doi:10.1016/0550-3213(83)90663-6}}.

\bibitem{Detar:1987kae}
C.~E. Detar, J.~B. Kogut, {Hadronic spectrum of the quark plasma}, Phys. Rev.
  Lett. 59 (1987) 399.
\newblock \href {https://doi.org/10.1103/PhysRevLett.59.399}
  {\path{doi:10.1103/PhysRevLett.59.399}}.

\bibitem{Detar:1988kn}
C.~E. Detar, T.~Kunihiro, {Linear sigma model with parity doubling}, Phys. Rev.
  D 39 (1989) 2805.
\newblock \href {https://doi.org/10.1103/PhysRevD.39.2805}
  {\path{doi:10.1103/PhysRevD.39.2805}}.

\bibitem{Weinberg:1990xn}
S.~Weinberg, {Mended symmetries}, Phys. Rev. Lett. 65 (1990) 1177--1180.
\newblock \href {https://doi.org/10.1103/PhysRevLett.65.1177}
  {\path{doi:10.1103/PhysRevLett.65.1177}}.

\bibitem{Nowak:1992um}
M.~A. Nowak, M.~Rho, I.~Zahed, {Chiral effective action with heavy quark
  symmetry}, Phys. Rev. D 48 (1993) 4370--4374.
\newblock \href {http://arxiv.org/abs/hep-ph/9209272}
  {\path{arXiv:hep-ph/9209272}}, \href
  {https://doi.org/10.1103/PhysRevD.48.4370}
  {\path{doi:10.1103/PhysRevD.48.4370}}.

\bibitem{Bardeen:1993ae}
W.~A. Bardeen, C.~T. Hill, {Chiral dynamics and heavy quark symmetry in a
  solvable toy field theoretic model}, Phys. Rev. D 49 (1994) 409--425.
\newblock \href {http://arxiv.org/abs/hep-ph/9304265}
  {\path{arXiv:hep-ph/9304265}}, \href
  {https://doi.org/10.1103/PhysRevD.49.409}
  {\path{doi:10.1103/PhysRevD.49.409}}.

\bibitem{Hatsuda:1994pi}
T.~Hatsuda, T.~Kunihiro, {QCD phenomenology based on a chiral effective
  Lagrangian}, Phys. Rept. 247 (1994) 221--367.
\newblock \href {http://arxiv.org/abs/hep-ph/9401310}
  {\path{arXiv:hep-ph/9401310}}, \href
  {https://doi.org/10.1016/0370-1573(94)90022-1}
  {\path{doi:10.1016/0370-1573(94)90022-1}}.

\bibitem{Leinweber:1994nm}
D.~B. Leinweber, {Nucleon properties from unconventional interpolating fields},
  Phys. Rev. D 51 (1995) 6383--6393.
\newblock \href {http://arxiv.org/abs/nucl-th/9406001}
  {\path{arXiv:nucl-th/9406001}}, \href
  {https://doi.org/10.1103/PhysRevD.51.6383}
  {\path{doi:10.1103/PhysRevD.51.6383}}.

\bibitem{Glozman:1995fu}
L.~Y. Glozman, D.~O. Riska, {The spectrum of the nucleons and the strange
  hyperons and chiral dynamics}, Phys. Rept. 268 (1996) 263--303.
\newblock \href {http://arxiv.org/abs/hep-ph/9505422}
  {\path{arXiv:hep-ph/9505422}}, \href
  {https://doi.org/10.1016/0370-1573(95)00062-3}
  {\path{doi:10.1016/0370-1573(95)00062-3}}.

\bibitem{Cohen:1996sb}
T.~D. Cohen, X.-D. Ji, {Chiral multiplets of hadron currents}, Phys. Rev. D 55
  (1997) 6870--6876.
\newblock \href {http://arxiv.org/abs/hep-ph/9612302}
  {\path{arXiv:hep-ph/9612302}}, \href
  {https://doi.org/10.1103/PhysRevD.55.6870}
  {\path{doi:10.1103/PhysRevD.55.6870}}.

\bibitem{Jido:2001nt}
D.~Jido, M.~Oka, A.~Hosaka, {Chiral symmetry of baryons}, Prog. Theor. Phys.
  106 (2001) 873--908.
\newblock \href {http://arxiv.org/abs/hep-ph/0110005}
  {\path{arXiv:hep-ph/0110005}}, \href {https://doi.org/10.1143/PTP.106.873}
  {\path{doi:10.1143/PTP.106.873}}.

\bibitem{Cohen:2002st}
T.~D. Cohen, L.~Y. Glozman, {Does one observe chiral symmetry restoration in
  baryon spectrum?}, Int. J. Mod. Phys. A 17 (2002) 1327--1354.
\newblock \href {http://arxiv.org/abs/hep-ph/0201242}
  {\path{arXiv:hep-ph/0201242}}, \href
  {https://doi.org/10.1142/S0217751X02009679}
  {\path{doi:10.1142/S0217751X02009679}}.

\bibitem{Harada:2003jx}
M.~Harada, K.~Yamawaki, {Hidden local symmetry at loop: A New perspective of
  composite gauge boson and chiral phase transition}, Phys. Rept. 381 (2003)
  1--233.
\newblock \href {http://arxiv.org/abs/hep-ph/0302103}
  {\path{arXiv:hep-ph/0302103}}, \href
  {https://doi.org/10.1016/S0370-1573(03)00139-X}
  {\path{doi:10.1016/S0370-1573(03)00139-X}}.

\bibitem{Nagata:2007di}
K.~Nagata, A.~Hosaka, V.~Dmitrasinovic, {Chiral properties of baryon
  interpolating fields}, Mod. Phys. Lett. A 23 (2008) 2381--2384.
\newblock \href {http://arxiv.org/abs/0705.1896} {\path{arXiv:0705.1896}},
  \href {https://doi.org/10.1142/S0217732308029423}
  {\path{doi:10.1142/S0217732308029423}}.

\bibitem{Gallas:2009qp}
S.~Gallas, F.~Giacosa, D.~H. Rischke, {Vacuum phenomenology of the chiral
  partner of the nucleon in a linear sigma model with vector mesons}, Phys.
  Rev. D 82 (2010) 014004.
\newblock \href {http://arxiv.org/abs/0907.5084} {\path{arXiv:0907.5084}},
  \href {https://doi.org/10.1103/PhysRevD.82.014004}
  {\path{doi:10.1103/PhysRevD.82.014004}}.

\bibitem{Dmitrasinovic:2012zz}
V.~Dmitrasinovic, {Chiral symmetry of heavy-light scalar mesons with $U_A(1)$
  symmetry breaking}, Phys. Rev. D 86 (2012) 016006.
\newblock \href {https://doi.org/10.1103/PhysRevD.86.016006}
  {\path{doi:10.1103/PhysRevD.86.016006}}.

\bibitem{Yoshida:2015tia}
T.~Yoshida, E.~Hiyama, A.~Hosaka, M.~Oka, K.~Sadato, {Spectrum of heavy baryons
  in the quark model}, Phys. Rev. D 92~(11) (2015) 114029.
\newblock \href {http://arxiv.org/abs/1510.01067} {\path{arXiv:1510.01067}},
  \href {https://doi.org/10.1103/PhysRevD.92.114029}
  {\path{doi:10.1103/PhysRevD.92.114029}}.

\bibitem{Aarts:2017rrl}
G.~Aarts, C.~Allton, D.~De~Boni, S.~Hands, B.~J\"ager, C.~Praki, J.-I.
  Skullerud, {Light baryons below and above the deconfinement transition:
  medium effects and parity doubling}, JHEP 06 (2017) 034.
\newblock \href {http://arxiv.org/abs/1703.09246} {\path{arXiv:1703.09246}},
  \href {https://doi.org/10.1007/JHEP06(2017)034}
  {\path{doi:10.1007/JHEP06(2017)034}}.

\bibitem{Ma:2017nik}
Y.-L. Ma, M.~Harada, {Chiral partner structure of doubly heavy baryons with
  heavy quark spin-flavor symmetry}, J. Phys. G 45~(7) (2018) 075006.
\newblock \href {http://arxiv.org/abs/1709.09746} {\path{arXiv:1709.09746}},
  \href {https://doi.org/10.1088/1361-6471/aac86e}
  {\path{doi:10.1088/1361-6471/aac86e}}.

\bibitem{Yamazaki:2018stk}
T.~Yamazaki, M.~Harada, {Chiral partner structure of light nucleons in an
  extended parity doublet model}, Phys. Rev. D 99~(3) (2019) 034012.
\newblock \href {http://arxiv.org/abs/1809.02359} {\path{arXiv:1809.02359}},
  \href {https://doi.org/10.1103/PhysRevD.99.034012}
  {\path{doi:10.1103/PhysRevD.99.034012}}.

\bibitem{Kawakami:2019hpp}
Y.~Kawakami, M.~Harada, {Singly heavy baryons with chiral partner structure in
  a three-flavor chiral model}, Phys. Rev. D 99~(9) (2019) 094016.
\newblock \href {http://arxiv.org/abs/1902.06774} {\path{arXiv:1902.06774}},
  \href {https://doi.org/10.1103/PhysRevD.99.094016}
  {\path{doi:10.1103/PhysRevD.99.094016}}.

\bibitem{Suenaga:2021qri}
D.~Suenaga, A.~Hosaka, {Novel pentaquark picture for singly heavy baryons from
  chiral symmetry}, Phys. Rev. D 104~(3) (2021) 034009.
\newblock \href {http://arxiv.org/abs/2101.09764} {\path{arXiv:2101.09764}},
  \href {https://doi.org/10.1103/PhysRevD.104.034009}
  {\path{doi:10.1103/PhysRevD.104.034009}}.

\bibitem{Isgur:1989vq}
N.~Isgur, M.~B. Wise, {Weak decays of heavy mesons in the static quark
  approximation}, Phys. Lett. B 232 (1989) 113--117.
\newblock \href {https://doi.org/10.1016/0370-2693(89)90566-2}
  {\path{doi:10.1016/0370-2693(89)90566-2}}.

\bibitem{Georgi:1990um}
H.~Georgi, {An effective field theory for heavy quarks at low energies}, Phys.
  Lett. B 240 (1990) 447--450.
\newblock \href {https://doi.org/10.1016/0370-2693(90)91128-X}
  {\path{doi:10.1016/0370-2693(90)91128-X}}.

\bibitem{Isgur:1991wq}
N.~Isgur, M.~B. Wise, {Spectroscopy with heavy-quark symmetry}, Phys. Rev.
  Lett. 66 (1991) 1130--1133.
\newblock \href {https://doi.org/10.1103/PhysRevLett.66.1130}
  {\path{doi:10.1103/PhysRevLett.66.1130}}.

\bibitem{Neubert:1993mb}
M.~Neubert, {Heavy-quark symmetry}, Phys. Rept. 245 (1994) 259--396.
\newblock \href {http://arxiv.org/abs/hep-ph/9306320}
  {\path{arXiv:hep-ph/9306320}}, \href
  {https://doi.org/10.1016/0370-1573(94)90091-4}
  {\path{doi:10.1016/0370-1573(94)90091-4}}.

\bibitem{Korner:1994nh}
J.~G. Korner, M.~Kramer, D.~Pirjol, {Heavy baryons}, Prog. Part. Nucl. Phys. 33
  (1994) 787--868.
\newblock \href {http://arxiv.org/abs/hep-ph/9406359}
  {\path{arXiv:hep-ph/9406359}}, \href
  {https://doi.org/10.1016/0146-6410(94)90053-1}
  {\path{doi:10.1016/0146-6410(94)90053-1}}.

\bibitem{Jenkins:1996de}
E.~E. Jenkins, {Heavy baryon masses in the $1/m_Q$ and $1/N_c$ expansions},
  Phys. Rev. D 54 (1996) 4515--4531.
\newblock \href {http://arxiv.org/abs/hep-ph/9603449}
  {\path{arXiv:hep-ph/9603449}}, \href
  {https://doi.org/10.1103/PhysRevD.54.4515}
  {\path{doi:10.1103/PhysRevD.54.4515}}.

\bibitem{Chen:2008qv}
H.-X. Chen, V.~Dmitrasinovic, A.~Hosaka, K.~Nagata, S.-L. Zhu, {Chiral
  properties of baryon fields with flavor $SU(3)$ symmetry}, Phys. Rev. D 78
  (2008) 054021.
\newblock \href {http://arxiv.org/abs/0806.1997} {\path{arXiv:0806.1997}},
  \href {https://doi.org/10.1103/PhysRevD.78.054021}
  {\path{doi:10.1103/PhysRevD.78.054021}}.

\bibitem{Chen:2009sf}
H.-X. Chen, V.~Dmitrasinovic, A.~Hosaka, {Baryon fields with $U_L(3) \times
  U_R(3)$ chiral symmetry: Axial currents of nucleons and hyperons}, Phys. Rev.
  D 81 (2010) 054002.
\newblock \href {http://arxiv.org/abs/0912.4338} {\path{arXiv:0912.4338}},
  \href {https://doi.org/10.1103/PhysRevD.81.054002}
  {\path{doi:10.1103/PhysRevD.81.054002}}.

\bibitem{Chen:2010ba}
H.-X. Chen, V.~Dmitrasinovic, A.~Hosaka, {Baryon fields with $U_L(3) \times
  U_R(3)$ chiral symmetry. III. Interactions with chiral $[(3,\bar{3}) \oplus
  (\bar{3},3)]$ spinless mesons}, Phys. Rev. D 83 (2011) 014015.
\newblock \href {http://arxiv.org/abs/1009.2422} {\path{arXiv:1009.2422}},
  \href {https://doi.org/10.1103/PhysRevD.83.014015}
  {\path{doi:10.1103/PhysRevD.83.014015}}.

\bibitem{Chen:2011rh}
H.-X. Chen, V.~Dmitrasinovic, A.~Hosaka, {Baryon fields with $U_L(3) \times
  U_R(3)$ chiral symmetry. IV. Interactions with chiral $(8,1) \oplus (1,8)$
  vector and axial-vector mesons and anomalous magnetic moments}, Phys. Rev. C
  85 (2012) 055205.
\newblock \href {http://arxiv.org/abs/1109.3130} {\path{arXiv:1109.3130}},
  \href {https://doi.org/10.1103/PhysRevC.85.055205}
  {\path{doi:10.1103/PhysRevC.85.055205}}.

\bibitem{Chen:2013aga}
H.-X. Chen, {Isospin Symmetry Breaking and Octet Baryon Masses due to Their
  Mixing with Decuplet Baryons} (12 2013).
\newblock \href {http://arxiv.org/abs/1312.1451} {\path{arXiv:1312.1451}}.

\bibitem{Dmitrasinovic:2016hup}
V.~Dmitra\v{s}inovi\'c, H.-X. Chen, A.~Hosaka, {Baryon fields with $U_L(3)
  \times U_R(3)$ chiral symmetry. V. Pion-nucleon and kaon-nucleon $\Sigma$
  terms}, Phys. Rev. C 93~(6) (2016) 065208.
\newblock \href {http://arxiv.org/abs/1812.03414} {\path{arXiv:1812.03414}},
  \href {https://doi.org/10.1103/PhysRevC.93.065208}
  {\path{doi:10.1103/PhysRevC.93.065208}}.

\bibitem{Dmitrasinovic:2020wye}
V.~Dmitra\v{s}inovi\'c, H.-X. Chen, {Chiral $SU_L(3) \times SU_R(3)$ symmetry
  of baryons with one charmed quark}, Phys. Rev. D 101~(11) (2020) 114016.
\newblock \href {https://doi.org/10.1103/PhysRevD.101.114016}
  {\path{doi:10.1103/PhysRevD.101.114016}}.

\bibitem{ATLAS:2026ubk}
G.~Aad, et~al., {Observation of a $B_c^{*+}$ Meson with the ATLAS Detector},
  Phys. Rev. Lett. 137~(14) (2026) 141901.
\newblock \href {http://arxiv.org/abs/2605.16228} {\path{arXiv:2605.16228}},
  \href {https://doi.org/10.1103/vrxp-zkbs} {\path{doi:10.1103/vrxp-zkbs}}.

\bibitem{LHCb:2019epo}
R.~Aaij, et~al., {Precision measurement of the $\Xi_{cc}^{++}$ mass}, JHEP 02
  (2020) 049.
\newblock \href {http://arxiv.org/abs/1911.08594} {\path{arXiv:1911.08594}},
  \href {https://doi.org/10.1007/JHEP02(2020)049}
  {\path{doi:10.1007/JHEP02(2020)049}}.

\bibitem{LHCb:2026pxn}
R.~Aaij, et~al., {Observation of the Doubly Charmed Baryon $\Xi_{cc}^+$ with
  the LHCb Run 3 Detector}, Phys. Rev. Lett. 137~(2) (2026) 021902.
\newblock \href {http://arxiv.org/abs/2603.28456} {\path{arXiv:2603.28456}},
  \href {https://doi.org/10.1103/dmv6-7gdv} {\path{doi:10.1103/dmv6-7gdv}}.

\bibitem{LHCb:2026lnj}
R.~Aaij, et~al., {Observation of the doubly charmed baryon $\varOmega^+_{cc}$}
  (9 2026).
\newblock \href {http://arxiv.org/abs/2609.21921} {\path{arXiv:2609.21921}}.

\end{thebibliography}

\end{document}